\documentclass[namedreferences]{solarphysics}

\usepackage[hyperref,optionalrh,showbiblabels]{spr-sola-addons} 
\usepackage{graphicx}        
\usepackage{amssymb}        
\usepackage{color}           
\usepackage{breakurl}        

\chardef\us=`\_

\begin{document}

\begin{article}
\begin{opening}

{\title{Properties and Energetics of Magnetic Reconnection: I. Evolution of Flare Ribbons}}

\runningtitle{Properties and Energetics of Magnetic Reconnection}

\author[addressref=aff1]{\inits{}\fnm{J.}~\lnm{Qiu}}
\author[addressref=aff2]{\inits{}\fnm{J.}~\lnm{Cheng}}
   
\address[id=aff1]{Department of Physics, Montana State University, Bozeman, MT, 59717, USA (email: qiu@montana.edu)}
\address[id=aff2]{CAS Key Laboratory of Planetary Sciences, Shanghai Astronomical Observatory, Chinese Academy of Sciences, Shanghai 200030, China}

\runningauthor{J. Qiu \& J. Cheng}

\begin{abstract}

In this article, we measure the mean magnetic shear from the morphological evolution of flare ribbons, and examine the evolution of flare thermal and non-thermal X-ray emissions during the progress of flare reconnection. We analyze three eruptive flares and three confined flares ranging from GOES class C8.0 to M7.0. They exhibit well-defined two ribbons along the magnetic polarity inversion line (PIL), and have been observed by the Atmospheric Imaging Assembly and the Ramaty High Energy Solar Spectroscopic Imager from the onset of the flare throughout the impulsive phase. The analysis confirms the strong-to-weak shear evolution in the core region of the flare, and the flare hard X-ray emission rises as the shear decreases. In eruptive flares in this sample, significant non-thermal hard X-ray emission lags the ultraviolet emission from flare ribbons, and rises rapidly when the shear is modest. In all flares, we observe that the plasma temperature rises in the early phase when the flare ribbons rapidly spread along the PIL and the shear is high. We compare these results with prior studies, and discuss their implications, as well as complications, related to physical mechanisms governing energy partition during flare reconnection.

\end{abstract}


\end{opening}

\section{INTRODUCTION}
\label{sec:intro}

Whereas flare energy release is governed by fast magnetic reconnection, of order 10 - 10000 V m$^{-1}$ in terms of the reconnection electric field \citep[see][and references therein]{Hinterreiter2018}, it is not clear what mechanism governs the efficiency of conversion of this energy into non-thermal electrons, plasma heat, or bulk kinetic energy. In the effort to answer this question, past observational studies have explored the temporal and/or spatial correlation between hard X-ray (HXR) and/or microwave emissions and reconnection rate in solar flares, following the early attempt by \citet{Poletto1986}. Flare HXR and microwave emissions are produced by non-thermal electrons interacting with plasmas and magnetic field \citep{Benz2017}, and the reconnection rate has been frequently inferred from imaging observations of the flare emission in the lower atmosphere, using the recipe provided by \citet{Forbes1984}. The rate of magnetic reconnection is characterized by the reconnection electric field $E = v_{in}B_{in}$, where $B_{in}$ is the magnetic field brought into the reconnection current sheet in the corona with the inflow speed $v_{in}$. Lacking direct measurements of magnetic field in the corona, $E$ can be estimated by $E \approx v_l B_l$, $v_l$ being the apparent motion speed of flare ribbons or kernels in the lower atmosphere, and $B_l$ the magnetic field there \citep{Forbes1984, Poletto1986}. The path integral of this electric field along the current sheet yields the total flux change rate $\dot{\psi}$, which, in practice, is equated by the time rate of the magnetic flux $\psi (t)$ swept up by newly brightened flare ribbons $\psi(t) = \int B_ldA_l$ \citep{Fletcher2001, Qiu2004, Saba2006, Miklenic2007, Kazachenko2017}. In the following text, we will refer to $\dot{\psi}$ as the global or total reconnection rate, and the reconnection electric field $E$ as the local or mean reconnection rate.

When inferring $E$ from the lower atmosphere observations, $B_l$ is usually measured as the vertical or longitudinal magnetic field in the photosphere, with a few exceptions that sometimes extrapolate the field to the chromosphere \citep[e.g.][]{Qiu2007}. The apparent motion speed $v_l$ is measured in a variety of ways, either as the mean speed $v_l \approx v_{\perp}$ of the perpendicular expansion of flare ribbon fronts or leading edges observed in optical or UV wavelengths \citep{Poletto1986, Qiu2004, Asai2004, Xu2004, Isobe2005, Jing2005, Saba2006, Temmer2007, Miklenic2007, Jing2008, Liu2009a, Qiu2010, Qiu2017, Wang2017a, Hinterreiter2018}, or as the proper motion speed $v_l \approx v_k$ of the brightest or prominent kernels observed in infrared, optical, UV, or hard X-ray wavelengths, regardless of the direction of the motion \citep{Qiu2002, Krucker2003, Fletcher2004, Krucker2005, Lee2006, Liu2009b, Yang2011, Inglis2013}. The interpretation of $v_l$ measured in different ways much depends on the magnetic configuration presumed or invoked for the given cases in the study \citep[see the discussion in][]{Fletcher2009}. These past studies have shown, with varying degrees of success, that HXR (or microwave) emissions are temporally, and sometimes spatially, correlated with the global or local reconnection rate. On the other hand, these studies did not verify a one-to-one coincidence between significant HXR emission and enhanced reconnection rate, however it is measured \citep[e.g. see the discussion in ][]{Fletcher2002, Miklenic2007, Inglis2013, Qiu2021}. In particular, a recent case study by \citet{Naus2021}, using high-resolution observations of flare ribbons by the Interface Region Imaging Spectrograph \citep[IRIS:][]{DePontieu2014}, has characterized in great detail the bursty nature of flare reconnection. The analysis has identified locations (of size 600 - 1200~km) on flare ribbons, where the width of the newly brightened ribbon is enhanced, and EUV emissions at these locations are correlated with the $\ge$25 keV HXR lightcurve. The study has also revealed enhanced ribbon front width at other locations a few minutes before the onset of the HXR emission. The ribbon front width may be translated into the local reconnection rate, and the findings by \citet{Naus2021} therefore suggest that the production of non-thermal electrons is strongly connected with, yet not solely dependent on, the locally enhanced reconnection rate. 

The mean or local reconnection rate is inferred with a 2D approximation following the proposal by \citet{Forbes1984}. The extension to a 2.5D configuration of the reconnection current sheet would require the consideration of the third dimension, the direction of the inferred electric field, along which the magnetic field component, or the guide field $B_g$, does not vanish. 
The presence of an appropriate guide field may play an important role in energizing electrons \citep[e.g.][]{Holman1985, Litvinenko1996}. A recent development of new computational models has produced power-law distributions of electrons in reconnection current layers with a modest guide field $B_g /B_{in} \le 1$ \citep{Arnold2021}. In flare observations, direct measurements of the magnetic field vector in the reconnection current sheet in the corona have been hard and rare \citep{Chen2020}, posing difficulties to test theoretical models. On the other hand, it has been widely recognized that flare brightening often starts near the polarity inversion line (PIL) of the photospheric magnetic field, forming flare loops much inclined toward the PIL, or strongly sheared. As flare arcades form at higher altitude with two ribbons moving away from the PIL, flare loops become less sheared. Such strong-to-weak shear evolution is most evident in eruptive two-ribbon flares, such as the SOL2000-07-14 X5.7 flare, or the Bastille Day flare \citep{Aschwanden2001, Qiu2010}. It indicates that flare reconnection likely starts with a strong guide field in the early phase, and reconnection then proceeds at higher altitudes with the ever weakening guide field. Using high-resolution 3D MHD simulations, \citet{Dahlin2021} have reproduced such strong-to-weak shear evolution observed in flare ribbons or loops \citep[e.g.][and references therein]{Qiu2017}, and confirmed its association with the variation of the guide field in the reconnection current sheet that cannot be directly measured in observations. The verification with high-resolution numerical models provides the crucial bridge between reconnection properties in the coronal current sheet and measurements with observations of flare ribbons.

In the past decade, magnetic shear has been measured in a number of studies \citep{Su2006, Ji2006, Su2007, Yang2009, Liu2009b, Qiu2010, Cheng2012, Inglis2013, Qiu2017, Sharykin2018, Zimovets2020}, but its relation with flare energetics is not yet elucidated. In this study, we examine the relationship between flare energetics and reconnection properties inferred from flare ribbon observations, taking into account the shear evolution during the flare. The study is attempted to explore the following questions: Upon reconnection, what happens first, particle acceleration or direct heating, or do they occur simultaneously?  Does magnetic shear, apart from the reconnection rate, affect heating or electron acceleration?  To answer these questions, we analyze six flares that exhibit coherent two ribbons and have $\ge 25$keV HXR emissions. These flares were well observed by the Atmospheric Imaging Assembly \citep[AIA:][]{Lemen2012} and Helioseismic and Magnetic Imager \cite[HMI:][]{Schou2012} on board the Solar Dynamics Observatories \citep[SDO:][]{Pesnell2012} and the Ramaty High Energy Solar Spectroscopic Imager \citep[RHESSI:][]{Lin2002} from the flare onset throughout the impulsive phase. In the next section, we will provide an overview of the selected flares. With these observations, in Section 3, we measure the evolution of the mean shear using the method developed in \citet{Qiu2009} and \citet{Qiu2010}, and analyze flare thermal and non-thermal properties using the X-ray spectral analysis. In the last section, results will be summarized, discussed, and compared with prior research. 

\section{Overview of Observations: Eruptive Flares and Confined Flares}
\label{sec:overview}
In this article, we present the analysis of six flares listed in Table~\ref{tab:info}. The sample includes three eruptive flares accompanied by coronal mass ejections (CMEs), and three confined flares without evident association with large-scale eruptions like filament eruptions or CMEs.
These flares are selected based on their geometry and the availability of the RHESSI X-ray observations. We have inspected $\approx$~400 flares from the flare ribbon database by \citet{Kazachenko2017}, together with the relevant CME and X-ray observations, and selected the six events listed in Table~\ref{tab:info} for the preliminary experiment. These flares have a relatively simple morphology, all exhibiting prominent and nearly coherent two ribbons aligned along the polarity inversion line (PIL) of the photospheric magnetic field. It is therefore justifiable to infer physical parameters with respect to an approximate 2.5-dimensional configuration \citep[e.g.][]{Qiu2010, Qiu2017}. To examine flare energetics in the early phase of the flare evolution, we select events observed by RHESSI from the start of the flare and throughout the impulsive phase, with a relatively clean background. The flares in Table~\ref{tab:info} have significant HXR emissions at photon energies larger than 20~keV, which are usually produced by non-thermal electrons.
With these observations, spectral analysis can be conducted to provide diagnostics of thermal and non-thermal properties of the flares. The stringent selection criterion leads to the small sample and also sets the limit on what we can conclude from this study.

\begin{table}
\caption{List of events}
\label{tab:info}
\begin{tabular}{cccccccccc }
 & Start time$^a$ & Mag.$^a$ & NOAA position$^a$ & CME &  \multicolumn{5}{c}{Shear index and energetics properties} \\
& & & & & $\rho ^b$ & $\langle \mathcal{R} \rangle (\mathcal{F}_p)^c$ & $\langle \gamma \rangle$  ($\mathcal{F}_p)^c$  & $\langle \mathcal{R}\rangle (T_p)^d$ & $\langle {\rm T_{p}}\rangle~ {\rm [MK]}^d$ \\
\hline

1 & 2014-08-25T14:46 & M2.0 & 12146 N06W39 & yes & -0.87 & $0.6\pm0.1$ & $4.2\pm0.5$ & $0.6\pm0.2$ & $18.2\pm0.6$\\
2 & 2014-12-18T21:41 & M6.9 & 12241 S11E10 & yes & -0.65 & $0.6\pm0.2$ & $5.6\pm0.7$ & $2.4\pm0.8$ & $20.0\pm2.2$\\
3 & 2013-08-12T10:21 & M1.5 & 11817 S21E17 & yes & -0.29 & $2.1\pm0.4$ & $5.7\pm0.4$ & $2.1\pm0.8$ & $21.7\pm0.8$\\
4 & 2015-01-29T05:15 & C8.2 & 12268 S13W03 & no & -0.89 & $1.7\pm0.4$ & $6.1\pm0.5$ & $2.5\pm0.8$ & $26.2\pm4.0$\\
5 & 2014-05-10T06:51 & C8.7 & 12056 N03E27 & no & -0.79 & $2.4\pm0.2$ & {\bf $4.3\pm0.1$} & {\bf $4.4\pm0.8$} & {\bf $21.4\pm5.2$} \\
6 & 2014-12-17T14:56 & C9.8 & 12242 S18E00 & no & -0.86 & $2.2\pm0.2$ & {\bf $5.7\pm0.4$} & $2.2\pm0.4$ & {\bf $23.3\pm1.6$}\\
\hline
\end{tabular}
$a$: Flare information provided by the Solar Monitor \url{https://solarmonitor.org/}; the flare magnitude and start time are based on GOES data.  \\ 
$b$: Coefficient of the linear cross correlation between the $25-50$~keV HXR counts rate $\mathcal{F}_{hxr}$ and the shear index $\mathcal{R}$ during the rise of the HXR emission. \\
$c$: The mean and standard deviation of the shear index $\mathcal{R}$ measured from flare ribbons and the mean and standard deviation of the power-law index $\gamma$ of the non-thermal HXR photon spectrum, derived during the time intervals when $\mathcal{F}_{hxr}$ is more than 70\% of its peak value.\\
$d$: The mean and standard deviation of the shear index $\mathcal{R}$ and of the plasma temperature during three time intervals around the peak plasma temperature. \\

\end{table}



\begin{figure}
\centering
\includegraphics[width=6.0cm]{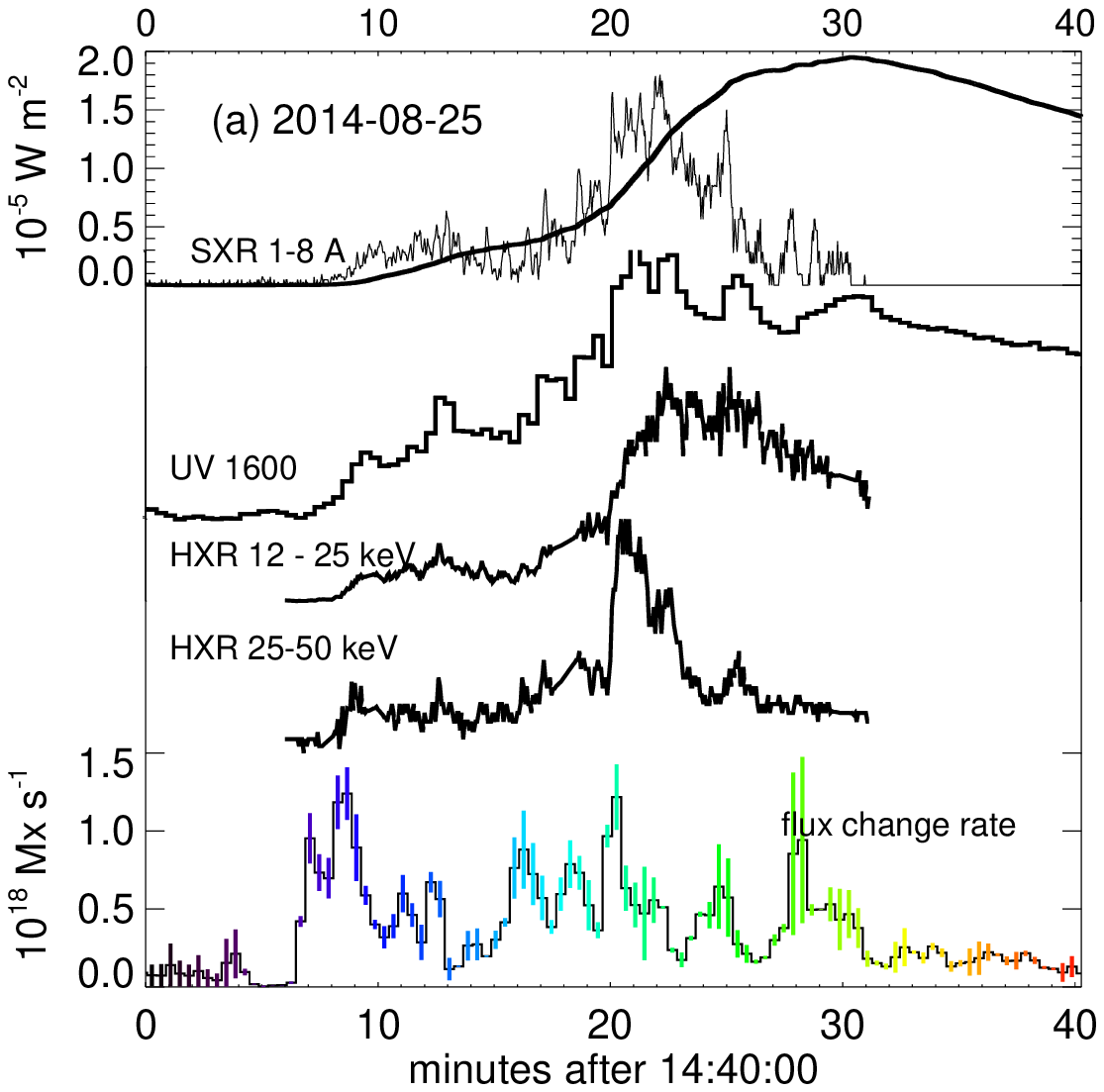} \includegraphics[width=6.0cm]{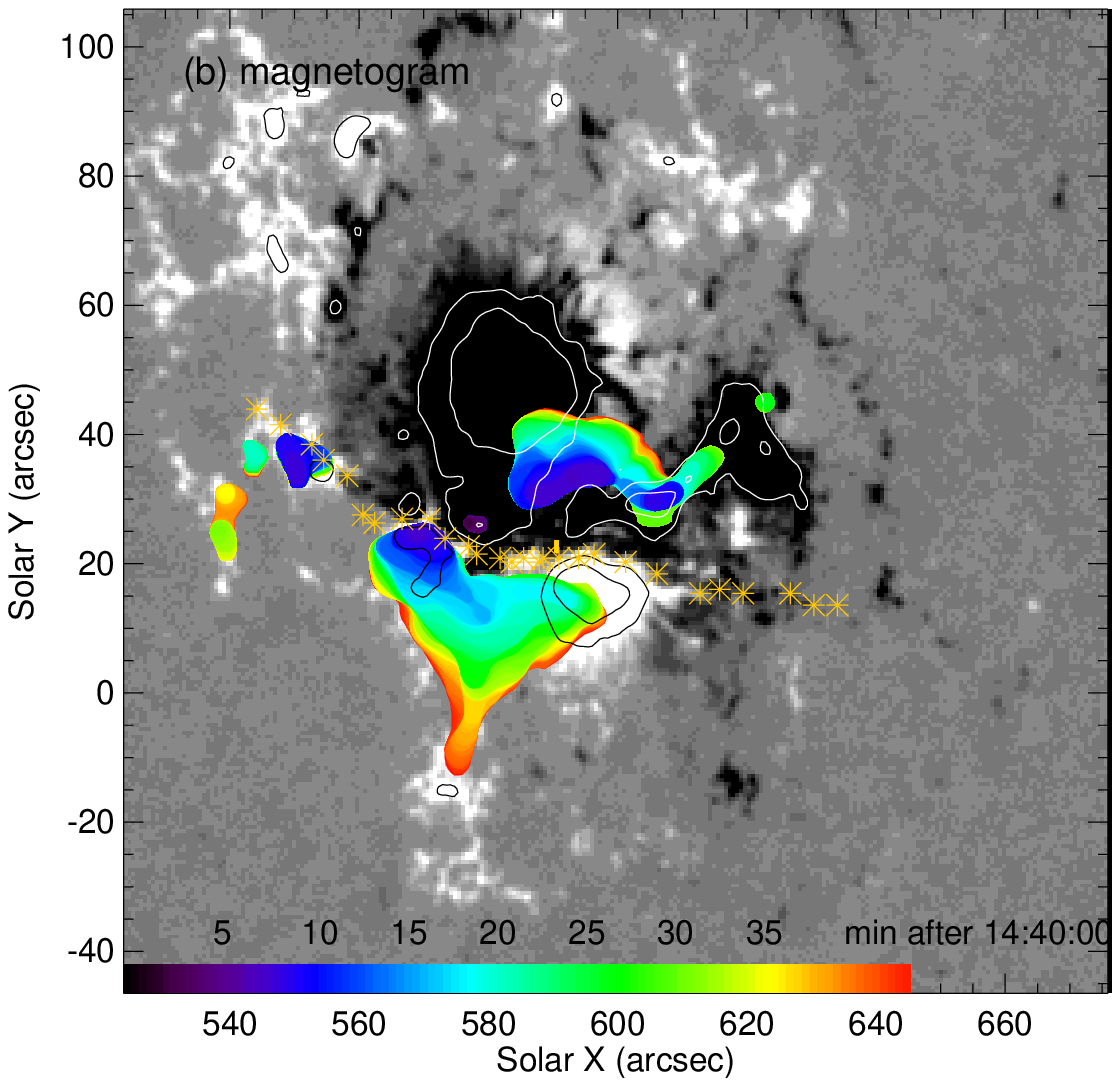}
\includegraphics[width=13.6cm]{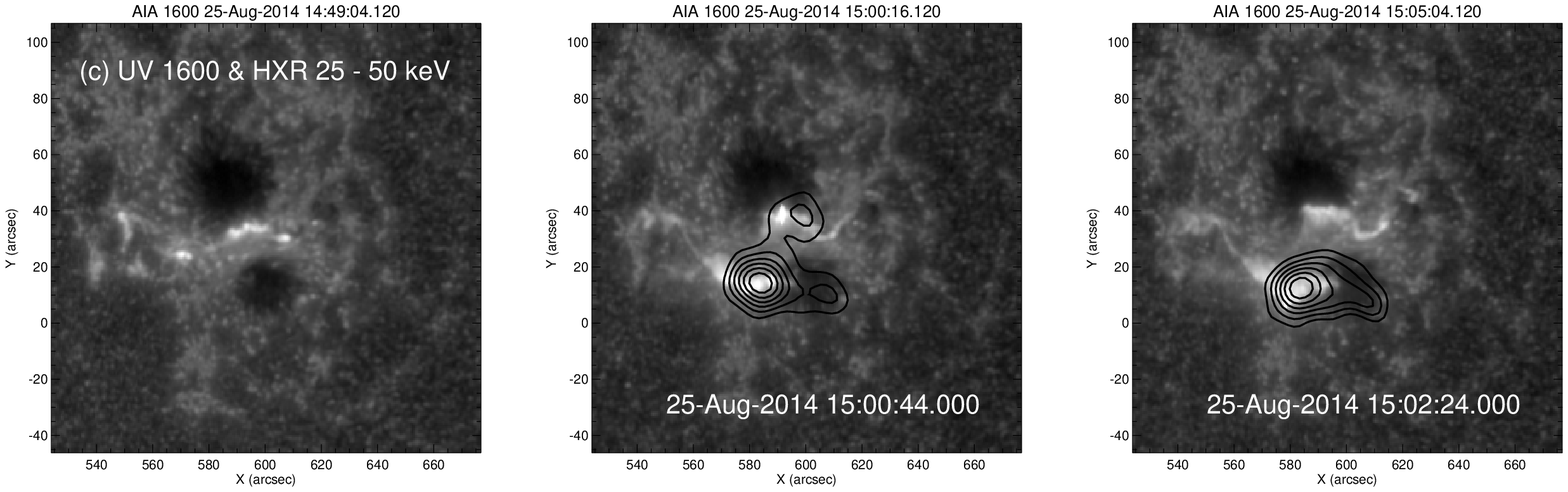}
\includegraphics[width=13.6cm]{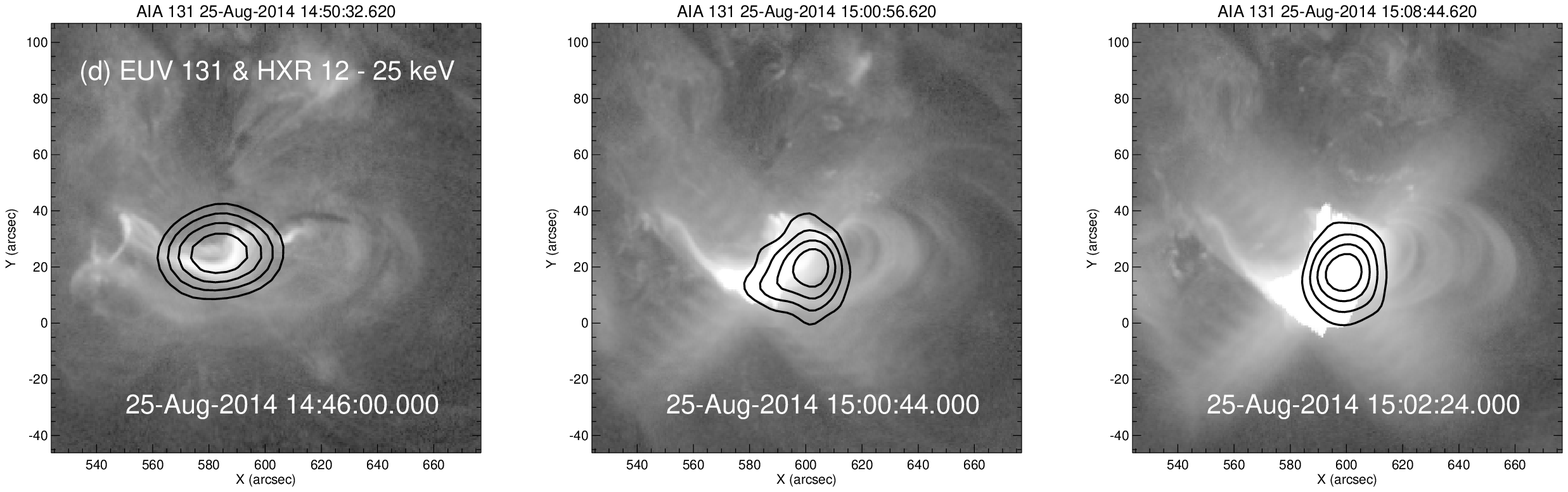}
\caption{Overview of the SOL2014-08-25 eruptive flare. (a) SXR (GOES) and its time derivative, HXR (RHESSI), and UV (AIA) light curves, and reconnection rate $\dot{\psi}$. (b) Evolution of flare ribbon fronts superimposed on a pre-flare longitudinal magnetogram by HMI. Contour levels indicate the magnetic flux density at $\pm$400, and 800 Mx~cm$^{-2}$. Orange symbols outline the PIL. (c) Snapshots of flare ribbon evolution in UV 1600~\AA\ by AIA, superimposed with HXR images at 25 - 50 keV from RHESSI. Contour levels indicate 40\%, 50\%, 60\%, 70\%, 80\%, and 90\% of the maximum intensity of each HXR image. (d) Snapshots of flare loops in EUV 131~\AA\ by AIA, superimposed with HXR images at 12 - 25 keV from RHESSI. Contour levels indicate 60\%, 70\%, 80\%, and 90\% of the maximum intensity of each HXR image.}
\label{fig:overview_140825}
\end{figure}

Figure~\ref{fig:overview_140825} gives an overview of the evolution of the SOL2014-08-25 M2.0 eruptive flare. The ultraviolet (UV~1600~\AA, AIA) and soft X-ray (1-8~\AA, GOES) emissions start to rise about 10 minutes before the rapid rise of the 25-50 keV HXR emission (in units of counts per second, RHESSI), hereafter denoted as $\mathcal{F}_{hxr}$ (Figure~\ref{fig:overview_140825}a). At the flare onset, the global reconnection rate, or the flux change rate $\dot{\psi}$, also rises rapidly. The flux change rate is measured by summing up the longitudinal magnetic flux enclosed in newly brightened flare ribbons per unit time \citep[see][for the discussion of the measurement method and uncertainties]{Qiu2010}. 
Evolution of the newly brightened flare ribbon fronts is presented in Figure~\ref{fig:overview_140825}b, with the convention that cold colors indicate the earlier times, whereas warm colors indicate later times. The flare ribbon fronts are derived from the time sequence of the UV 1600~\AA\ images obtained by AIA, shown in Figure~\ref{fig:overview_140825}c. Two flare ribbons are aligned with the PIL, which is outlined by orange symbols. Flare ribbons in magnetic fields of opposite polarities map the conjugate feet of reconnection formed flare loops, as confirmed in AIA 131~\AA\ images (Figure~\ref{fig:overview_140825}d) that exhibit flare loops of 10~MK plasmas connecting the two ribbons. From the morphological evolution of the flare, it is apparent that earlier formed flare loops, connecting the violet-blue patches of the two ribbons, are more sheared with respect to the PIL than later formed loops connecting the green-orange ribbon patches. Such a pattern of the strong-to-weak shear evolution has been widely reported in observations of eruptive two-ribbon flares \citep[e.g.][and references therein]{Su2007, Yang2009}, and also demonstrated in magnetohydrodynamic simulations \citep[e.g.][]{Aulanier2012, Dahlin2021}. 
Finally, HXR maps have been obtained using the Pixon method at two energies, 25 - 50~keV and 12 - 25~keV, and are superimposed with the AIA images in Figure~\ref{fig:overview_140825}c and Figure~\ref{fig:overview_140825}d, respectively. Despite the coarse resolution of the HXR maps, it is shown that 25 - 50 keV HXR emission tracks the UV emission on the two ribbons, whereas 12 - 25 keV HXRs are likely from the loop connecting the ribbons, and the strong-to-weak shear evolution is evident from the morphology of HXR sources.

\begin{figure}
\centering
\includegraphics[width=6.5cm]{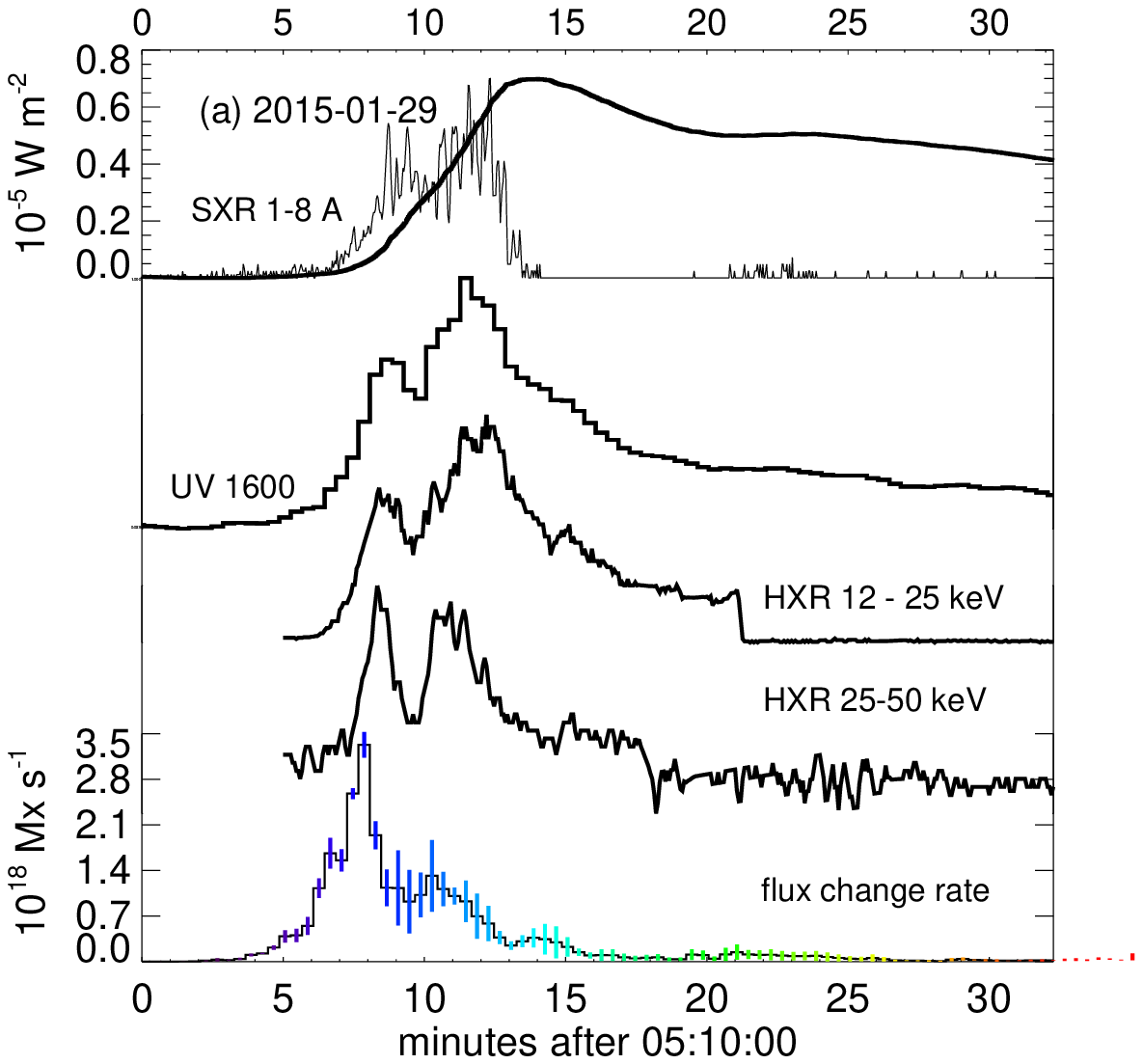}\includegraphics[width=6.0cm]{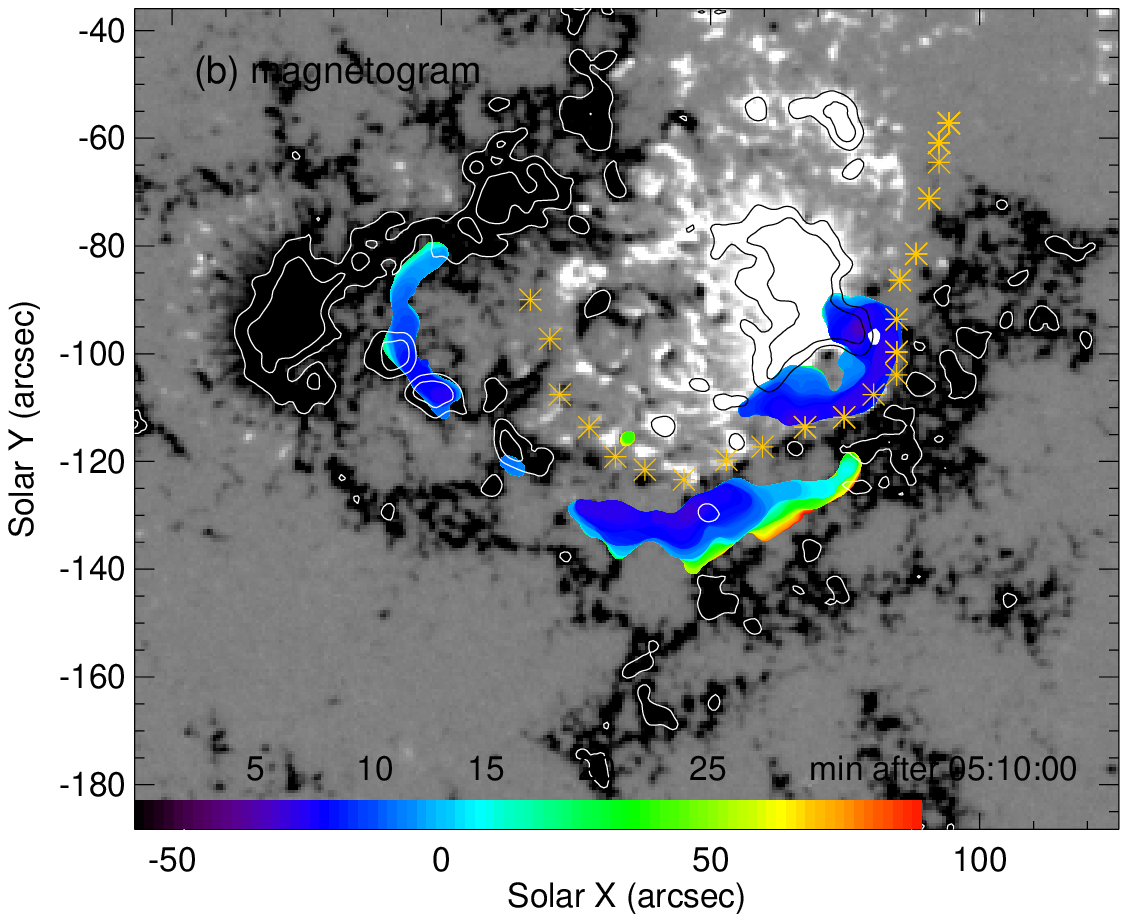}
\includegraphics[width=14.5cm]{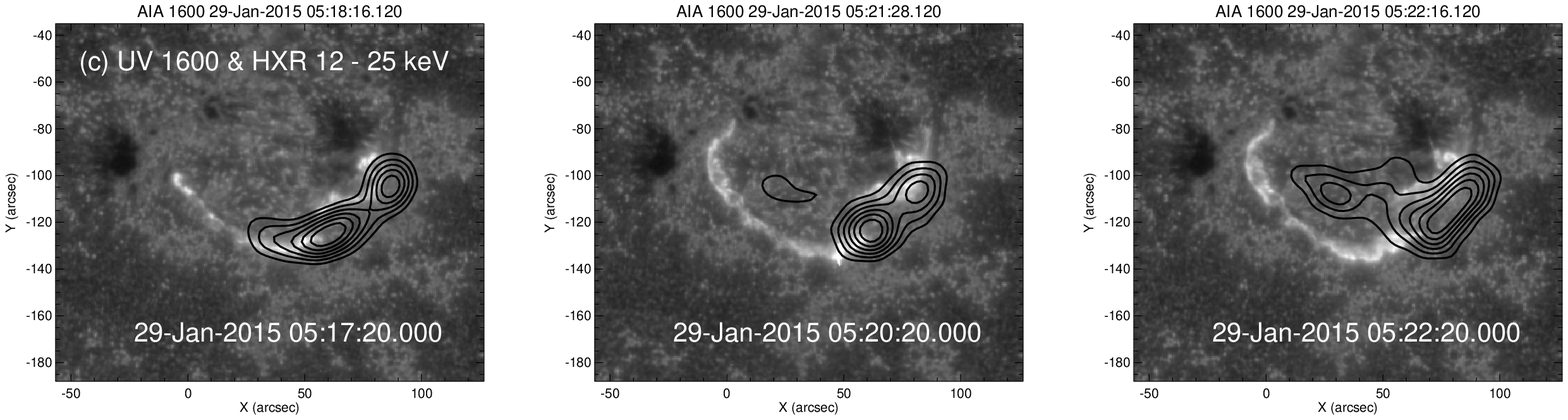} \vspace{-0cm}\includegraphics[width=14.5cm]{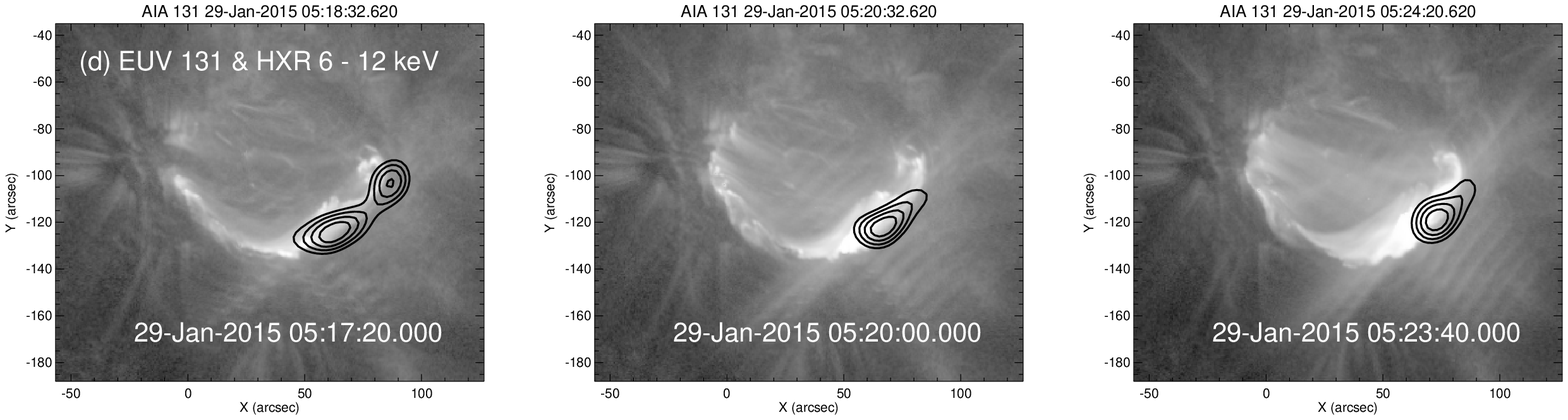}
\caption{Overview of the SOL2015-01-29  confined flare. (a) SXR, HXR, and UV light curves, and reconnection rate $\dot{\psi}$. (b) Evolution of flare ribbon fronts superimposed on a pre-flare longitudinal magnetogram. Orange symbols outline the PIL. (c) Snapshots of flare ribbon evolution in UV 1600~\AA\ by AIA, superimposed with HXR images at 12 - 25 keV obtained by RHESSI. Contour levels indicate 40\%, 50\%, 60\%, 70\%, 80\%, and 90\% of the maximum intensity of each HXR image. (d)  Snapshots of flare loops in EUV 131~\AA\ by AIA, superimposed with HXR images at 6 - 12 keV obtained by RHESSI, and contour levels indicate 60\%, 70\%, 80\%, and 90\% of the maximum intensity of each HXR image.}
\label{fig:overview_150129}
\end{figure}

Figure~\ref{fig:overview_150129} gives an overview of the SOL2015-01-29 C8.2 confined flare, which evolves much faster without an apparent lead time of the enhanced UV emission (from flare ribbons) ahead of the rapid rise of the $\ge 25$ keV HXR emission. This confined flare exhibits more than two ribbons; here we focus on the most energetic part of the flare that form two major ribbons next to the PIL in the west of Figure~\ref{fig:overview_150129}b. As shown in Figure~\ref{fig:overview_150129}c and \ref{fig:overview_150129}d, these two ribbons are conjugate feet of flare loops, where HXR emissions are concentrated. The connectivity between these two ribbons is also confirmed in the non-linear-force-free (NLFF) extrapolation by \citet{Zhong2019}. The ribbon in the negative magnetic field to the east of the major ribbons appears to be connected with a remote brightening in a different active region not shown in this figure, forming long loops overlying the loop systems of the two major ribbons, as also demonstrated in \citet{Zhong2019}. The overlying long loops are visible in the EUV and X-ray images in Figure~\ref{fig:overview_150129}c and d after 5:20~UT. It is apparent that the flare loops connecting the two major ribbons near the PIL are strongly sheared at the onset of the flare; then both ribbons spread westward with different speeds, illustrating the strong-to-weak shear variation during the flare evolution; on the other hand, the extent of the ribbon expansion in the direction perpendicular to the PIL is less significant compared with the eruptive flare.

\begin{figure}
\centering
\includegraphics[width=6cm]{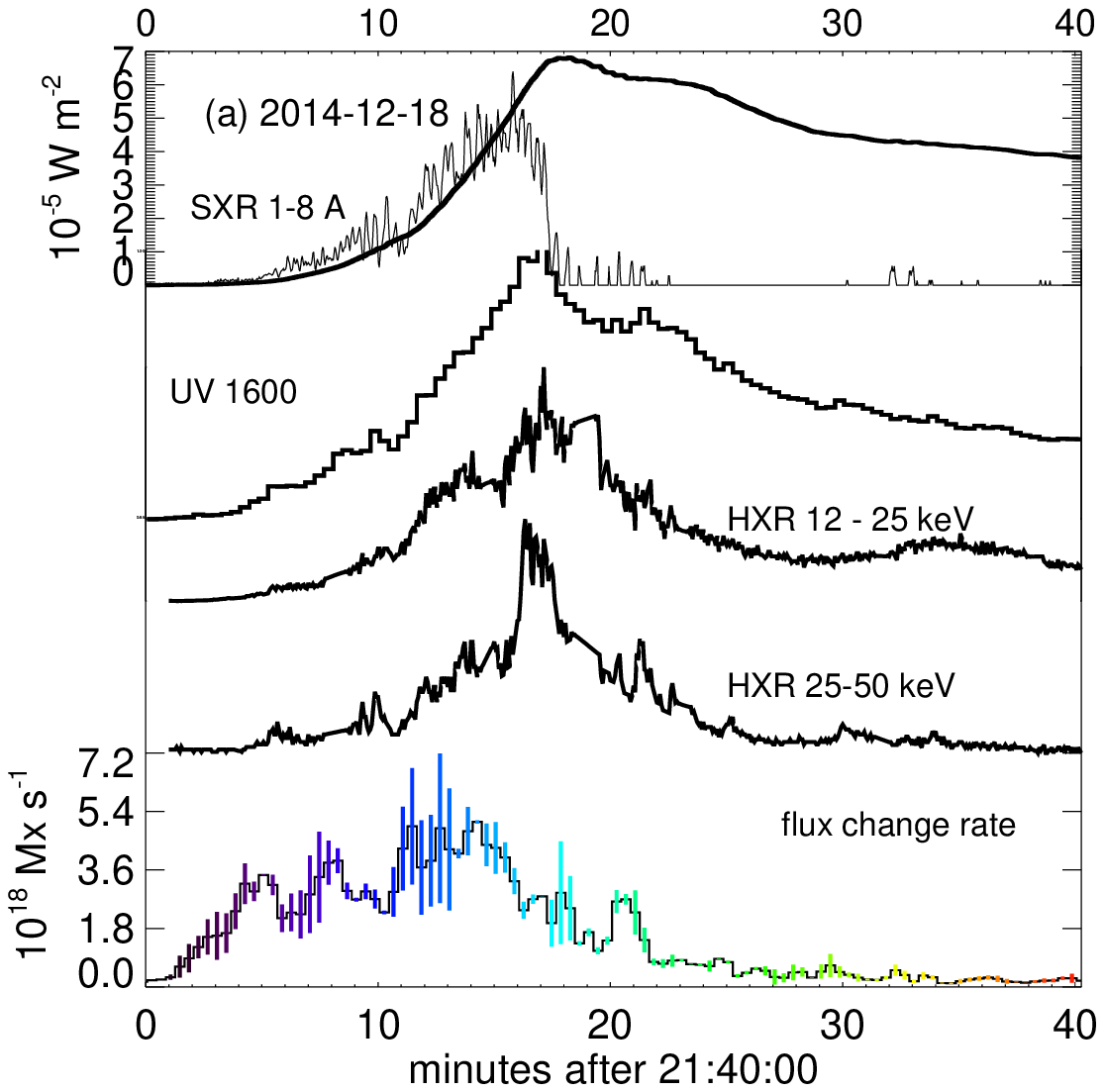} 
\includegraphics[width=6.cm]{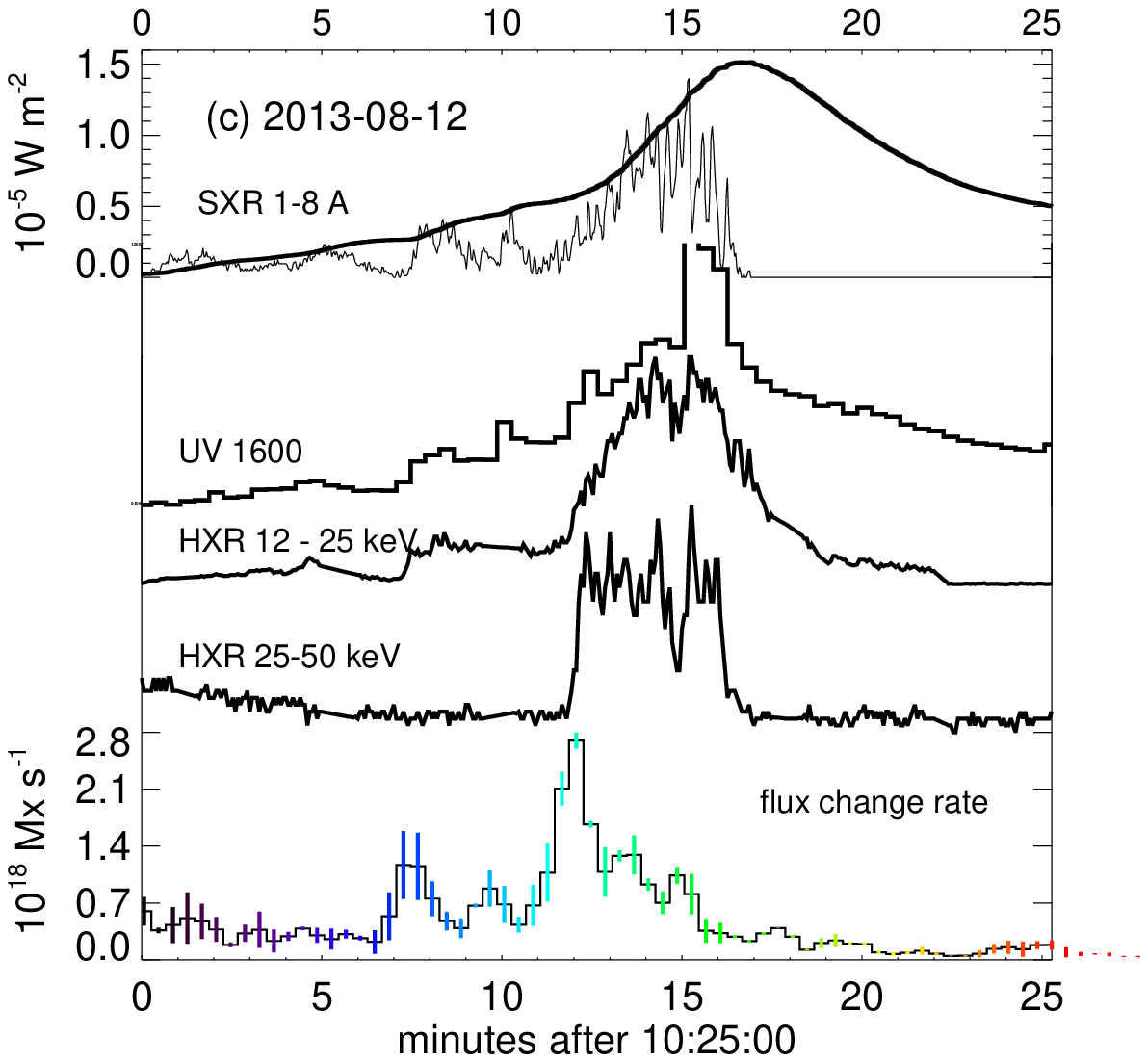} 
\includegraphics[width=6.cm]{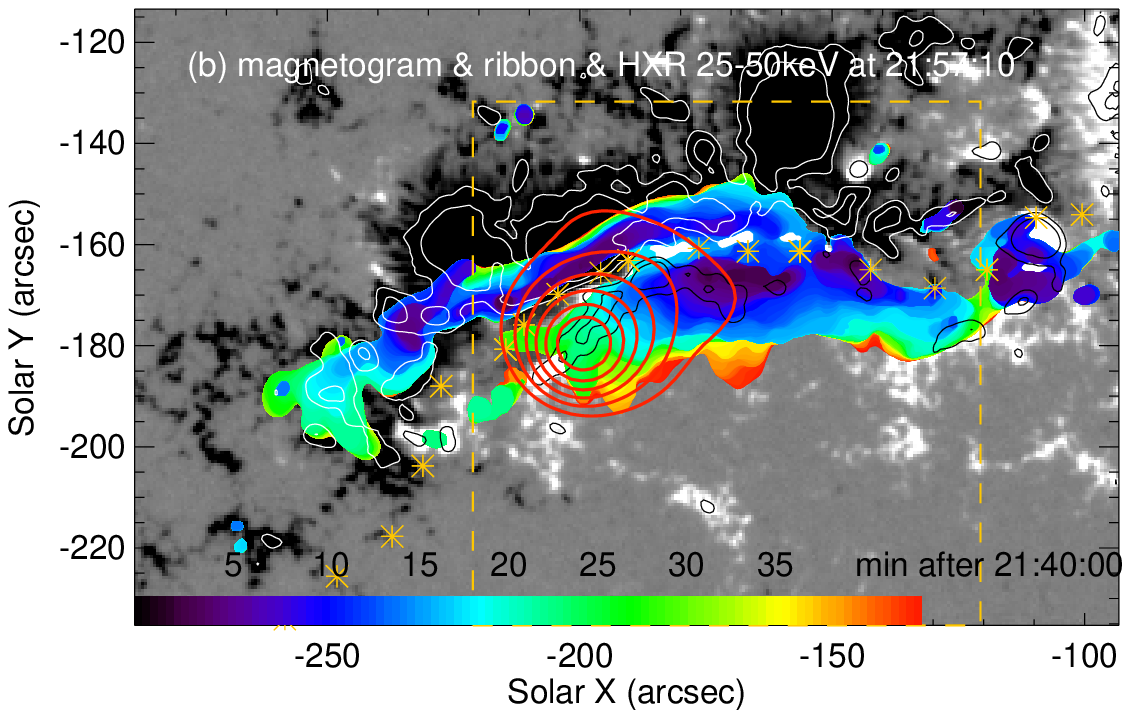}
\includegraphics[width=5.6cm]{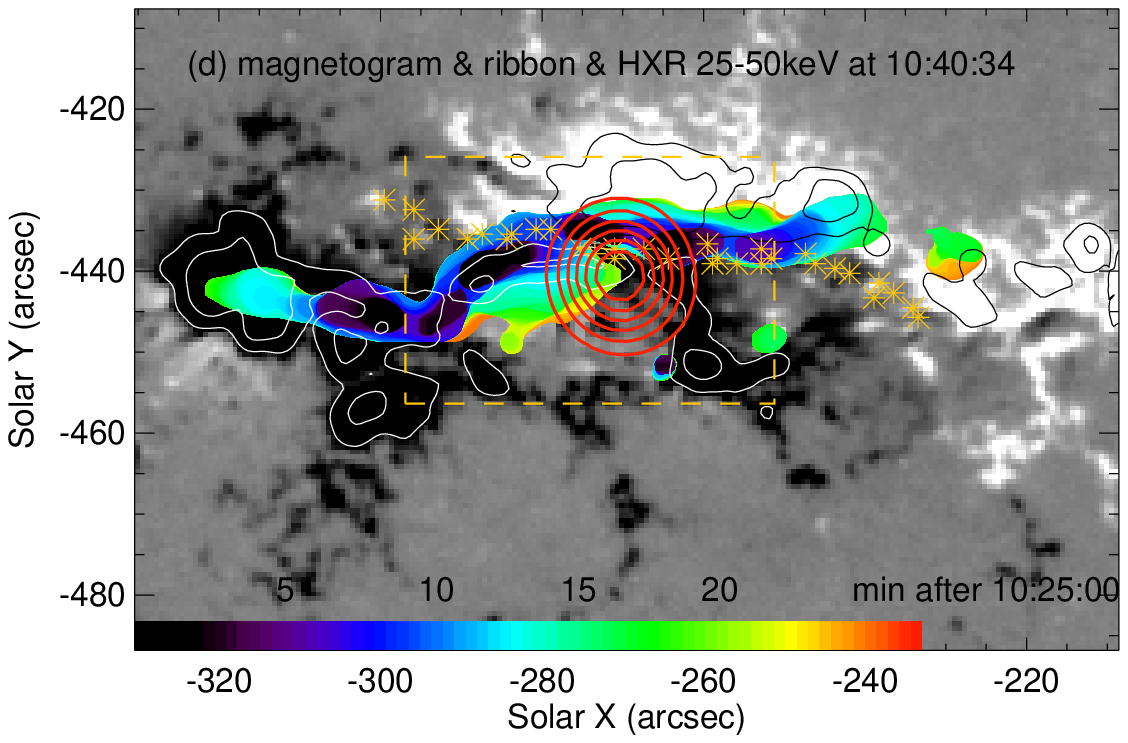}
\caption{Overview of the other two eruptive flares. (a) SXR, HXR, and UV light curves, and flux change rate $\dot{\psi}$ for the SOL2014-12-18 flare. (b) Evolution of flare ribbon fronts superimposed on a pre-flare longitudinal magnetogram. Orange symbols outline the PIL. Overlaid in red contours is the HXR maps in 25-50 keV at the peak of the emission, with the contour levels at 40\%, 50\%, 60\%, 70\%, 80\%, and 90\% of the maximum intensity. The orange box encloses the parts of the ribbons where shear index is measured. (c-d) Same as (a-b) but for the SOL2013-08-12 flare.}
\label{fig:overview_eruptive}
\end{figure}

\begin{figure}
\centering
\includegraphics[width=6.cm]{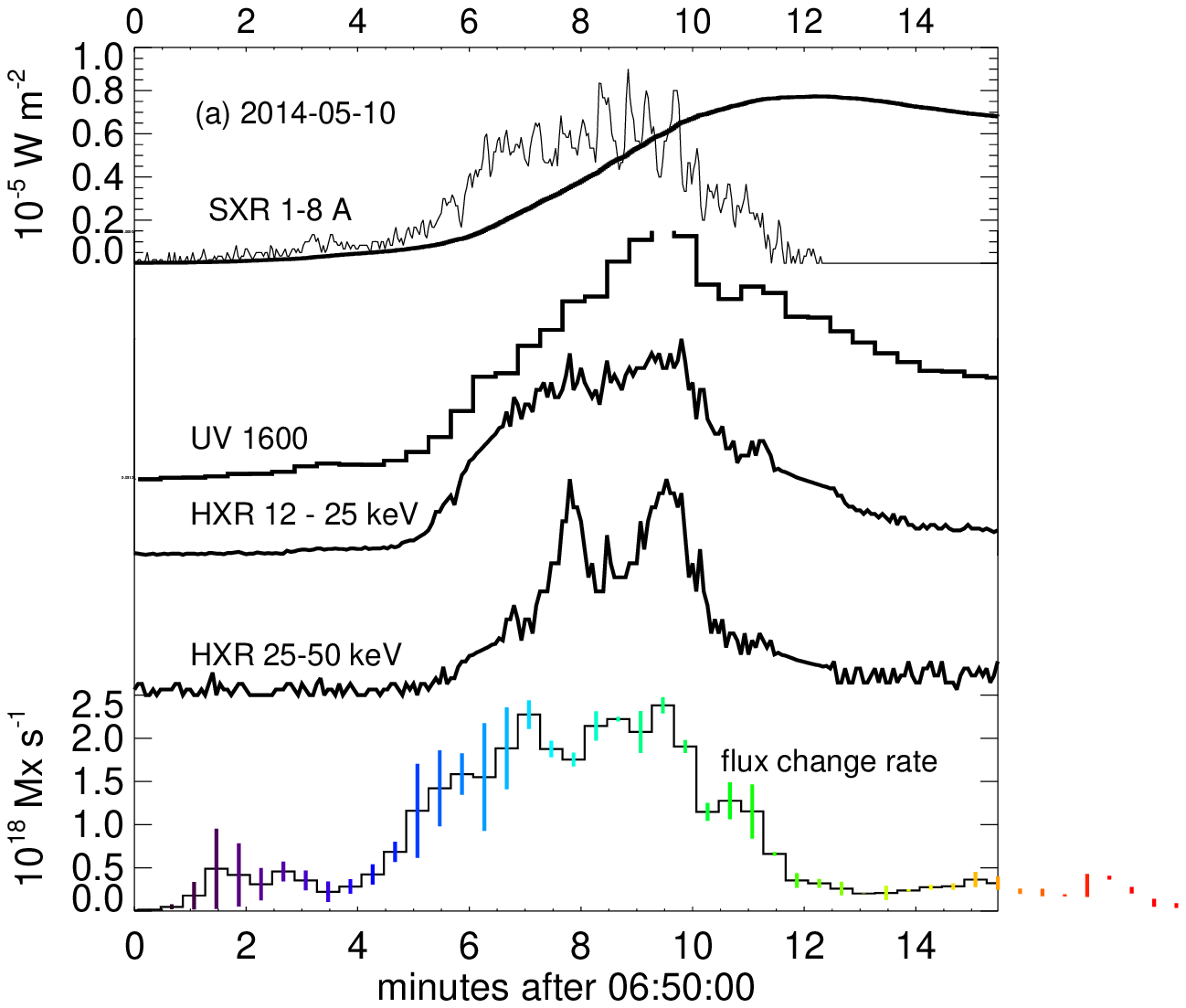} 
\includegraphics[width=6.cm]{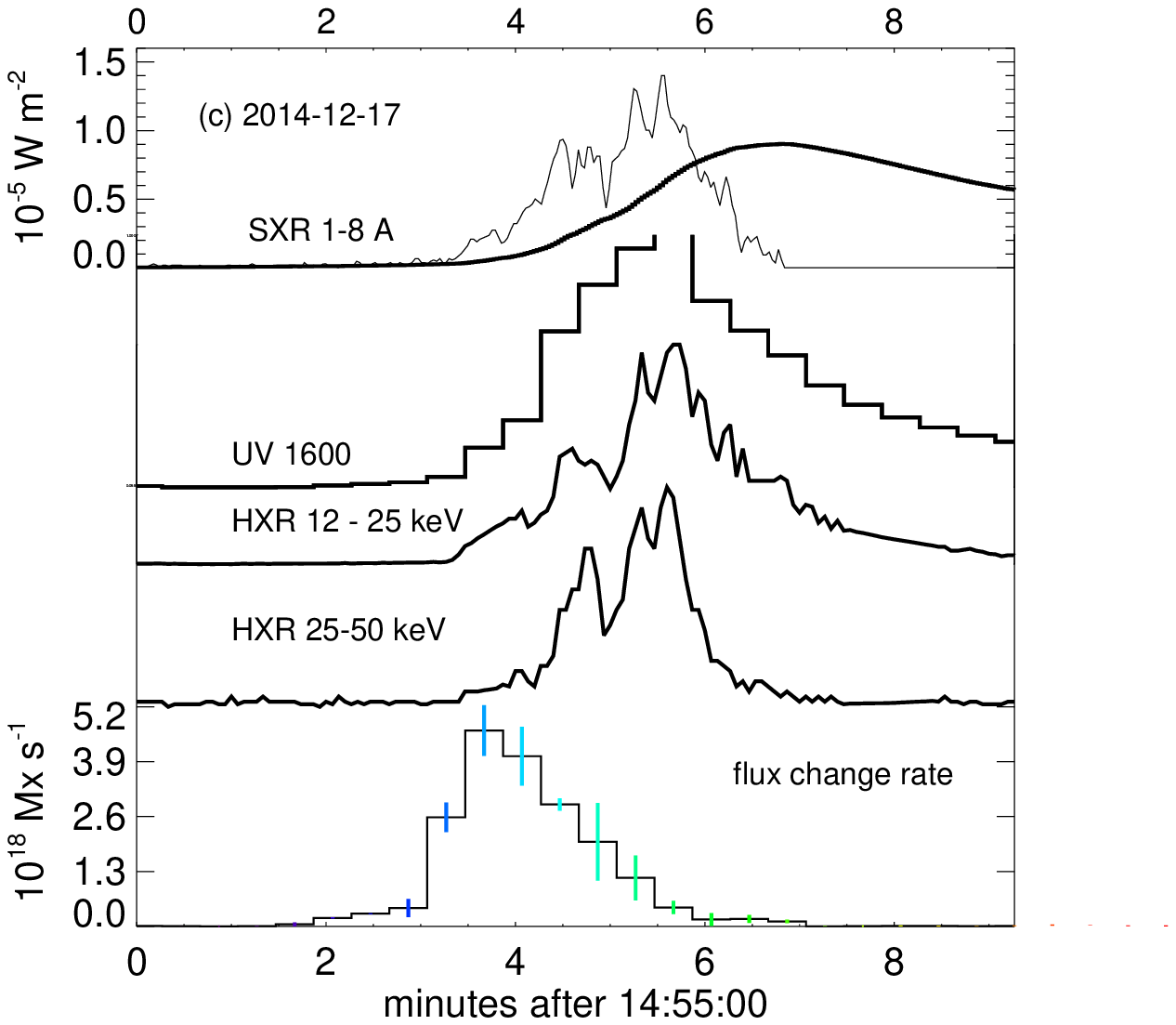}
\includegraphics[width=7.5cm]{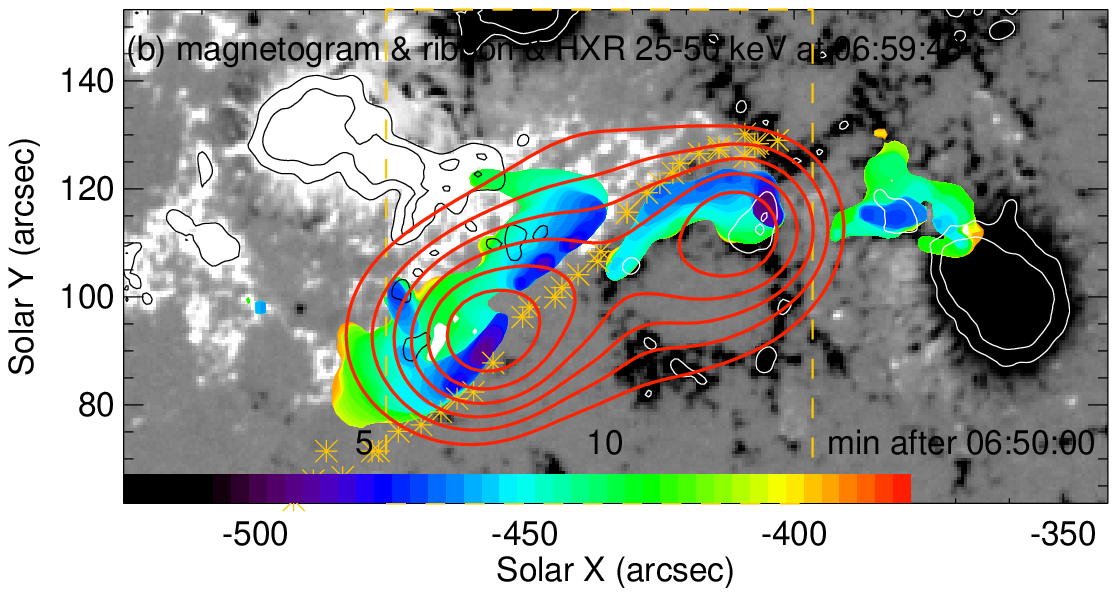}\includegraphics[width=6.0cm]{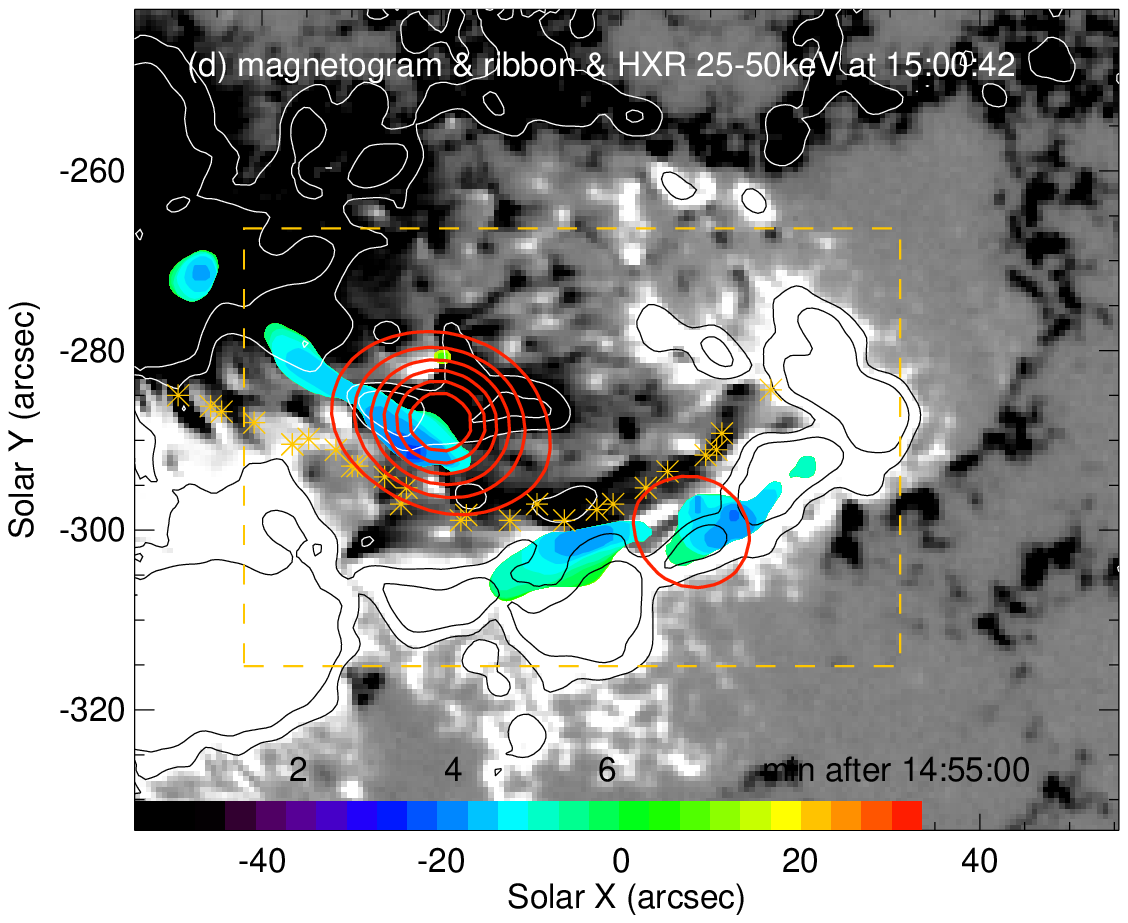}
\caption{Same as Figure~\ref{fig:overview_eruptive}, but for the confined flares SOL2014-05-10 (a-b) and SOL2014-12-17 (c-d). }
\label{fig:overview_confined}
\end{figure}

Figure~\ref{fig:overview_eruptive} shows the evolution of the other two eruptive flares SOL2014-12-18 M6.9 and SOL2013-08-12 M1.5, and Figure~\ref{fig:overview_confined} shows the evolution of the other two confined flares SOL2014-05-10 C8.7 and SOL2014-12-17 C9.8. For eruptive flares, we observe a 5-10 min lead time of enhanced UV and soft X-ray emission before the rapid rise of the $\ge 25$ keV HXR emission $\mathcal{F}_{hxr}$, whereas the confined flares evolve on a much shorter timescale, with a less than 2~min lag of $\mathcal{F}_{hxr}$ with respect to the rise of the soft X-ray and UV emissions. In these events, because of the close proximity of the two ribbons and the limited spatial resolution and dynamic range of the HXR maps, we cannot distinguish whether the $\ge 25$ keV HXR is emitted at the flare footpoints or from the flare loop tops; nevertheless, significant HXR emissions appear to concentrate at the major flare ribbons or loops enclosed in the orange boxes in each figure. 

Observations shown in these overview figures reveal the following interesting facts. First, emission in the UV 1600~\AA\ passband, and the inferred flux change rate, $\dot{\psi}$, exhibit multiple spikes, indicative of multiple episodes of energy release. It is also noted that flare UV emission and $\dot{\psi}$ start to rise before the rise of significant $\ge 25$ keV HXR emissions. Particularly in the eruptive flares in this study, $\dot{\psi}$ has several spikes
before $\mathcal{F}_{hxr}$, suggesting episodes of energy release that do not produce significant non-thermal emission. This has been observed in a number of other flares, such as those reported by \citet{Warren2001} and more recently by \citet{Naus2021} and \citet{Qiu2021}. Second, all these flares, eruptive or confined, exhibit the strong-to-weak shear evolution when we consider the connectivity between the major pair of conjugate ribbons. We note that the initial large enhancement of the flux change rate $\dot{\psi}$ is primarily contributed by the fast spreading of flare ribbons along the PIL, when the shear also decreases rapidly. In a 2.5D approximation, the local reconnection rate, characterized by the mean reconnection electric field $\langle E \rangle = V_{\perp} B$ \citep[e.g.][]{Forbes1984}, is small despite the large $\dot{\psi}$ in the early phase of the flare.

In this article, we discuss the energetic behavior along with the different patterns of reconnection reflected in the evolution of flare ribbons, following the previous study by \citet{Qiu2010}. We track the geometric evolution of conjugate flare ribbons to infer the shear of post-reconnection flare loops. Flare energetics, namely plasma heating and electron acceleration, will be learned from spectroscopic analysis of RHESSI (and GOES) X-ray observations. We compare flare energetics with the evolution of the mean magnetic shear during the progress of flare reconnection, particularly during the {\em rise} of the flare. For consistency in the following text, we define the {\em rise} of the flare as the {\em rise} of the flare soft X-ray emission in the GOES 1-8 ~\AA\ channel. The {\em impulsive phase} of the flare, typically characterized by flare HXR emissions, often coincides with the {\em rise} of the flare \citep{Fletcher2011}.

It is also noted that, in this study, we focus on these most energetic parts of the flare ribbons that exhibit continuous and coherent apparent motions. The SOL2014-12-18 flare was studied by \citet{Wang2017b} and \citet{Joshi2017}, both suggesting that the ribbon sections at the two ends outside the orange box in Figure~\ref{fig:overview_eruptive}b are likely the feet of a flux rope, whereas the major ribbons enclosed inside the box are formed by tether-cutting reconnection below the flux rope following a standard picture \citep{Moore2001}. Assuming the connectivity of the two ends, the evolution pattern shown in Figure~\ref{fig:overview_eruptive}b suggests that they are moving apart, consistent with the flux rope expanding outward. With the expansion, the length of field lines and the magnetic shear would grow, and such pattern has been revealed in recent MHD simulations \citep{Aulanier2019,Jiang2021a, Dahlin2021}. Similarly, the evolution of the active region hosting the SOL2013-08-12 flare has been thoroughly studied by \citet{Liu2016}, suggesting a low-lying flux rope being disturbed and erupting, with two-ribbons below the rope from tether-cutting reconnection. Again, the two ends of the two ribbons outside the box well represent the feet of a flux rope identified by Wang et al. (2021, in preparation) from combined coronal dimming and vertical current analysis. This study focuses on flare energetics, so we do not analyze the two ends of the ribbons reflecting the flux rope signatures. Each of the other two confined flares also exhibit more than two ribbons; we focus on the major ribbons enclosed in the box, where HXR emissions appear to concentrate. In the subsequent analysis, we measure the shear index $\mathcal{R}$ and reconnection rate $\dot{\psi}$ only on these major ribbons that are close to and aligned with the PIL.

\section{Evolution of Reconnection Properties and Flare Energetics}
\label{sec:evo}

\subsection{Methodology}
\label{ssec:method}

The spatial evolution of flare ribbon brightening may be translated to magnetic shear of reconnecting field lines. Lacking direct measurements of magnetic field in the corona during the flare, it has not been possible to determine the 3D magnetic field participating in magnetic reconnection in the current sheet. Conventional and regular magnetic field measurements have been provided in the photosphere, at which the reconnection formed flare loops are anchored; therefore, properties of magnetic reconnection in the corona may be inferred by tracking the flare ribbon evolution across the magnetic field in the lower atmosphere.

The six flares selected in this study display two conjugate ribbons aligned along the PIL, as well as globally organized motion patterns; these allow us to approximate the flare geometry to a 2.5D configuration, with the presumed macroscopic reconnection current sheet along the PIL, or the direction of the elongation of the flare ribbons, which we consider as the guide direction. 
Using the method by \citet{Qiu2009} and \citet{Qiu2010, Qiu2017}, we measure reconnection properties
following the evolution of newly brightened flare ribbons. In this study, UV 1600~\AA\ images by AIA have been used to analyze flare ribbons. Newly brightened flare ribbon pixels are identified if the pixel brightness stays $N$ times greater than the median brightness of the pre-flare quiescent regions for 10 consecutive time frames, or 4 minutes. Based on statistics of flare ribbon brightness in the UV 1600~\AA\ passband, we choose $N$ to be in the range of 4 to 6, which effectively distinguishes flare brightness from plages. The requirement for the pixel to stay bright for a few minutes is based on the statistics of heating and cooling timescales at the footpoints of reconnection formed flare loops, and it helps to minimize effects such as saturation or brief brightening due to the {\bf projection effect of bright ejecta in the corona}. The methodology and uncertainties are discussed in \citet[][and references therein]{Naus2021}. 
We then project the positions of newly brightened ribbon pixels in the direction along the curved PIL that is smoothed by a low-degree polynomial function, and measure the time-dependent mean parallel distance $\langle d_{||} \rangle (t)$ of newly brightened flare ribbons. We also measure the mean distance of the ribbon fronts perpendicular to the PIL, by $\langle d_{\perp}\rangle (t) = A(t)/l_{||}(t)$, where $A(t)$ is the area covered between the PIL and newly brightened ribbon fronts, and $l_{||}(t)$ the total length of the newly brightened ribbon projected along the PIL. Measurements are made for ribbons in the positive and negative magnetic fields separately. Uncertainties are estimated by using varying thresholds to determine newly brightened ribbon pixels. 

We then infer the time dependent reconnection geometry in two-ribbon flares. In particular, we define the shear index $\mathcal{R} \equiv |\Delta \langle d_{||} \rangle/ \sum {\langle d_{\perp}\rangle}|,$ the difference and sum derived with the measurements of the positive and negative ribbons. The shear index may be related with the ratio of the reconnection guide field to the reconnection outflow field $\mathcal{R} \sim B_{g}/B_o$ \citep{Qiu2017}. We note that the exact mapping of the current sheet structure to the ribbon evolution is non-trivial, and much depends on the dynamic evolution of the post-reconnection magnetic field in a truly 3D manner; on the other hand, with a 2.5D approximation, $\mathcal{R}$ provides a qualitative proxy of the variation of the guide field during the flare reconnection \citep{Dahlin2021}. Qualitatively, a large $\mathcal{R}$ is indicative of a relatively strong guide field, which often decreases during the rise of the flare. 
Such measurements of the shear variation from the ribbon geometry have been made previously on a few two-ribbon flares, and the inferred shear evolution pattern is found to be consistent with the measurements of the inclination of observed post-reconnection flare loops with respect to the extrapolated PIL in the corona \citep{Qiu2017}. In this study, we measure the shear index exclusively with the UV 1600~\AA\ observations of flare ribbons, and do not extend the analysis to flare loops. The cadence of the measurements is 24~s.

Readers are reminded that, first, the method only returns the average shear, by assuming the connectivity of newly brightened ribbon patches in opposite polarities. Second, flare ribbon brightening usually starts with a few bright kernels adjacent to the PIL, and in a few minutes, the ribbon starts to form connecting the bright kernels. This is the case with several flares analyzed here, and the measurements only apply when a continuous section of the ribbon has properly formed and exhibits a coherent motion. Finally, we track the newly brightened ribbons, defined by \citet{Qiu2010}; it is different from tracking the brightest kernels in optical, UV, or HXR emissions \citep{Qiu2002, Krucker2003, Fletcher2004,  Lee2008, Yang2009, Inglis2013}. On the other hand, Figures~1-4 suggest that $>$25~keV HXR emissions tend to be coincident with newly brightened ribbons. Therefore, in the following analysis, we track the evolution of the two major conjugate ribbons and examine how it is related to flare energetics reflected in flare X-ray emissions.

For each flare, we collect HXR observations by RHESSI, and employ the Object Spectral Executive (OSPEX) from the Solar SoftWare (SSW) to model the X-ray photon spectra \citep{Freeland1998, Freeland2012}. Each spectrum, with the integration of 20~s, is fitted to the model with an isothermal component and a broken power-law component, with the OSPEX fitting procedure ${\rm vth+bpow+drm\_mod}$. The photon energy range for the fitting is determined based on the attenuator status and the signal to noise level. For the six flares in this study, the lower limit of the photon energy is at 6~keV, and the upper limit varies from about 15~keV to up to 50~keV at the peak of the HXR emission. During these flares, the background X-ray emission is low and mostly flat; we select the background from the pre- or post- flare period. Spectral fitting is conducted with data obtained with a single detector; we have also experimented fitting with data from two or more well-performing detectors, typically detector 8 or 9, which return consistent fitting results. These flares are moderate, with their magnitude ranging between C8.0 and M7.0, so the pile-up effect is not significant, and there is only trivial difference between the fitting results with and without pile-up correction. For each flare, fitting starts from the peak time of the flare HXR emission, and proceeds to earlier or later times. For each interval, when necessary, the initial guesses have been further adjusted to randomize the residual pattern to the maximal extent. Finally, we discard fitting results for time intervals with $\chi^2 \ge 3$. 

From the spectral fitting, we derive the time evolution of the power-law index $\gamma$ of the photon spectrum, the low-energy cutoff of the power-law distribution $E_{c}$, and the temperature and emission measure of thermal plasmas. In these flares, $E_{c}$ is found to be around 15~keV; therefore, $\ge 25 $keV HXRs are produced by non-thermal electrons. Additionally, we also derive the temperature and emission measure of flare plasmas using the GOES two-channel X-ray observations, to be compared with the flare parameters obtained from RHESSI observations. The flares selected for this study were observed by RHESSI from the onset throughout the rise phase, so that the evolution of the flare temperature and non-thermal spectral index can be studied in comparison with reconnection properties inferred from flare ribbon evolution. On the other hand, given the uncertainties associated with the electron low-energy cutoff, the isothermal assumption of the fitting model, and lack of exact knowledge of whether the non-thermal emission is from thick-target or thin-target, in this study, we do not calculate thermal or non-thermal energies using the fitting parameters \citep[see discussions in][]{Aschwanden2016, Aschwanden2019}. Energies of individual flares will be estimated in a separate study, taking advantage of the UV Footpoint Calorimeter (UFC) method \citep{Qiu2012, Liu2013, Zhu2018, Qiu2021}, and the model of Thin Flux Tubes (TFTs) heated by slow magnetosonic shocks \citep{Longcope2016}. These methods have been developed to model multi-loop heating in observed flare events, and are free from the isothermal assumption.
 


\subsection{Eruptive Flares}
\label{ssec:eruptive}

\begin{figure}
\centering
\includegraphics[width=3.9cm]{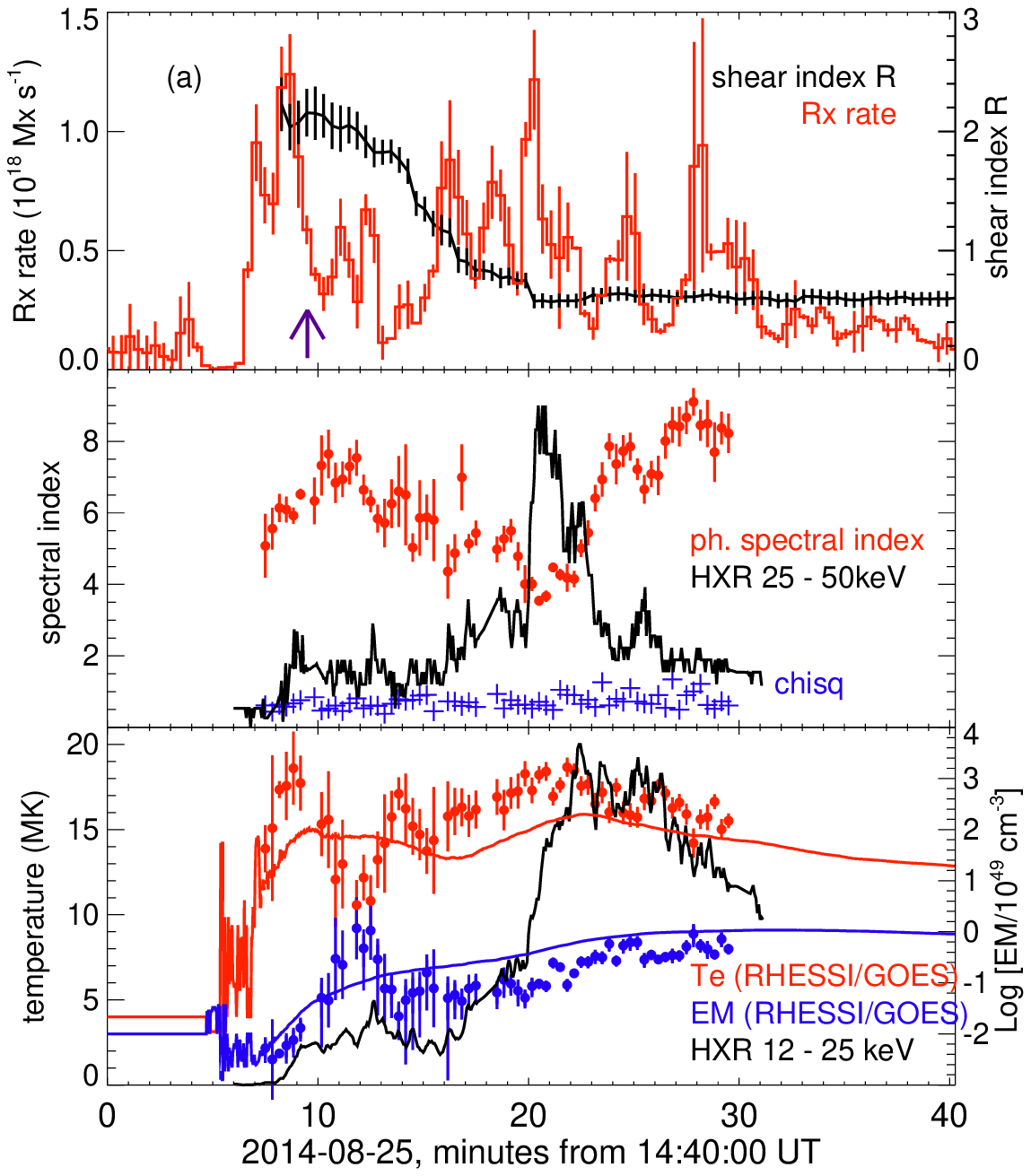}
\includegraphics[width=3.9cm]{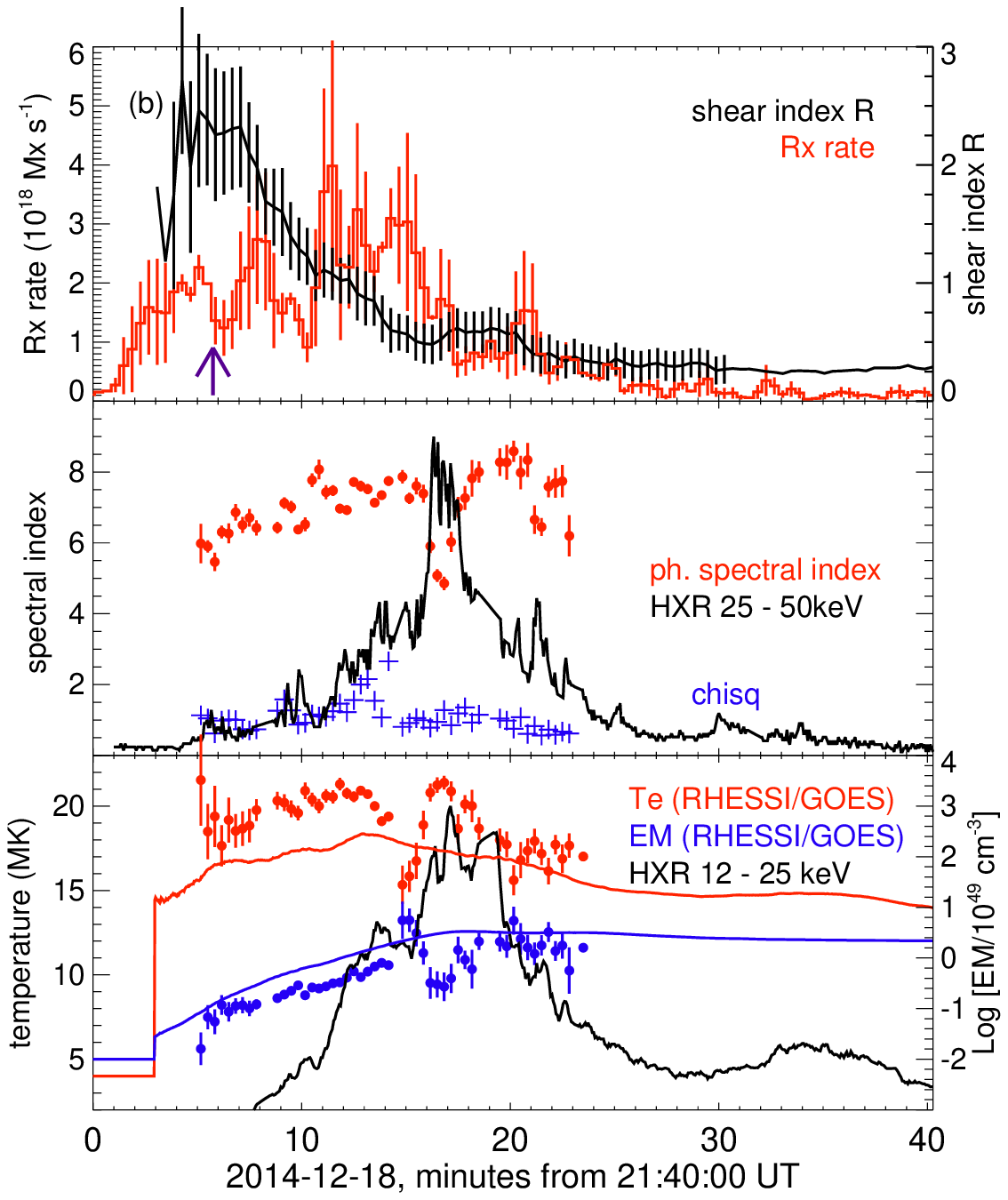}
\includegraphics[width=3.9cm]{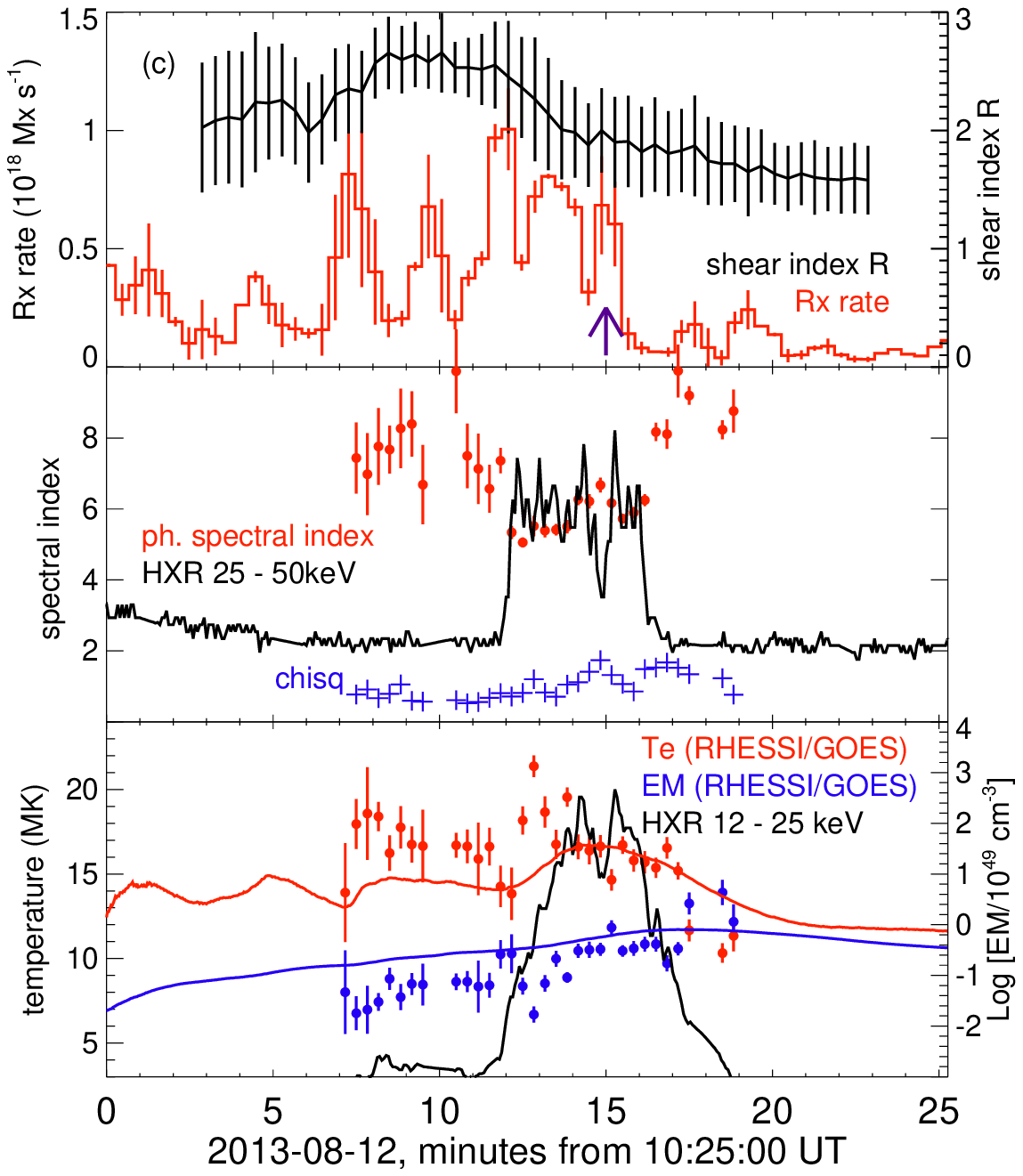}
\caption{Evolution of the eruptive flares SOL2014-08-25 (a), SOL2014-12-18 (b), and SOL2013-08-12 (c). Top: evolution of the mean shear index $\mathcal{R}$ (black) and the reconnection rate $\dot{\psi}$ (red) measured in the two major ribbons (or ribbon sections). Purple arrows indicate the onset times of eruptions observed in EUV images obtained by AIA (see Figure~\ref{fig:eruption}).
Middle: time evolution of the HXR count rates at $25-50$ keV $\mathcal{F}_{hxr}$ (black) and the power-law index $\gamma$ of the non-thermal photon spectrum (red symbols), and the $\chi^2$ of the spectral fitting (blue plus symbols). Bottom: temperature (red) and emission measure (blue) of the isothermal component derived from RHESSI (symbols) and GOES (lines). Also shown are the HXR count rates at $12-25$~keV (black). Vertical bars in properties derived from RHESSI spectral analysis indicate the 1$--\sigma$ uncertainty of the fitting parameters. }
\label{fig:eruptive}
\end{figure}

We examine the time evolution of reconnection properties and flare energetics in eruptive flares (Figure~\ref{fig:eruptive}) and confined flares (Figure~\ref{fig:confined}). 
For the SOL2014-08-25 event (Figure~\ref{fig:eruptive}a), during the first 12 minutes from around 14:48~UT to 15:00~UT, $\mathcal{R}$ varies from 2.0 to 0.6, which reflects the strong-to-weak shear evolution of the post-reconnection arcade. Note that the rapid elongation of the brightening in the early phase of the flare leads to a large $\dot{\psi}$; the mean reconnection electric field, on the other hand, is more significant when the perpendicular motion dominates later in the flare progress.
From 14:52~UT, the non-thermal HXR emission $\mathcal{F}_{hxr}$ gradually grows as the shear decreases, and the photon spectrum exhibits the well-known soft-hard-soft behavior with the power-law spectral index $\gamma$ roughly anti-correlated with the HXR count rates. $\mathcal{F}_{hxr}$ starts to rise rapidly at 15:00~UT when the mean magnetic shear has dropped to less than 1, and peaks at $\mathcal{R} \approx 0.6$. At this time, there is a significant non-thermal component in the flare HXR photon spectrum from 15 to 40~keV, with the hardest spectrum $\gamma \approx 4.0$.

Overall, from the flare onset to the peak of $\mathcal{F}_{hxr}$, the non-thermal photon spectrum gradually hardens as the shear $\mathcal{R}$ decreases. In the following 7 minutes, whereas the mean shear stays low, $\mathcal{F}_{hxr}$ starts to diminish and $\gamma$ increases, as reconnection is diminishing. These observations suggest that the combination of a strong reconnection rate and relatively low shear are in favor of producing non-thermal electrons in this event.

It is worth noting that, before 14:50~UT, prior to the significant enhancement of the $>25$ keV HXRs, 
spectroscopic analysis reveals the presence of a hot plasma component of temperature $T_e \approx 18$~MK with a low emission measure $EM \approx 8\times10^{46}$ cm$^{-3}$, coincident with the spike in the $\dot{\psi}$ profile at the start of the flare. The rise of the plasma temperature prior to significant non-thermal emission in this event is also confirmed with GOES data. Such a hot onset has been reported in recent studies \citep[e.g.][]{Hudson2021}. Note that in this event, at the onset, both the temperature and emission measure grow, different from \citet{Hudson2021} where the temperature is nearly constant at 10 - 15~MK. In the first few minutes of the flare, the rise of $T_e$ along with the rise of $\dot{\psi}$ is an indication of continuous and enhanced heating. EUV and HXR images before 14:50~UT suggest that this high temperature emission may come from sheared loops (Figure~\ref{fig:overview_140825}d).

In the same manner, we have analyzed the other two eruptive flares, the SOL2014-12-18 M6.9 flare, and the SOL2013-08-12 M1.5 flare. Both flares exhibit a slow rise of the UV emission starting 5-10 min before the rapid rise of $\mathcal{F}_{hxr}$, when the mean shear $\mathcal{R}$ has decreased substantially. Evolution of the SOL2014-12-18 M6.9 flare follows a very similar pattern to that of the SOL2014-08-25 flare, with the mean shear variation from $\mathcal{R} \approx  2.5$ at the onset of the flare to $\mathcal{R} < 1$ in 10 minutes, when $\mathcal{F}_{hxr}$ starts to rise rapidly. The last event, the SOL2013-08-12 M1.5 flare, exhibits strong shear throughout the flare, which varies from 2.5 to 1.7 at the rise of $\mathcal{F}_{hxr}$. In all three flares, prior to the rapid rise of $\mathcal{F}_{hxr}$, high temperature ($T_e \le 20$~MK), but not super-hot, plasmas are produced, and the plasma temperature and emission measure derived from RHESSI observations are generally consistent with those derived using the GOES two-channel observations.


\begin{figure}
\centering
\includegraphics[width=13cm]{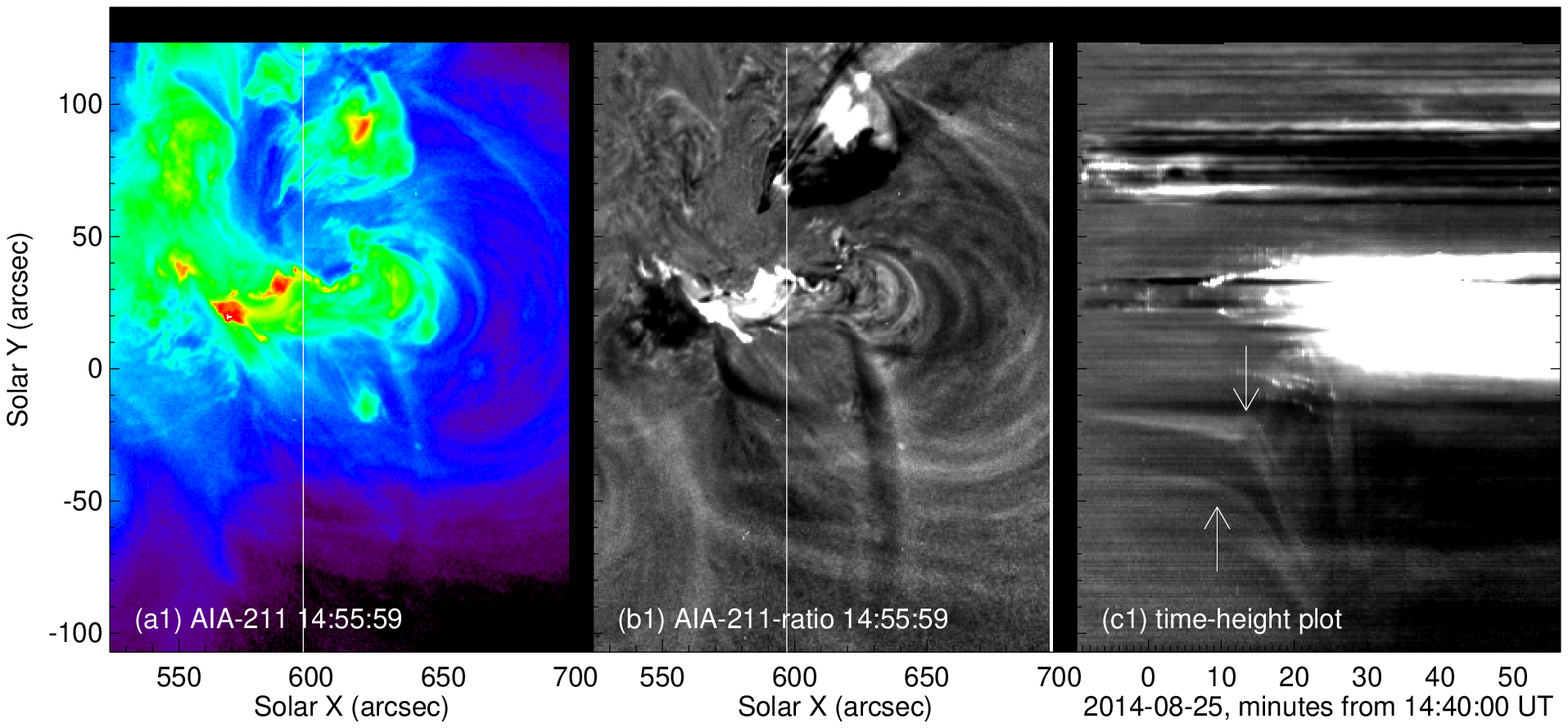}
\includegraphics[width=13cm]{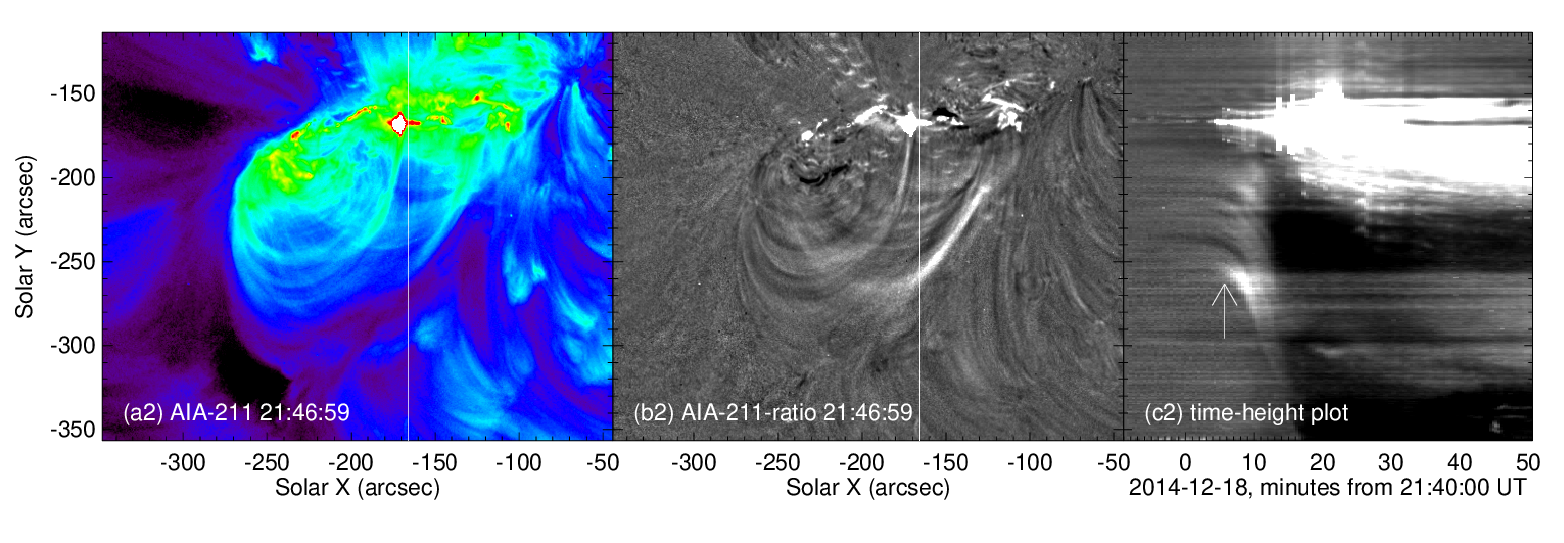}
\includegraphics[width=13cm]{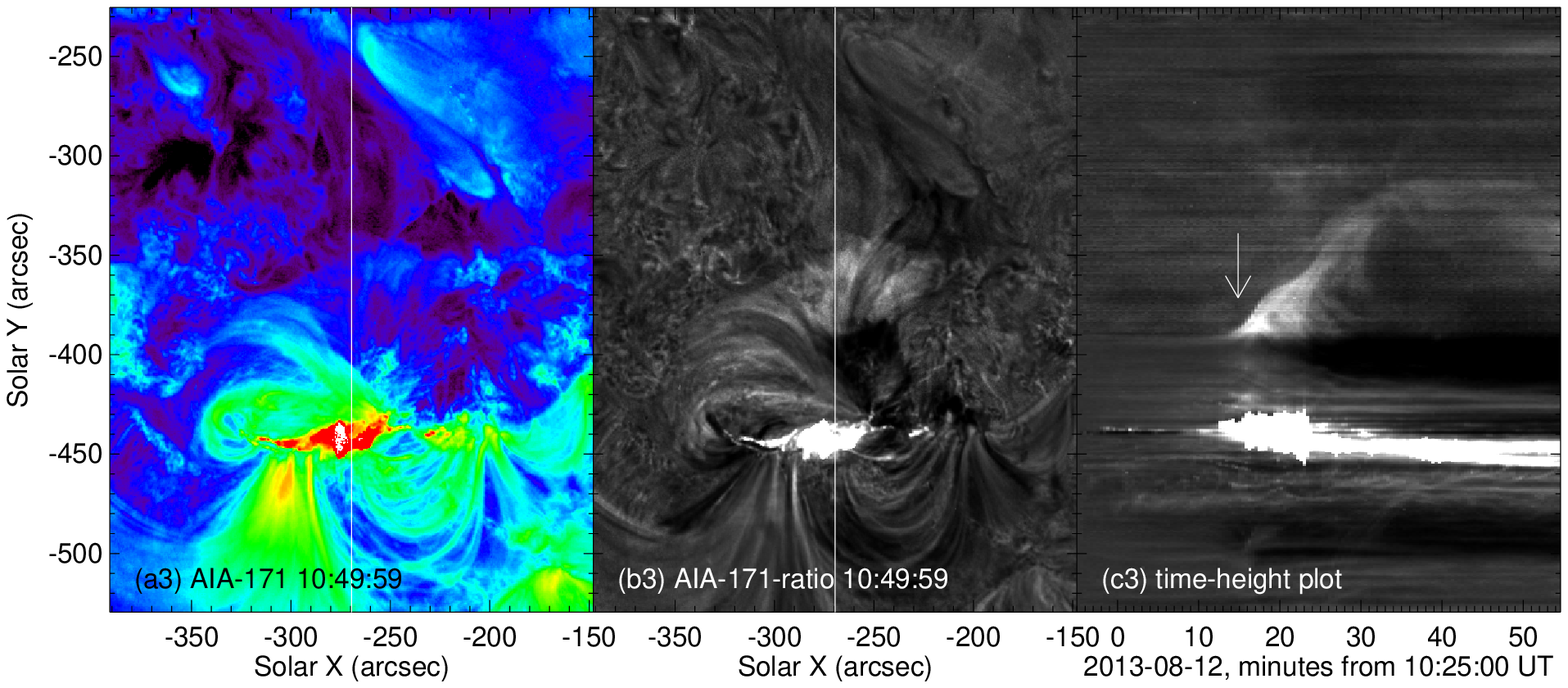}
\caption{Snapshots of coronal structures observed in (a) AIA EUV images, (b) the ratio images, and (c) the time-distance diagram along the slit in panels b, for the three eruptive flares SOL2014-08-25 (top), SOL2014-12-18 (middle), and SOL2013-08-12 (bottom). Arrows in panels c1-c3 indicate the times when coronal structures are observed to erupt.}
\label{fig:eruption}
\end{figure}

Furthermore, in eruptive flares, the eruption can significantly change the magnetic configuration. We therefore examine the evolution of the mean shear, as well as flare energetics, with respect to the timing of the onset of the eruption in the three eruptive flares, identified from high-cadence high-resolution observations of coronal structures in AIA EUV images. Figure~\ref{fig:eruption} shows EUV images (panels a), and the ratio images, which are the same images normalized to the pre-flare base image (panels b), of the active-region corona during the flare. Erupting coronal structures can be tracked in the time-distance diagrams constructed using ratio images along the slits shown in panels a and b. Arrows in panels c indicate the identified times of the eruption onset, which are also marked by purple arrows in Figure~\ref{fig:eruptive}. For the first two flares, the onset of the eruption occurs in the early phase of flare reconnection, when $\mathcal{R}$ starts to decrease, and before the perpendicular expansion of flare ribbons. Significant HXR emission occurs several minutes after the eruption onset, when the mean shear has decreased substantially. Whereas in the third event, the observed onset of the eruption occurs in the phase of ribbon expansion, a few minutes after the rise of $\mathcal{F}_{hxr}$ when the shear has already decreased. The relative timing of the shear evolution with respect to the eruption onset is different in these three flares, which may be related to the configuration of overlying fields. On the other hand, the timing of significant HXRs seems to be more sensitive to the shear decrease than to the eruption onset.

\subsection{Confined Flares}
\label{ssec:confined}



\begin{figure}
\centering
\includegraphics[width=3.9cm]{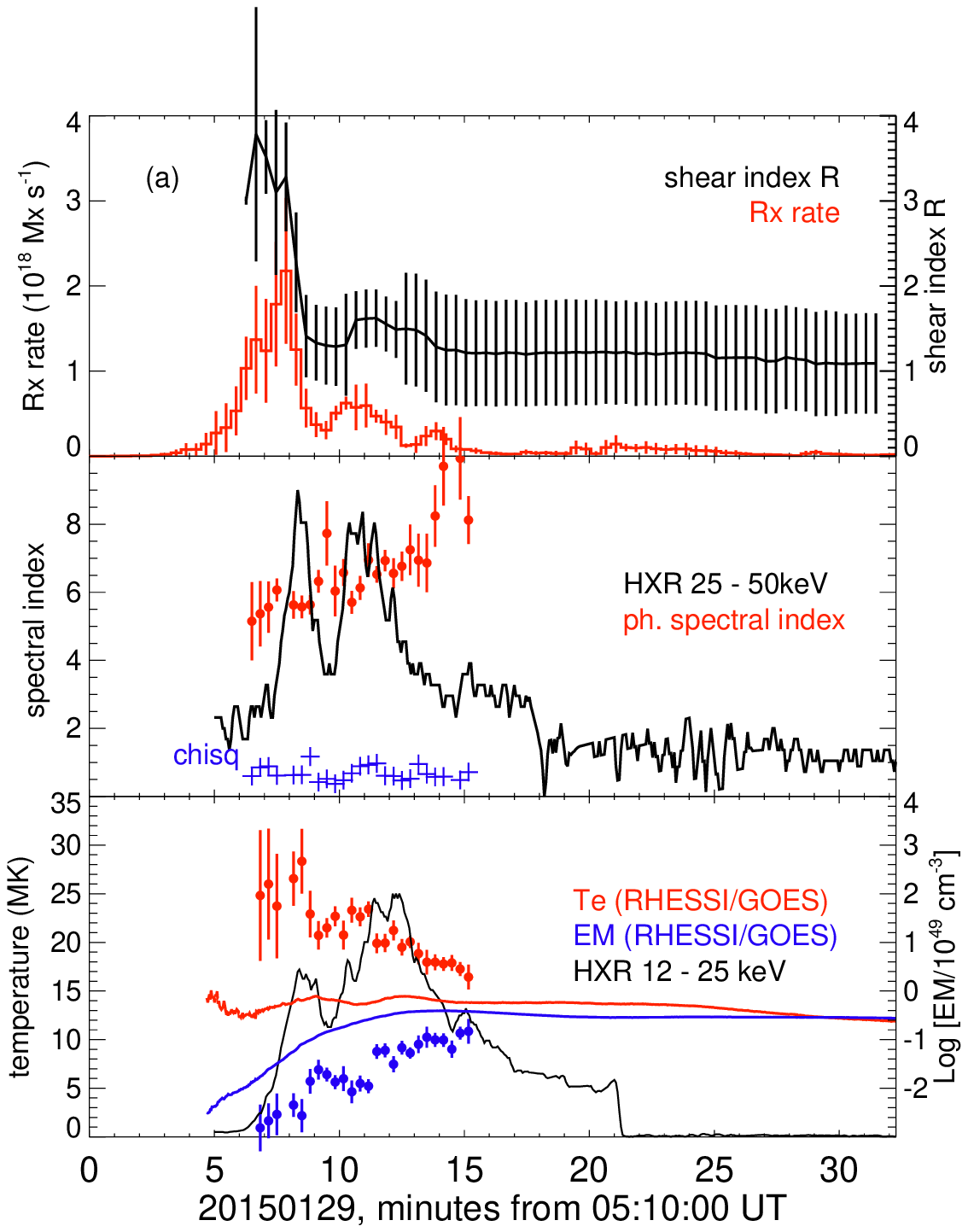}
\includegraphics[width=3.9cm]{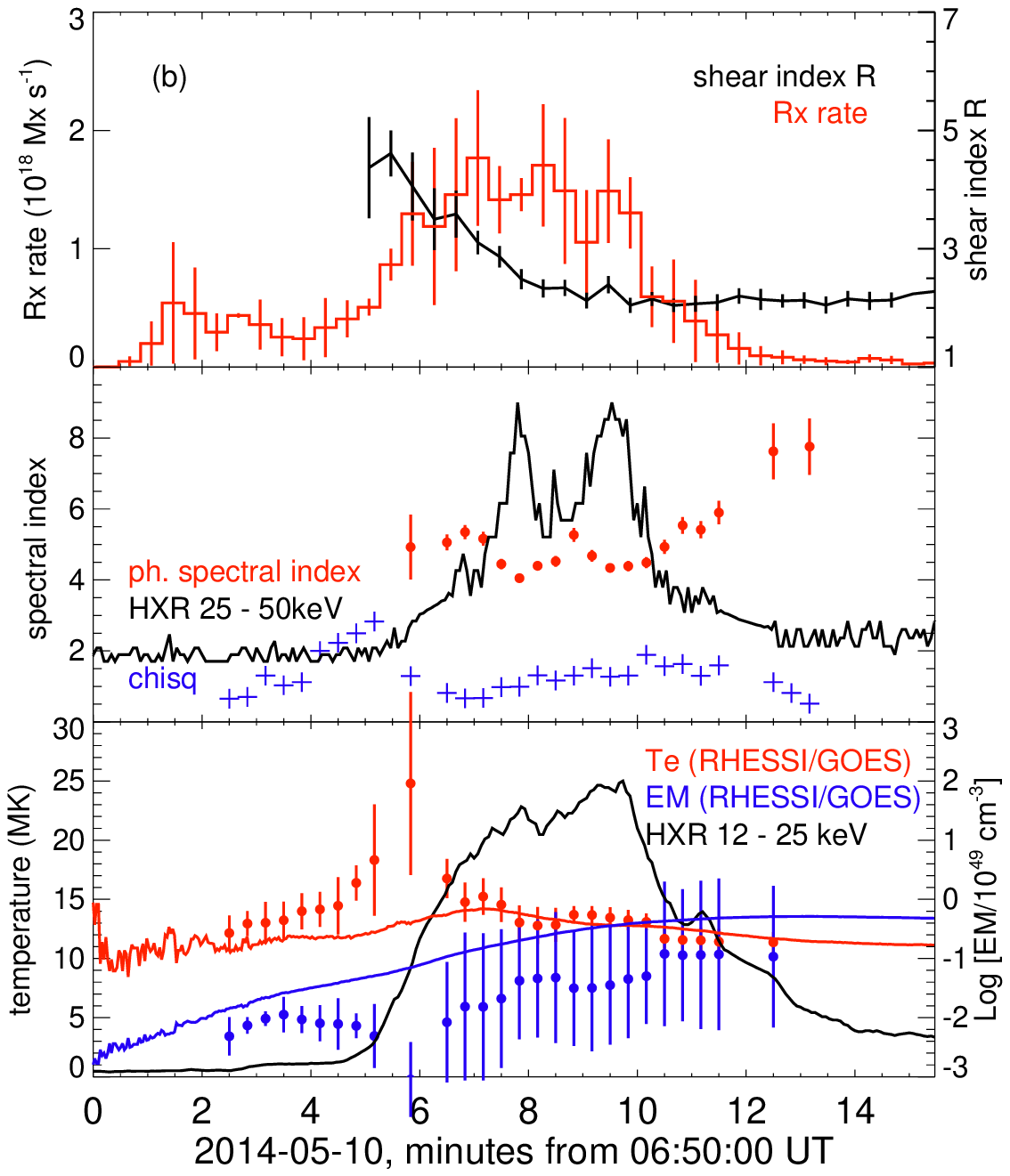}
\includegraphics[width=3.9cm]{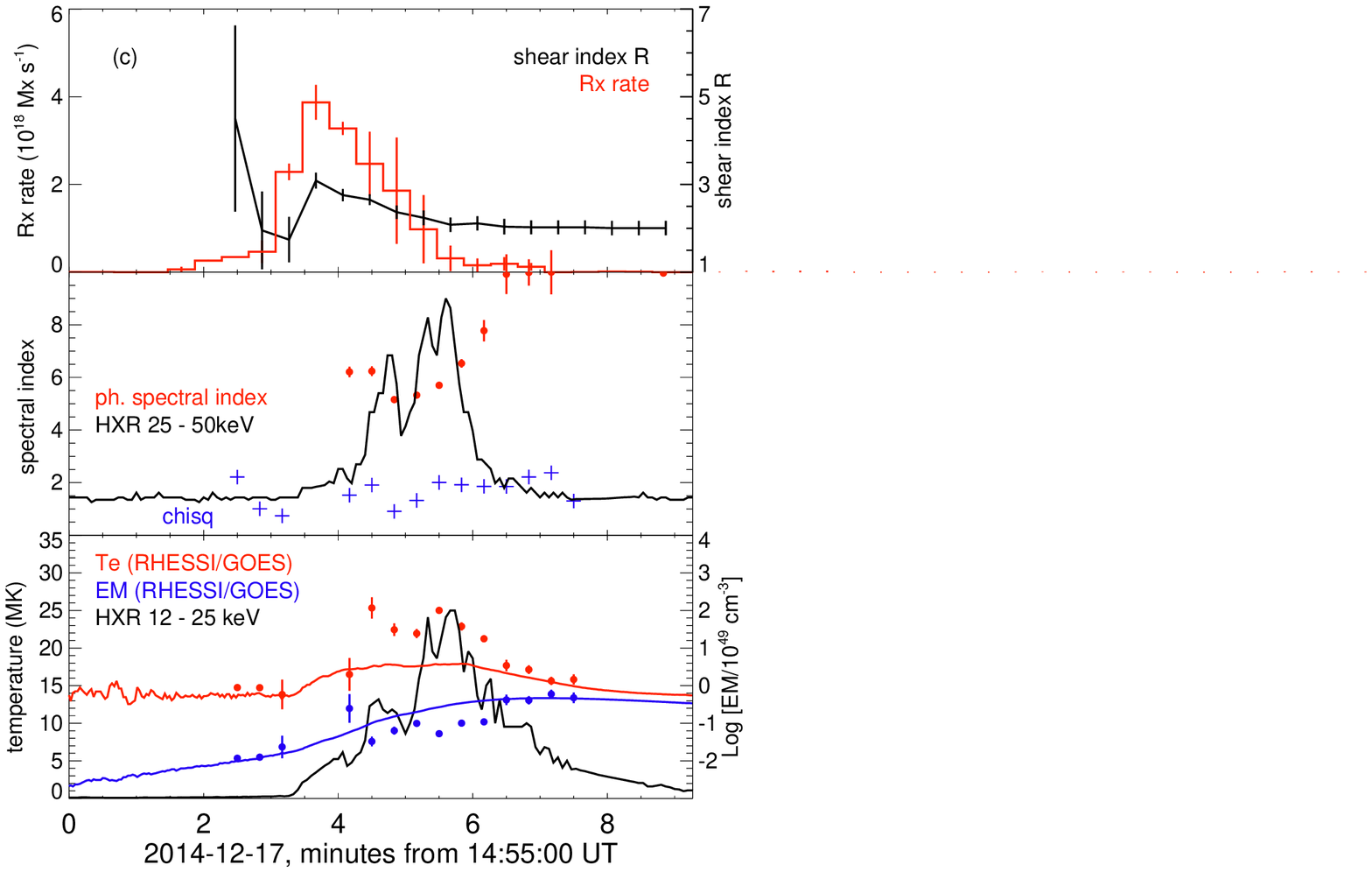}
\caption{Same as Figure~\ref{fig:eruptive}, but for the confined flares SOL2015-01-29 (a), SOL2014-05-10 (b), and SOL2014-12-17 (c). For the SOL2014-05-10 flare, we take the mean of the fitting results from two different detectors.}
\label{fig:confined}
\end{figure}

Analyses of the shear evolution and X-ray energetics are also conducted with the confined flares (Figure~\ref{fig:confined}).
The SOL2015-01-29 C8.9 flare (Figure~\ref{fig:confined}a) is fast evolving, exhibiting HXR emissions in photon energy $\ge$25~keV nearly coincident with the enhanced UV emission. During the rise of $\mathcal{F}_{hxr}$, the inferred magnetic shear decreases rapidly from $\mathcal{R} \approx 3.0$ to $\mathcal{R} \approx 1.5$.
Spectroscopic analysis reveals the presence of a weak non-thermal component with a rather soft photon spectrum, $\gamma \approx 5.0 - 6.5$. Throughout the flare evolution, the non-thermal photon spectrum is gradually softening. Notably, at the onset of the flare when the shear index is large $\mathcal{R} \approx 3.0$, the X-ray spectrum is dominated by a super-hot component, characterized by a mean temperature over 25~MK and tenuous emission measure $3\times10^{47}$ cm$^{-3}$. The plasma temperature subsequently decreases as the flare progresses in the following 10 minutes. Different from the eruptive flares, analysis with the GOES observations does not yield the high temperature component at the onset of the flare, perhaps because of the low sensitivity of GOES detectors to $>20 $~MK plasmas \citep{hannah2011}. 

The properties of the other two confined flares are presented in Figure~\ref{fig:confined}b and \ref{fig:confined}c, respectively. The lead time between the flare onset and production of significant non-thermal HXR emission is very short, within a minute. Although all confined flares also exhibit the strong-to-weak shear evolution, the mean shear of confined flares is larger than that of eruptive flares in this sample. The mean shear of the SOL2014-05-10 event varies from 4.5 at the onset to 2.5 at the peak of the HXR, and the HXR photon spectral index $\gamma$ varies between 4.0 and 4.5. For the SOL2014-12-17 event, the shear varies sharply from more than 5.0 at the onset to 2.0 at the HXR peak, when the HXR spectral index varies between 6.5 and 5.0. 
The two flares also exhibit a hot plasma component during the rise of the HXR emission, with the peak temperature up to 25~MK. This is a prominent difference from eruptive flares in this study. On the other hand, we caution that, for the SOL2014-05-10 flare, the presence of this hot component is brief, and there are large uncertainties in the fitting results. These uncertainties are possibly related to the choices of the initial guess or fitting model, such as the low-energy cutoff and the isothermal assumption, which can affect the results derived from the non-linear inversion method.

\subsection{Evolution of Magnetic Shear and Flare Energetics}
\label{ssec:all}

Comparison between Figures~\ref{fig:eruptive} and \ref{fig:confined} suggests differences in the evolution of flare energetics between eruptive and confined flares studied in this sample. For the eruptive flares, there appears a long, of up to 10 minutes, warm-up phase before the onset of impulsive and significant non-thermal HXR emission. During this warm-up phase, as reconnection proceeds, the mean shear decreases, X-ray and UV emissions ramp up gradually, and the temperature and emission measure of X-ray emitting plasmas also grow slowly. These signatures indicate gradual energy release at a low level, with the lower atmosphere being heated by either thermal conduction or small amounts of non-thermal flux. Significant HXR and UV emissions take place afterwards, with a much enhanced rate of energy release, in particular non-thermal energy release, per unit reconnection flux. It is also noted that, in this limited sample, the progress of the shear evolution and flare energetics does not seem to depend on when the eruption takes place. The confined flares, on the other hand, evolve on shorter timescales, and significant HXR and UV emissions, or qualitatively the energy release rate, tracks the reconnection rate more closely from the start of the flare. 

\begin{figure}
\centering
\includegraphics[width=15cm]{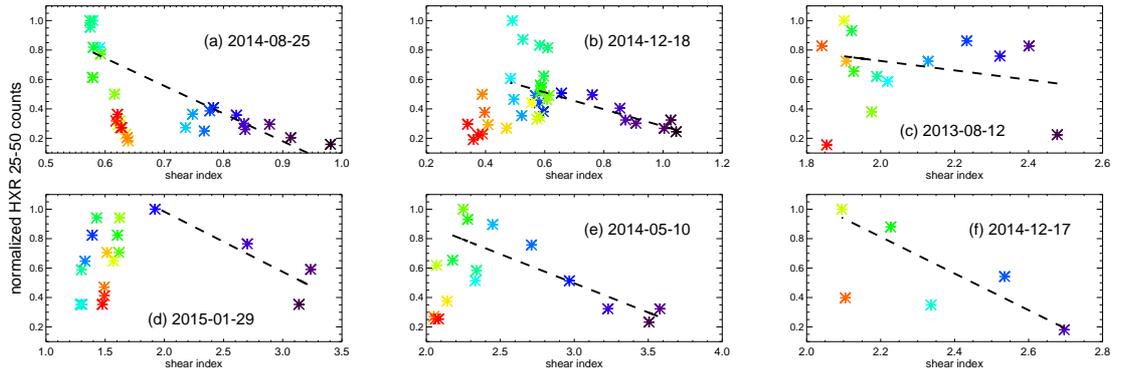}
\caption{HXR 25-50 keV count rates $\mathcal{F}_{hxr}$ (normalized to its peak) against the shear index $\mathcal{R}$ for erutpive flares (top) and confined flares (bottom). Here the color indicates the time evolution from the rise (violet-blue) to the peak (cyan-green) and then the decay (yellow-red) of the HXRs. Only time intervals when $\mathcal{F}_{hxr}$ is above 20\% of its peak are plotted.}
\label{fig:allevent}
\end{figure}

In spite of these differences, in both groups of flares, the mean shear decreases as the flare proceeds. We further probe whether, and how, non-thermal and/or thermal properties of the flare are related to the shear evolution. Figure~\ref{fig:allevent} illustrates the flare HXR emission $\mathcal{F}_{hxr}$ with respect to the shear variation during the progress of flare reconnection. In each panel, $\mathcal{F}_{hxr}$ is plotted against the shear index $\mathcal{R}$ in asterisk symbols. 
The color code indicates the time evolution from the rise (violet-blue) to the peak (cyan-green) and then to the decay (orange-red) of $\mathcal{F}_{hxr}$. For each event, only time intervals with $\mathcal{F}_{hxr}$ above 20\% of its peak are plotted. 

From these plots, it is immediately evident that, in most of the events, there is nearly an anti-correlation between $\mathcal{F}_{hxr}$ and $\mathcal{R}$ during the rise of the HXR emission; this is just the representation that the shear decreases as the HXR rises, as demonstrated in Figures~\ref{fig:eruptive} and ~\ref{fig:confined}. The $R-\mathcal{F}_{hxr}$ pairs from the onset to the peak of $\mathcal{F}_{hxr}$ in most events can be well fitted to a straight line, as indicated by the dashed lines. The coefficient $\rho$ of the linear cross-correlation between these two parameters is provided in Table~\ref{tab:info}, which ranges between $-0.60$ and $-0.90$ in the five events. Given that the number of time intervals for the correlation analysis is rather small, between 5 and 10, the presented anti-correlation should be taken as a conservative result. 

The analysis is not extended to the decay of the HXR emission, because $\mathcal{R}$ stays low and does not vary much after the peak of the HXR emission; on the other hand, the reconnection rate $\dot{\psi}$ starts to decrease. These observations suggest that production of non-thermal HXRs should be related with the combination of the reconnection rate $\dot{\psi}$ and the shear index $\mathcal{R}$, although it is unclear in what way.

The evolution of the non-thermal spectral index $\gamma$ with respect to the HXR flux $\mathcal{F}_{hxr}$ generally follows the well-known soft-hard-soft pattern, with the hardest spectrum around the peak of $\mathcal{F}_{hxr}$ (and also the lowest shear)
for each flare. A correlation analysis between $\gamma$ and $\mathcal{R}$, however, returns only an insignificant correlation coefficient; therefore, the quantitative relation between these two parameters is still remote from this limited sample study.
We may also compare the $\gamma - \mathcal{R}$ relation between different events. Here, for each event, we derive $\langle \gamma \rangle$ and $\langle \mathcal{R} \rangle$ averaged over the time intervals when $\mathcal{F}_{hxr}$ is above 70\% of its peak. The mean and deviation of $\gamma$ and $\mathcal{R}$ are provided in Table~\ref{tab:info}. From this small sample, we do not observe any meaningful relation between $\langle \mathcal{R} \rangle$ and $\langle \gamma \rangle$ across different flares. 

Apart from non-thermal properties, we also examine the plasma thermal properties. 
For each flare, the mean temperature and mean shear index averaged over three time intervals around the peak temperature are presented in Table~\ref{tab:info}. It is noted that, pending some uncertainties in the fitting procedure, the confined flares exhibit a hot plasma component at the start of the flare when the shear is high, with the temperature up to 25~MK, whereas none of the eruptive flares have produced hot plasmas beyond 20~MK. 
Confined flares in general have a larger shear index than eruptive flares in this sample. Beyond these observations, we find no further quantitative relation between plasma temperatures and shear indices.  

\section{Discussion}
\label{sec:discussion}

To understand the results from this study and their implications, we discuss assumptions underlying the measurements, the similarities and differences of these results in comparison with earlier studies, and how our studies may be further improved in the future.

\subsection{Measurements of Shear and Inference of Reconnection Guide Field}

To measure the shear, this study uses an approximate approach that projects the apparent motion of newly brightened ribbon fronts
along (parallel with) and across (perpendicular to) the curved PIL. The approach in this experiment applies to a relatively simple configuration characterized by a single major PIL separating nearly bipolar magnetic fields, which can be approximated to a 2.5D geometry, with the guiding direction along the PIL. The key to establish the shear is connectivity. In this study \citep[and also in][]{Qiu2010}, the connectivity is assumed between newly brightened ribbon fronts, and is qualitatively confirmed by post-flare loops observed in EUV images and/or maps of the thick-target HXR emission. Using a different method, \citet{Qiu2017} also measured the shear as the angle between the top of a set of post-reconnection flare loops outlined in EUV images and the PIL of the magnetic field extrapolated to the height of the flare loops, and showed that the general pattern of shear variation is consistent with that measured from flare ribbons. Again, we emphasize that the measurements in this article are approximate and return the average shear. \citet{Qiu2009} applied a more refined flare ribbon analysis to the SOL2004-11-07 X-2.0 flare, by considering the connectivity between multiple patches of flare ribbons in both polarities; a reconnection sequence analysis was conducted to derive dynamically evolving connectivity between eight ribbon patches in the positive magnetic field and six patches in the negative field. The more involved approach has provided a more comprehensive description of the connectivity, yet the pattern of the shear evolution is similar: as the flare progresses, the shear decreases. In some other studies, the shear was measured with a simpler method, as the angle of a straight line connecting two prominent and conjugate footpoints identified in either H$\alpha$ and UV/EUV images \citep{Su2006, Ji2006, Su2007} or HXR maps \citep{Yang2009, Inglis2013} with respect to the PIL also approximated as a straight line. However the shear is measured, the strong-to-weak shear evolution is found for the majority of flare events reported in the literature; for example, \citet{Su2007} measured the shear from conjugate UV footpoints at two times during a flare, and found the strong-to-weak shear change in 86\% of the 50 M- and X- class flares. Such shear evolution is often accompanied by the two-phase motion pattern of flare ribbon fronts or kernels, with the ribbon brightenings first elongating along the PIL, followed by the apparent perpendicular expansion of the two ribbons \citep[][and references therein]{Qiu2017}. The shear index $\mathcal{R}$ defined and measured in the selected samples in this study exhibits a behavior consistent with that reported in the literature, and such evolution pattern reflects the sequence of magnetic reconnection driven in the magnetic configuration that builds up free magnetic energy with strong magnetic shear near the PIL \citep{Dahlin2021}.

On the other hand, we recognize that although the strong-to-weak shear evolution is commonly observed and/or measured, it is not universal. A small number of {\em measurable} events do exhibit different patterns of shear evolution \citep{Su2007, Yang2009, Cheng2012}. A lot many flare events occur in regions of complicated magnetic configuration characterized by complex-shaped PILs, for which it is not always feasible to apply a simplified approach to measure, or even define, the shear \citep{Bogachev2005, Grigis2005}. Furthermore, published methods, including the one used in this study, measure the shear assuming the connectivity between two sources. In different studies, these two sources are selected in different ways. In this article, we measure the average positions of newly brightened ribbons in the central or core region, and the mean shear derived for this core region tends to follow the strong-to-weak evolution. It is noted that, in some eruptive flares, if the connectivity is established between the far ends of the two ribbons, which may indicate the feet of an erupting flux rope, the shear appears to increase, as also shown in numerical simulations \citep{Aulanier2019, Dahlin2021}. Many other studies assume connectivity between two kernels of strong emissions on the two sides of the PIL, and estimate the magnetic shear tracking this specific connectivity. During the flare, a large number of flare loops, of order a few hundred, are produced. Improved measurement of the shear requires a more refined connectivity analysis applied to high-resolution observations, such as by \citet{Fletcher2009} and \citet{Qiu2009}, and examine the tempo-spatial variations of magnetic shear in contrast to the average shear measured in prior and present studies.

Next, shear measured from the morphological evolution of flare ribbons or kernels reflects topology of reconnection in the corona. However, establishing the mapping of the coronal magnetic field down to the ribbon is not trivial. Approximating the coronal magnetic field with a 2.5D configuration, $\mathcal{R}$ may be related to the ratio of the reconnection guide field to the outflow field $\mathcal{R} \sim B_g/B_o$. We emphasize that the value of $B_g/B_o$ inferred in this way should not be taken on face value, although the trend of the evolution is evident, and is also broadly consistent with the $B_g/B_{in}$ evolution derived using high-resolution MHD simulations that do provide the mapping of the coronal magnetic field to the ribbons \citep{Dahlin2021}. In our future effort, observational measurements of the shear and inference of the guide field $B_g/B_o$ will also be explored together with microwave observations when available \citep{Gary2018, Chen2020}, and be complemented with magnetic field modeling guided by observed flare loop geometry. Ultimately, the guide field relative to the inflow magnetic field $B_{in}$ at the reconnection current sheet could be estimated incorporating advanced and high-resolution numerical models of flare reconnection \citep[e.g.][]{Jiang2021, Dahlin2021}.

\subsection{Reconnection Properties and Flare Eenergetics}

In this article, the preliminary study of the sample suggests that a significant production of non-thermal electrons may take place a few minutes after the onset of the flare, when the inferred mean shear is decreasing. Previous studies have reported delayed HXR emission with respect to the UV or optical emission. To cite a few prominent examples, \citet{Warren2001} collected nine M-class flares observed 
with very high cadence (2-3~s) and from the beginning of the flare by
the Transition Region and Coronal Explorer \citep[TRACE:][]{Handy1999}, the Hard X-ray Telescope \citep[HXT;][]{Kosugi1991} on board Yohkoh, and/or the Burst And Transient Source Experiment
on board the Compton Gamma-Ray Observatory \citep[BATSE/CGRO:][]{Schwartz1992}, and all events exhibit evident delay of the onset of the HXR emission with respect to the UV emission by 1 -- 10 minutes. In the SOL2006-12-06 X6.5 flare analyzed in \citet{Krucker2011}, the onset of $\ge 25$ keV HXRs lags the flare SXR emission by as long as 10 minutes. These studies did not analyze the shear evolution using UV or optical observations. In {\em eruptive} flares in the present study, HXR emission starts to rise rapidly when the shear index has decreased to $\mathcal{R} \le 2$. Previously, \citet{Qiu2010} analyzed the SOL2000-07-14 X5.7 flare \citep[the Bastille-day flare, also see][]{Aschwanden2001} with the same method that decomposes the motion of flare UV kernels in the directions along and across the PIL, and suggested that significant non-thermal emission occurs when shear is modest. Similarly, \citet{Su2006} measured the shear of the conjugate EUV footpoints of the SOL2003-10-28 X17 flare, and showed that the shear angle, defined as the angle of the footpoint line with respect to the vertical of the PIL, decreased from 75$^{\circ}$ (equivalent to $\mathcal{R} \approx 3.7$) two minutes before the onset of the $\ge 150$~keV HXR emission to 55$^{\circ}$ ($\mathcal{R} \approx 1.4$) when $\ge 150$~keV HXR emission rises. \citet{Ji2006} measured the shear tracking H$\alpha$ emission centroids using the same method as \citet{Su2006}, and found that the shear starts to decrease, $\mathcal{R} \le 1$, at the onset of the $\ge$25 keV HXR emission of the SOL2004-11-01 M1.1 flare. If $\mathcal{R}$ is translated to the relative guide field $B_g/B_o$, these observations would support recent findings from the numerical modeling that efficient production of non-thermal electrons occurs with a modest guide field $B_g/B_{in} \approx 1$ \citep{Arnold2021}.

On the other hand, a number of other studies have measured shear by tracking the HXR foot-points, and showed that, although most events exhibit the strong-to-weak shear evolution, non-thermal HXRs can be produced in the early phase when the shear angle is as large as 60 - 70$^{\circ}$, or equivalently $\mathcal{R} \approx 1.7 - 2.7$ \citep{Yang2009, Cheng2012, Inglis2013, Sharykin2018}.
Furthermore, in all flares in the present paper and in previous studies, shear remains low as the HXR emission diminishes. These observations suggest that shear is not the only parameter related to HXR emissions.  

Previous studies have extensively explored the connection between reconnection rate and production of non-thermal emissions (see Section 1). In particular, \citet{Qiu2010} showed that $\ge 20$keV HXR light curve correlates more strongly with the flux change rate taking into account only the perpendicular motion $\dot{\psi}_{\perp} \approx \int B_l v_{\perp}dl$ than with the total flux change rate $\dot{\psi}$. For the events in this study, we have observed a large flux change rate $\dot{\psi}$ prior to the significant increase of HXRs; since in these events the strong-to-weak shear evolution is coincident with the transition from parallel to perpendicular motion of the ribbons, $\dot{\psi}_{\perp}$ would be much reduced from $\dot{\psi}$ prior to the HXR emission and therefore is better correlated with $\mathcal{F}_{hxr}$ than $\dot{\psi}$, similar to that shown in \citet{Qiu2010}. \citet{Qiu2017} have measured shear evolution as well as the mean parallel and perpendicular motion of flare ribbons; it is suggestive from the properties listed in Table~1 of that study, that flares or episodes of flares with significant HXR emissions tend to exhibit large perpendicular motion $v_{\perp}$ in strong magnetic fields $B_l \ge 500$~G. These observations indicate that the reconnection rate in terms of $\dot{\psi}_{\perp}$ or the reconnection electric field $\langle E \rangle \approx v_{\perp} B_l$ is important, though not always sufficient, for the production of non-thermal emissions \citep[e.g.][]{Naus2021}.

Finally, although strong-to-weak shear evolution is observed in confined flares as well, the energetic behavior appears to be different. HXR emission occurs when the shear is strong $\mathcal{R} \approx 2 - 3$, and a high-temperature (up to $25$~MK), low density plasma component is produced at the onset of the confined flares in this sample. Super-hot emissions have been studied in a number of flares, a good sample having been presented by \citet{Caspi2014}. It remains to be investigated what kind of flares, confined or eruptive, tend to produce hot and super-hot plasma emissions, and what role magnetic shear plays in this regime \citep{Dahlin2015, Warmuth2020}. It is also noted that the shear measurement is based on a 2.5D approximation, and the direction of the PIL is assumed to be the direction of the (macroscopic) electric field in the reconnection current sheet in the corona. This same assumption has been adopted for the confined flare in this study, which presents rather well defined two ribbons along the PIL, although many confined flares are envisioned as (semi-)circular ribbon flares developed around a null point and circular fan-shaped quasi-separatrix layers \citep[QSLs, see][and references thereafter]{Masson2009}. It warrants further research how to locate the reconnection current sheet and determine the reconnection rate $\langle E \rangle$ and guide field in a truly 3D configuration.

From available observations and measurements, it is evident that the production of non-thermal electrons is not singularly or even primarily dependent on one parameter. The disparity presented in prior and present observations is partly due to large differences in the spatio-temporal resolutions, coverage in time and space domains, observing wavelengths, and analysis methods. On the other hand, it is anticipated that in different flares and different magnetic configurations, the governing mechanisms for flare energetics, for example, the particle acceleration mechanism, can be different as well. How, and which, key parameters collectively control particle acceleration and generation of the power-law energy spectrum remains to be clarified from both theoretical investigation and improved observational analysis that can provide measurements of temporally and spatially resolved properties of flare reconnection and energetics.

\section{Summary}
This article presents an experiment that infers the mean magnetic shear from the evolution of flare ribbons observed in the lower atmosphere, and examines properties of flare energy release during the progress of magnetic reconnection. The sample includes three eruptive flares and three confined flares ranging from C8.0 - M7.0 class. All these flares exhibit well-defined two ribbons aligned with a relatively simple PIL. We define the shear index $\mathcal{R}$ as the ratio of the mean parallel distance between newly brightened conjugate flare ribbons projected along the PIL to their mean distance in the direction perpendicular to the PIL. 
All these flares have been observed by RHESSI from the onset of the flare throughout the rise phase. At the peak time, these flares have non-thermal HXRs at photon energies above 25 keV. Spectral analysis provides flare non-thermal properties such as the power-law slope of the photon spectrum, as well as the thermal properties including the plasma temperature. Our analysis yields the following results.

\begin{itemize}
\item In the six events in this study, $\mathcal{R}$ varies from more than 3 at the onset of the flare to below 1 at the peak of the HXR emission, confirming the widely reported strong-to-weak shear variation during the progress of flare reconnection. The mean shear does not vary much after the peak of the HXR emission. On average, confined flares exhibit stronger shear.

\item In most of the events, there is an indication of anti-correlation between the $\ge 25$ keV HXR count rates $\mathcal{F}_{hxr}$ and the shear index $\mathcal{R}$ during the rise of the HXR emission. In eruptive flares in this sample, the UV emission and global reconnection rate $\dot{\psi}$ lead the significant and impulsive HXR emission by up to 10 minutes, suggesting the presence of a gradual warm-up phase when the efficiency of (non-thermal) energy release is low. Significant non-thermal emission occurs later when the shear is modest, $\mathcal{R} \le 2$. In this limited sample, the production of significant non-thermal emissions appears more sensitive to the shear decrease than the timing of the eruption onset.


\item In all flares, we observe the temperature rise in the early phase when the flare ribbons rapidly spread along its length. RHESSI X-ray spectral analysis reveals a high-temperature component (up to 25~MK) at the onset of the confined flares, lasting for up to one minute. This high-temperature component is not observed in the eruptive flares, although the eruptive flares all have higher magnitude than the confined flares in this sample.

\end{itemize}

These observations bear some indications of the role magnetic shear possibly plays in the production of non-thermal electrons and hot plasmas during the progress of flare reconnection and in different magnetic configurations, namely in eruptive versus confined flares. Improved diagnostics of flare energetics combined with modeling will help with the progress towards elucidating physical mechanisms underlying these results.

\begin{acks}
This work was started from the discussion with Dr. Paul Cassak, and its completion was motivated by the Solar Flare Energy Release (SolFER) collaboration and conversation with Dr. James Drake. We thank the anonymous referee for insightful suggestions and the editor (Cristina H. Mandrini) for editing the manuscript. We thank Ms. Lilly Bralts-Kelly and Heather Mei for helping prepare observations during the Research Experiences for Undergraduates (REU) program at Montana State University. J.C. is funded by the B-type Strategic Priority Program of the Chinese Academy of Sciences, Grant No. XDB41000000, NSFC grants 11673048. SDO is a mission of NASA's Living With a Star Program. The authors also thank the RHESSI Mission Archive  \url{https://hesperia.gsfc.nasa.gov/rhessi\_extras/flare\_images/hsi\_flare\_image\_archive.html}
for the data support.

\end{acks}

\section*{Disclosure of Potential Conflicts of Interest}
The authors declare that they have no conflicts of interest.
\bibliographystyle{spr-mp-sola}
\bibliography{shear}

\begin{thebibliography}{84}
\ifx\bisbn     \undefined \def\bisbn  #1{ISBN #1}\fi
\ifx\binits    \undefined \def\binits#1{#1}\fi
\ifx\bauthor   \undefined \def\bauthor#1{#1}\fi
\ifx\batitle   \undefined \def\batitle#1{#1}\fi
\ifx\bjtitle   \undefined \def\bjtitle#1{\textit{#1}}\fi
\ifx\bvolume   \undefined \def\bvolume#1{\textbf{#1}}\fi
\ifx\byear     \undefined \def\byear#1{#1}\fi
\ifx\bissue    \undefined \def\bissue#1{#1}\fi
\ifx\bfpage    \undefined \def\bfpage#1{#1}\fi
\ifx\blpage    \undefined \def\blpage #1{#1}\fi
\ifx\burl      \undefined \def\burl#1{\textsf{#1}}\fi
\ifx\href      \undefined \def\href#1#2{\textsf{#2}}\fi
\ifx\betal     \undefined \def\betal{\textit{et al.}}\fi
\ifx\bctitle   \undefined \def\bctitle#1{#1}\fi
\ifx\beditor   \undefined \def\beditor#1{#1}\fi
\ifx\bbtitle   \undefined \def\bbtitle#1{\textit{#1}}\fi
\ifx\bedition  \undefined \def\bedition#1{#1}\fi
\ifx\bseriesno \undefined \def\bseriesno#1{\textbf{#1}}\fi
\ifx\blocation \undefined \def\blocation#1{#1}\fi
\ifx\bsertitle \undefined \def\bsertitle#1{\textit{#1}}\fi
\ifx\bsnm      \undefined \def\bsnm#1{#1}\fi
\ifx\bsuffix   \undefined \def\bsuffix#1{#1}\fi
\ifx\bparticle \undefined \def\bparticle#1{#1}\fi
\ifx\barticle  \undefined \def\barticle#1{}\fi
\ifx\binstitute  \undefined \def\binstitute#1{#1}\fi
\ifx\bpublisher  \undefined \def\bpublisher#1{#1}\fi
\ifx\doiurl    \undefined
  \def\doiurl#1{\href{http://dx.doi.org/#1}{\textsf{DOI}}}\fi
\ifx\arxivurl  \undefined
  \def\arxivurl#1{\href{http://arxiv.org/abs/#1}{\textsf{arXiv}}}\fi
\ifx\adsurl    \undefined
  \def\adsurl#1{\href{http://adsabs.harvard.edu/abs/#1}{\textsf{ADS}}}\fi
\ifx\botherref \undefined \def\botherref#1{}\fi
\ifx\url       \undefined \def\url#1{\textsf{#1}}\fi
\ifx\bchapter  \undefined \def\bchapter#1{}\fi
\ifx\bbook     \undefined \def\bbook#1{}\fi
\ifx\bcomment  \undefined \def\bcomment#1{#1}\fi
\ifx\oauthor   \undefined \def\oauthor#1{#1}\fi
\ifx\citeauthoryear \undefined\def \citeauthoryear#1{#1}\fi
\ifx\endbibitem\undefined \def\endbibitem{}\fi
\ifx\bconflocation  \undefined \def\bconflocation#1{#1} \fi

\bibitem[\protect\citeauthoryear{{Arnold} \textit{et~al.}}{2021}]{Arnold2021}
\begin{barticle}
\bauthor{\bsnm{{Arnold}}, \binits{H.}},
\bauthor{\bsnm{{Drake}}, \binits{J.F.}},
\bauthor{\bsnm{{Swisdak}}, \binits{M.}},
\bauthor{\bsnm{{Guo}}, \binits{F.}},
\bauthor{\bsnm{{Dahlin}}, \binits{J.T.}},
\bauthor{\bsnm{{Chen}}, \binits{B.}},
\bauthor{\bsnm{{Fleishman}}, \binits{G.}},
\bauthor{\bsnm{{Glesener}}, \binits{L.}},
\bauthor{\bsnm{{Kontar}}, \binits{E.}},
\bauthor{\bsnm{{Phan}}, \binits{T.}},
\bauthor{\bsnm{{Shen}}, \binits{C.}}:
\byear{2021},
\batitle{{Electron Acceleration during Macroscale Magnetic Reconnection}}.
\bjtitle{\prl}
\bvolume{126}(\bissue{13}),
\bfpage{135101}.
\doiurl{10.1103/PhysRevLett.126.135101}.
\adsurl{https://ui.adsabs.harvard.edu/abs/2021PhRvL.126m5101A}.
\end{barticle}
\endbibitem

\bibitem[\protect\citeauthoryear{{Asai} \textit{et~al.}}{2004}]{Asai2004}
\begin{barticle}
\bauthor{\bsnm{{Asai}}, \binits{A.}},
\bauthor{\bsnm{{Yokoyama}}, \binits{T.}},
\bauthor{\bsnm{{Shimojo}}, \binits{M.}},
\bauthor{\bsnm{{Masuda}}, \binits{S.}},
\bauthor{\bsnm{{Kurokawa}}, \binits{H.}},
\bauthor{\bsnm{{Shibata}}, \binits{K.}}:
\byear{2004},
\batitle{{Flare Ribbon Expansion and Energy Release Rate}}.
\bjtitle{\apj}
\bvolume{611}(\bissue{1}),
\bfpage{557}.
\doiurl{10.1086/422159}.
\adsurl{https://ui.adsabs.harvard.edu/abs/2004ApJ...611..557A}.
\end{barticle}
\endbibitem

\bibitem[\protect\citeauthoryear{{Aschwanden} and
  {Alexander}}{2001}]{Aschwanden2001}
\begin{barticle}
\bauthor{\bsnm{{Aschwanden}}, \binits{M.J.}},
\bauthor{\bsnm{{Alexander}}, \binits{D.}}:
\byear{2001},
\batitle{{Flare Plasma Cooling from 30 MK down to 1 MK modeled from Yohkoh,
  GOES, and TRACE observations during the Bastille Day Event (14 July 2000)}}.
\bjtitle{\solphys}
\bvolume{204},
\bfpage{91}.
\doiurl{10.1023/A:1014257826116}.
\adsurl{2001SoPh..204...91A}.
\end{barticle}
\endbibitem

\bibitem[\protect\citeauthoryear{{Aschwanden}, {Kontar}, and
  {Jeffrey}}{2019}]{Aschwanden2019}
\begin{barticle}
\bauthor{\bsnm{{Aschwanden}}, \binits{M.J.}},
\bauthor{\bsnm{{Kontar}}, \binits{E.P.}},
\bauthor{\bsnm{{Jeffrey}}, \binits{N.L.S.}}:
\byear{2019},
\batitle{{Global Energetics of Solar Flares. VIII. The Low-energy Cutoff}}.
\bjtitle{\apj}
\bvolume{881}(\bissue{1}),
\bfpage{1}.
\doiurl{10.3847/1538-4357/ab2cd4}.
\adsurl{https://ui.adsabs.harvard.edu/abs/2019ApJ...881....1A}.
\end{barticle}
\endbibitem

\bibitem[\protect\citeauthoryear{{Aschwanden}
  \textit{et~al.}}{2016}]{Aschwanden2016}
\begin{barticle}
\bauthor{\bsnm{{Aschwanden}}, \binits{M.J.}},
\bauthor{\bsnm{{Holman}}, \binits{G.}},
\bauthor{\bsnm{{O'Flannagain}}, \binits{A.}},
\bauthor{\bsnm{{Caspi}}, \binits{A.}},
\bauthor{\bsnm{{McTiernan}}, \binits{J.M.}},
\bauthor{\bsnm{{Kontar}}, \binits{E.P.}}:
\byear{2016},
\batitle{{Global Energetics of Solar Flares. III. Nonthermal Energies}}.
\bjtitle{\apj}
\bvolume{832}(\bissue{1}),
\bfpage{27}.
\doiurl{10.3847/0004-637X/832/1/27}.
\adsurl{https://ui.adsabs.harvard.edu/abs/2016ApJ...832...27A}.
\end{barticle}
\endbibitem

\bibitem[\protect\citeauthoryear{{Aulanier} and
  {Dud{\'\i}k}}{2019}]{Aulanier2019}
\begin{barticle}
\bauthor{\bsnm{{Aulanier}}, \binits{G.}},
\bauthor{\bsnm{{Dud{\'\i}k}}, \binits{J.}}:
\byear{2019},
\batitle{{Drifting of the line-tied footpoints of CME flux-ropes}}.
\bjtitle{\aap}
\bvolume{621},
\bfpage{A72}.
\doiurl{10.1051/0004-6361/201834221}.
\adsurl{https://ui.adsabs.harvard.edu/abs/2019A&A...621A..72A}.
\end{barticle}
\endbibitem

\bibitem[\protect\citeauthoryear{{Aulanier}, {Janvier}, and
  {Schmieder}}{2012}]{Aulanier2012}
\begin{barticle}
\bauthor{\bsnm{{Aulanier}}, \binits{G.}},
\bauthor{\bsnm{{Janvier}}, \binits{M.}},
\bauthor{\bsnm{{Schmieder}}, \binits{B.}}:
\byear{2012},
\batitle{{The standard flare model in three dimensions. I. Strong-to-weak shear
  transition in post-flare loops}}.
\bjtitle{\aap}
\bvolume{543},
\bfpage{A110}.
\doiurl{10.1051/0004-6361/201219311}.
\adsurl{2012A\%26A...543A.110A}.
\end{barticle}
\endbibitem

\bibitem[\protect\citeauthoryear{{Benz}}{2017}]{Benz2017}
\begin{barticle}
\bauthor{\bsnm{{Benz}}, \binits{A.O.}}:
\byear{2017},
\batitle{{Flare Observations}}.
\bjtitle{Living Reviews in Solar Physics}
\bvolume{14}(\bissue{1}),
\bfpage{2}.
\doiurl{10.1007/s41116-016-0004-3}.
\adsurl{https://ui.adsabs.harvard.edu/abs/2017LRSP...14....2B}.
\end{barticle}
\endbibitem

\bibitem[\protect\citeauthoryear{{Bogachev}
  \textit{et~al.}}{2005}]{Bogachev2005}
\begin{barticle}
\bauthor{\bsnm{{Bogachev}}, \binits{S.A.}},
\bauthor{\bsnm{{Somov}}, \binits{B.V.}},
\bauthor{\bsnm{{Kosugi}}, \binits{T.}},
\bauthor{\bsnm{{Sakao}}, \binits{T.}}:
\byear{2005},
\batitle{{The Motions of the Hard X-Ray Sources in Solar Flares: Images and
  Statistics}}.
\bjtitle{\apj}
\bvolume{630},
\bfpage{561}.
\doiurl{10.1086/431918}.
\adsurl{2005ApJ...630..561B}.
\end{barticle}
\endbibitem

\bibitem[\protect\citeauthoryear{{Caspi}, {Krucker}, and
  {Lin}}{2014}]{Caspi2014}
\begin{barticle}
\bauthor{\bsnm{{Caspi}}, \binits{A.}},
\bauthor{\bsnm{{Krucker}}, \binits{S.}},
\bauthor{\bsnm{{Lin}}, \binits{R.P.}}:
\byear{2014},
\batitle{{Statistical Properties of Super-hot Solar Flares}}.
\bjtitle{\apj}
\bvolume{781}(\bissue{1}),
\bfpage{43}.
\doiurl{10.1088/0004-637X/781/1/43}.
\adsurl{https://ui.adsabs.harvard.edu/abs/2014ApJ...781...43C}.
\end{barticle}
\endbibitem

\bibitem[\protect\citeauthoryear{{Chen} \textit{et~al.}}{2020}]{Chen2020}
\begin{barticle}
\bauthor{\bsnm{{Chen}}, \binits{B.}},
\bauthor{\bsnm{{Shen}}, \binits{C.}},
\bauthor{\bsnm{{Gary}}, \binits{D.E.}},
\bauthor{\bsnm{{Reeves}}, \binits{K.K.}},
\bauthor{\bsnm{{Fleishman}}, \binits{G.D.}},
\bauthor{\bsnm{{Yu}}, \binits{S.}},
\bauthor{\bsnm{{Guo}}, \binits{F.}},
\bauthor{\bsnm{{Krucker}}, \binits{S.}},
\bauthor{\bsnm{{Lin}}, \binits{J.}},
\bauthor{\bsnm{{Nita}}, \binits{G.M.}},
\bauthor{\bsnm{{Kong}}, \binits{X.}}:
\byear{2020},
\batitle{{Measurement of magnetic field and relativistic electrons along a
  solar flare current sheet}}.
\bjtitle{Nature Astronomy}
\bvolume{4},
\bfpage{1140}.
\doiurl{10.1038/s41550-020-1147-7}.
\adsurl{https://ui.adsabs.harvard.edu/abs/2020NatAs...4.1140C}.
\end{barticle}
\endbibitem

\bibitem[\protect\citeauthoryear{{Cheng}, {Kerr}, and {Qiu}}{2012}]{Cheng2012}
\begin{barticle}
\bauthor{\bsnm{{Cheng}}, \binits{J.X.}},
\bauthor{\bsnm{{Kerr}}, \binits{G.}},
\bauthor{\bsnm{{Qiu}}, \binits{J.}}:
\byear{2012},
\batitle{{Hard X-Ray and Ultraviolet Observations of the 2005 January 15
  Two-ribbon Flare}}.
\bjtitle{\apj}
\bvolume{744},
\bfpage{48}.
\doiurl{10.1088/0004-637X/744/1/48}.
\adsurl{2012ApJ...744...48C}.
\end{barticle}
\endbibitem

\bibitem[\protect\citeauthoryear{{Dahlin}, {Drake}, and
  {Swisdak}}{2015}]{Dahlin2015}
\begin{barticle}
\bauthor{\bsnm{{Dahlin}}, \binits{J.T.}},
\bauthor{\bsnm{{Drake}}, \binits{J.F.}},
\bauthor{\bsnm{{Swisdak}}, \binits{M.}}:
\byear{2015},
\batitle{{Electron acceleration in three-dimensional magnetic reconnection with
  a guide field}}.
\bjtitle{Physics of Plasmas}
\bvolume{22}(\bissue{10}),
\bfpage{100704}.
\doiurl{10.1063/1.4933212}.
\adsurl{https://ui.adsabs.harvard.edu/abs/2015PhPl...22j0704D}.
\end{barticle}
\endbibitem

\bibitem[\protect\citeauthoryear{{Dahlin} \textit{et~al.}}{2021}]{Dahlin2021}
\begin{botherref}
\oauthor{\bsnm{{Dahlin}}, \binits{J.T.}},
\oauthor{\bsnm{{Antiochos}}, \binits{S.K.}},
\oauthor{\bsnm{{Qiu}}, \binits{J.}},
\oauthor{\bsnm{{DeVore}}, \binits{C.R.}}:
2021,
{Variability of the Reconnection Guide Field in Solar Flares}.
\textit{arXiv e-prints},
arXiv:2110.04132.
\adsurl{https://ui.adsabs.harvard.edu/abs/2021arXiv211004132D}.
\end{botherref}
\endbibitem

\bibitem[\protect\citeauthoryear{{De Pontieu}
  \textit{et~al.}}{2014}]{DePontieu2014}
\begin{barticle}
\bauthor{\bsnm{{De Pontieu}}, \binits{B.}},
\bauthor{\bsnm{{Title}}, \binits{A.M.}},
\bauthor{\bsnm{{Lemen}}, \binits{J.R.}},
\bauthor{\bsnm{{Kushner}}, \binits{G.D.}},
\bauthor{\bsnm{{Akin}}, \binits{D.J.}},
\bauthor{\bsnm{{Allard}}, \binits{B.}},
\bauthor{\bsnm{{Berger}}, \binits{T.}},
\bauthor{\bsnm{{Boerner}}, \binits{P.}},
\bauthor{\bsnm{{Cheung}}, \binits{M.}},
\bauthor{\bsnm{{Chou}}, \binits{C.}},
\bauthor{\bsnm{{Drake}}, \binits{J.F.}},
\bauthor{\bsnm{{Duncan}}, \binits{D.W.}},
\bauthor{\bsnm{{Freeland}}, \binits{S.}},
\bauthor{\bsnm{{Heyman}}, \binits{G.F.}},
\bauthor{\bsnm{{Hoffman}}, \binits{C.}},
\bauthor{\bsnm{{Hurlburt}}, \binits{N.E.}},
\bauthor{\bsnm{{Lindgren}}, \binits{R.W.}},
\bauthor{\bsnm{{Mathur}}, \binits{D.}},
\bauthor{\bsnm{{Rehse}}, \binits{R.}},
\bauthor{\bsnm{{Sabolish}}, \binits{D.}},
\bauthor{\bsnm{{Seguin}}, \binits{R.}},
\bauthor{\bsnm{{Schrijver}}, \binits{C.J.}},
\bauthor{\bsnm{{Tarbell}}, \binits{T.D.}},
\bauthor{\bsnm{{W{\"u}lser}}, \binits{J.-P.}},
\bauthor{\bsnm{{Wolfson}}, \binits{C.J.}},
\bauthor{\bsnm{{Yanari}}, \binits{C.}},
\bauthor{\bsnm{{Mudge}}, \binits{J.}},
\bauthor{\bsnm{{Nguyen-Phuc}}, \binits{N.}},
\bauthor{\bsnm{{Timmons}}, \binits{R.}},
\bauthor{\bsnm{{van Bezooijen}}, \binits{R.}},
\bauthor{\bsnm{{Weingrod}}, \binits{I.}},
\bauthor{\bsnm{{Brookner}}, \binits{R.}},
\bauthor{\bsnm{{Butcher}}, \binits{G.}},
\bauthor{\bsnm{{Dougherty}}, \binits{B.}},
\bauthor{\bsnm{{Eder}}, \binits{J.}},
\bauthor{\bsnm{{Knagenhjelm}}, \binits{V.}},
\bauthor{\bsnm{{Larsen}}, \binits{S.}},
\bauthor{\bsnm{{Mansir}}, \binits{D.}},
\bauthor{\bsnm{{Phan}}, \binits{L.}},
\bauthor{\bsnm{{Boyle}}, \binits{P.}},
\bauthor{\bsnm{{Cheimets}}, \binits{P.N.}},
\bauthor{\bsnm{{DeLuca}}, \binits{E.E.}},
\bauthor{\bsnm{{Golub}}, \binits{L.}},
\bauthor{\bsnm{{Gates}}, \binits{R.}},
\bauthor{\bsnm{{Hertz}}, \binits{E.}},
\bauthor{\bsnm{{McKillop}}, \binits{S.}},
\bauthor{\bsnm{{Park}}, \binits{S.}},
\bauthor{\bsnm{{Perry}}, \binits{T.}},
\bauthor{\bsnm{{Podgorski}}, \binits{W.A.}},
\bauthor{\bsnm{{Reeves}}, \binits{K.}},
\bauthor{\bsnm{{Saar}}, \binits{S.}},
\bauthor{\bsnm{{Testa}}, \binits{P.}},
\bauthor{\bsnm{{Tian}}, \binits{H.}},
\bauthor{\bsnm{{Weber}}, \binits{M.}},
\bauthor{\bsnm{{Dunn}}, \binits{C.}},
\bauthor{\bsnm{{Eccles}}, \binits{S.}},
\bauthor{\bsnm{{Jaeggli}}, \binits{S.A.}},
\bauthor{\bsnm{{Kankelborg}}, \binits{C.C.}},
\bauthor{\bsnm{{Mashburn}}, \binits{K.}},
\bauthor{\bsnm{{Pust}}, \binits{N.}},
\bauthor{\bsnm{{Springer}}, \binits{L.}},
\bauthor{\bsnm{{Carvalho}}, \binits{R.}},
\bauthor{\bsnm{{Kleint}}, \binits{L.}},
\bauthor{\bsnm{{Marmie}}, \binits{J.}},
\bauthor{\bsnm{{Mazmanian}}, \binits{E.}},
\bauthor{\bsnm{{Pereira}}, \binits{T.M.D.}},
\bauthor{\bsnm{{Sawyer}}, \binits{S.}},
\bauthor{\bsnm{{Strong}}, \binits{J.}},
\bauthor{\bsnm{{Worden}}, \binits{S.P.}},
\bauthor{\bsnm{{Carlsson}}, \binits{M.}},
\bauthor{\bsnm{{Hansteen}}, \binits{V.H.}},
\bauthor{\bsnm{{Leenaarts}}, \binits{J.}},
\bauthor{\bsnm{{Wiesmann}}, \binits{M.}},
\bauthor{\bsnm{{Aloise}}, \binits{J.}},
\bauthor{\bsnm{{Chu}}, \binits{K.-C.}},
\bauthor{\bsnm{{Bush}}, \binits{R.I.}},
\bauthor{\bsnm{{Scherrer}}, \binits{P.H.}},
\bauthor{\bsnm{{Brekke}}, \binits{P.}},
\bauthor{\bsnm{{Martinez-Sykora}}, \binits{J.}},
\bauthor{\bsnm{{Lites}}, \binits{B.W.}},
\bauthor{\bsnm{{McIntosh}}, \binits{S.W.}},
\bauthor{\bsnm{{Uitenbroek}}, \binits{H.}},
\bauthor{\bsnm{{Okamoto}}, \binits{T.J.}},
\bauthor{\bsnm{{Gummin}}, \binits{M.A.}},
\bauthor{\bsnm{{Auker}}, \binits{G.}},
\bauthor{\bsnm{{Jerram}}, \binits{P.}},
\bauthor{\bsnm{{Pool}}, \binits{P.}},
\bauthor{\bsnm{{Waltham}}, \binits{N.}}:
\byear{2014},
\batitle{{The Interface Region Imaging Spectrograph (IRIS)}}.
\bjtitle{\solphys}
\bvolume{289}(\bissue{7}),
\bfpage{2733}.
\doiurl{10.1007/s11207-014-0485-y}.
\adsurl{https://ui.adsabs.harvard.edu/abs/2014SoPh..289.2733D}.
\end{barticle}
\endbibitem

\bibitem[\protect\citeauthoryear{{Fletcher}}{2009}]{Fletcher2009}
\begin{barticle}
\bauthor{\bsnm{{Fletcher}}, \binits{L.}}:
\byear{2009},
\batitle{{Ultra-violet footpoints as tracers of coronal magnetic connectivity
  and restructuring during a solar flare}}.
\bjtitle{\aap}
\bvolume{493}(\bissue{1}),
\bfpage{241}.
\doiurl{10.1051/0004-6361:20077972}.
\adsurl{https://ui.adsabs.harvard.edu/abs/2009A&A...493..241F}.
\end{barticle}
\endbibitem

\bibitem[\protect\citeauthoryear{{Fletcher} and {Hudson}}{2001}]{Fletcher2001}
\begin{barticle}
\bauthor{\bsnm{{Fletcher}}, \binits{L.}},
\bauthor{\bsnm{{Hudson}}, \binits{H.}}:
\byear{2001},
\batitle{{The Magnetic Structure and Generation of EUV Flare Ribbons}}.
\bjtitle{\solphys}
\bvolume{204},
\bfpage{69}.
\doiurl{10.1023/A:1014275821318}.
\adsurl{2001SoPh..204...69F}.
\end{barticle}
\endbibitem

\bibitem[\protect\citeauthoryear{{Fletcher} and {Hudson}}{2002}]{Fletcher2002}
\begin{barticle}
\bauthor{\bsnm{{Fletcher}}, \binits{L.}},
\bauthor{\bsnm{{Hudson}}, \binits{H.S.}}:
\byear{2002},
\batitle{{Spectral and Spatial Variations of Flare Hard X-ray Footpoints}}.
\bjtitle{\solphys}
\bvolume{210}(\bissue{1}),
\bfpage{307}.
\doiurl{10.1023/A:1022479610710}.
\adsurl{https://ui.adsabs.harvard.edu/abs/2002SoPh..210..307F}.
\end{barticle}
\endbibitem

\bibitem[\protect\citeauthoryear{{Fletcher}, {Pollock}, and
  {Potts}}{2004}]{Fletcher2004}
\begin{barticle}
\bauthor{\bsnm{{Fletcher}}, \binits{L.}},
\bauthor{\bsnm{{Pollock}}, \binits{J.A.}},
\bauthor{\bsnm{{Potts}}, \binits{H.E.}}:
\byear{2004},
\batitle{{Tracking of TRACE Ultraviolet Flare Footpoints}}.
\bjtitle{\solphys}
\bvolume{222},
\bfpage{279}.
\doiurl{10.1023/B:SOLA.0000043580.89730.4d}.
\adsurl{2004SoPh..222..279F}.
\end{barticle}
\endbibitem

\bibitem[\protect\citeauthoryear{{Fletcher}
  \textit{et~al.}}{2011}]{Fletcher2011}
\begin{barticle}
\bauthor{\bsnm{{Fletcher}}, \binits{L.}},
\bauthor{\bsnm{{Dennis}}, \binits{B.R.}},
\bauthor{\bsnm{{Hudson}}, \binits{H.S.}},
\bauthor{\bsnm{{Krucker}}, \binits{S.}},
\bauthor{\bsnm{{Phillips}}, \binits{K.}},
\bauthor{\bsnm{{Veronig}}, \binits{A.}},
\bauthor{\bsnm{{Battaglia}}, \binits{M.}},
\bauthor{\bsnm{{Bone}}, \binits{L.}},
\bauthor{\bsnm{{Caspi}}, \binits{A.}},
\bauthor{\bsnm{{Chen}}, \binits{Q.}},
\bauthor{\bsnm{{Gallagher}}, \binits{P.}},
\bauthor{\bsnm{{Grigis}}, \binits{P.T.}},
\bauthor{\bsnm{{Ji}}, \binits{H.}},
\bauthor{\bsnm{{Liu}}, \binits{W.}},
\bauthor{\bsnm{{Milligan}}, \binits{R.O.}},
\bauthor{\bsnm{{Temmer}}, \binits{M.}}:
\byear{2011},
\batitle{{An Observational Overview of Solar Flares}}.
\bjtitle{\ssr}
\bvolume{159}(\bissue{1-4}),
\bfpage{19}.
\doiurl{10.1007/s11214-010-9701-8}.
\adsurl{https://ui.adsabs.harvard.edu/abs/2011SSRv..159...19F}.
\end{barticle}
\endbibitem

\bibitem[\protect\citeauthoryear{{Forbes} and {Priest}}{1984}]{Forbes1984}
\begin{bchapter}
\bauthor{\bsnm{{Forbes}}, \binits{T.G.}},
\bauthor{\bsnm{{Priest}}, \binits{E.R.}}:
\byear{1984},
\bctitle{{Solar Terrestrial Physics: Present and Future}}.
In: \beditor{\bsnm{{Butler}}, \binits{D.M.}},
\beditor{\bsnm{{Paradupoulous}}, \binits{K.}} (eds.)
\bbtitle{Solar Terrestrial Physics: Present and Future (NASA)},
\bfpage{1}.
\end{bchapter}
\endbibitem

\bibitem[\protect\citeauthoryear{{Freeland} and {Handy}}{1998}]{Freeland1998}
\begin{barticle}
\bauthor{\bsnm{{Freeland}}, \binits{S.L.}},
\bauthor{\bsnm{{Handy}}, \binits{B.N.}}:
\byear{1998},
\batitle{{Data Analysis with the SolarSoft System}}.
\bjtitle{\solphys}
\bvolume{182}(\bissue{2}),
\bfpage{497}.
\doiurl{10.1023/A:1005038224881}.
\adsurl{https://ui.adsabs.harvard.edu/abs/1998SoPh..182..497F}.
\end{barticle}
\endbibitem

\bibitem[\protect\citeauthoryear{{Freeland} and {Handy}}{2012}]{Freeland2012}
\begin{botherref}
\oauthor{\bsnm{{Freeland}}, \binits{S.L.}},
\oauthor{\bsnm{{Handy}}, \binits{B.N.}}:
2012,
\textit{{SolarSoft: Programming and data analysis environment for solar
  physics}}.
\adsurl{https://ui.adsabs.harvard.edu/abs/2012ascl.soft08013F}.
\end{botherref}
\endbibitem

\bibitem[\protect\citeauthoryear{{Gary} \textit{et~al.}}{2018}]{Gary2018}
\begin{bchapter}
\bauthor{\bsnm{{Gary}}, \binits{D.E.}},
\bauthor{\bsnm{{Bastian}}, \binits{T.S.}},
\bauthor{\bsnm{{Chen}}, \binits{B.}},
\bauthor{\bsnm{{Fleishman}}, \binits{G.D.}},
\bauthor{\bsnm{{Glesener}}, \binits{L.}}:
\byear{2018},
\bctitle{{Radio Observations of Solar Flares}}.
In: \beditor{\bsnm{{Murphy}}, \binits{E.}} (ed.)
\bbtitle{Science with a Next Generation Very Large Array},
\bsertitle{Astronomical Society of the Pacific Conference Series}
\bseriesno{517},
\bfpage{99}.
\adsurl{https://ui.adsabs.harvard.edu/abs/2018ASPC..517...99G}.
\end{bchapter}
\endbibitem

\bibitem[\protect\citeauthoryear{{Grigis} and {Benz}}{2005}]{Grigis2005}
\begin{barticle}
\bauthor{\bsnm{{Grigis}}, \binits{P.C.}},
\bauthor{\bsnm{{Benz}}, \binits{A.O.}}:
\byear{2005},
\batitle{{The Evolution of Reconnection along an Arcade of Magnetic Loops}}.
\bjtitle{\apjl}
\bvolume{625},
\bfpage{L143}.
\doiurl{10.1086/431147}.
\adsurl{2005ApJ...625L.143G}.
\end{barticle}
\endbibitem

\bibitem[\protect\citeauthoryear{{Handy} \textit{et~al.}}{1999}]{Handy1999}
\begin{barticle}
\bauthor{\bsnm{{Handy}}, \binits{B.N.}},
\bauthor{\bsnm{{Acton}}, \binits{L.W.}},
\bauthor{\bsnm{{Kankelborg}}, \binits{C.C.}},
\bauthor{\bsnm{{Wolfson}}, \binits{C.J.}},
\bauthor{\bsnm{{Akin}}, \binits{D.J.}},
\bauthor{\bsnm{{Bruner}}, \binits{M.E.}},
\bauthor{\bsnm{{Caravalho}}, \binits{R.}},
\bauthor{\bsnm{{Catura}}, \binits{R.C.}},
\bauthor{\bsnm{{Chevalier}}, \binits{R.}},
\bauthor{\bsnm{{Duncan}}, \binits{D.W.}},
\bauthor{\bsnm{{Edwards}}, \binits{C.G.}},
\bauthor{\bsnm{{Feinstein}}, \binits{C.N.}},
\bauthor{\bsnm{{Freeland}}, \binits{S.L.}},
\bauthor{\bsnm{{Friedlaender}}, \binits{F.M.}},
\bauthor{\bsnm{{Hoffmann}}, \binits{C.H.}},
\bauthor{\bsnm{{Hurlburt}}, \binits{N.E.}},
\bauthor{\bsnm{{Jurcevich}}, \binits{B.K.}},
\bauthor{\bsnm{{Katz}}, \binits{N.L.}},
\bauthor{\bsnm{{Kelly}}, \binits{G.A.}},
\bauthor{\bsnm{{Lemen}}, \binits{J.R.}},
\bauthor{\bsnm{{Levay}}, \binits{M.}},
\bauthor{\bsnm{{Lindgren}}, \binits{R.W.}},
\bauthor{\bsnm{{Mathur}}, \binits{D.P.}},
\bauthor{\bsnm{{Meyer}}, \binits{S.B.}},
\bauthor{\bsnm{{Morrison}}, \binits{S.J.}},
\bauthor{\bsnm{{Morrison}}, \binits{M.D.}},
\bauthor{\bsnm{{Nightingale}}, \binits{R.W.}},
\bauthor{\bsnm{{Pope}}, \binits{T.P.}},
\bauthor{\bsnm{{Rehse}}, \binits{R.A.}},
\bauthor{\bsnm{{Schrijver}}, \binits{C.J.}},
\bauthor{\bsnm{{Shine}}, \binits{R.A.}},
\bauthor{\bsnm{{Shing}}, \binits{L.}},
\bauthor{\bsnm{{Strong}}, \binits{K.T.}},
\bauthor{\bsnm{{Tarbell}}, \binits{T.D.}},
\bauthor{\bsnm{{Title}}, \binits{A.M.}},
\bauthor{\bsnm{{Torgerson}}, \binits{D.D.}},
\bauthor{\bsnm{{Golub}}, \binits{L.}},
\bauthor{\bsnm{{Bookbinder}}, \binits{J.A.}},
\bauthor{\bsnm{{Caldwell}}, \binits{D.}},
\bauthor{\bsnm{{Cheimets}}, \binits{P.N.}},
\bauthor{\bsnm{{Davis}}, \binits{W.N.}},
\bauthor{\bsnm{{Deluca}}, \binits{E.E.}},
\bauthor{\bsnm{{McMullen}}, \binits{R.A.}},
\bauthor{\bsnm{{Warren}}, \binits{H.P.}},
\bauthor{\bsnm{{Amato}}, \binits{D.}},
\bauthor{\bsnm{{Fisher}}, \binits{R.}},
\bauthor{\bsnm{{Maldonado}}, \binits{H.}},
\bauthor{\bsnm{{Parkinson}}, \binits{C.}}:
\byear{1999},
\batitle{{The transition region and coronal explorer}}.
\bjtitle{\solphys}
\bvolume{187},
\bfpage{229}.
\doiurl{10.1023/A:1005166902804}.
\adsurl{1999SoPh..187..229H}.
\end{barticle}
\endbibitem

\bibitem[\protect\citeauthoryear{{Hannah} \textit{et~al.}}{2011}]{hannah2011}
\begin{barticle}
\bauthor{\bsnm{{Hannah}}, \binits{I.G.}},
\bauthor{\bsnm{{Hudson}}, \binits{H.S.}},
\bauthor{\bsnm{{Battaglia}}, \binits{M.}},
\bauthor{\bsnm{{Christe}}, \binits{S.}},
\bauthor{\bsnm{{Ka{\v{s}}parov{\'a}}}, \binits{J.}},
\bauthor{\bsnm{{Krucker}}, \binits{S.}},
\bauthor{\bsnm{{Kundu}}, \binits{M.R.}},
\bauthor{\bsnm{{Veronig}}, \binits{A.}}:
\byear{2011},
\batitle{{Microflares and the Statistics of X-ray Flares}}.
\bjtitle{\ssr}
\bvolume{159}(\bissue{1-4}),
\bfpage{263}.
\doiurl{10.1007/s11214-010-9705-4}.
\adsurl{https://ui.adsabs.harvard.edu/abs/2011SSRv..159..263H}.
\end{barticle}
\endbibitem

\bibitem[\protect\citeauthoryear{{Hinterreiter}
  \textit{et~al.}}{2018}]{Hinterreiter2018}
\begin{barticle}
\bauthor{\bsnm{{Hinterreiter}}, \binits{J.}},
\bauthor{\bsnm{{Veronig}}, \binits{A.M.}},
\bauthor{\bsnm{{Thalmann}}, \binits{J.K.}},
\bauthor{\bsnm{{Tschernitz}}, \binits{J.}},
\bauthor{\bsnm{{P{\"o}tzi}}, \binits{W.}}:
\byear{2018},
\batitle{{Statistical Properties of Ribbon Evolution and Reconnection Electric
  Fields in Eruptive and Confined Flares}}.
\bjtitle{\solphys}
\bvolume{293}(\bissue{3}),
\bfpage{38}.
\doiurl{10.1007/s11207-018-1253-1}.
\adsurl{https://ui.adsabs.harvard.edu/abs/2018SoPh..293...38H}.
\end{barticle}
\endbibitem

\bibitem[\protect\citeauthoryear{{Holman}}{1985}]{Holman1985}
\begin{barticle}
\bauthor{\bsnm{{Holman}}, \binits{G.D.}}:
\byear{1985},
\batitle{{Acceleration of runaway electrons and Joule heating in solar
  flares}}.
\bjtitle{\apj}
\bvolume{293},
\bfpage{584}.
\doiurl{10.1086/163263}.
\adsurl{https://ui.adsabs.harvard.edu/abs/1985ApJ...293..584H}.
\end{barticle}
\endbibitem

\bibitem[\protect\citeauthoryear{{Hudson} \textit{et~al.}}{2021}]{Hudson2021}
\begin{barticle}
\bauthor{\bsnm{{Hudson}}, \binits{H.S.}},
\bauthor{\bsnm{{Sim{\~o}es}}, \binits{P.J.A.}},
\bauthor{\bsnm{{Fletcher}}, \binits{L.}},
\bauthor{\bsnm{{Hayes}}, \binits{L.A.}},
\bauthor{\bsnm{{Hannah}}, \binits{I.G.}}:
\byear{2021},
\batitle{{Hot X-ray onsets of solar flares}}.
\bjtitle{\mnras}
\bvolume{501}(\bissue{1}),
\bfpage{1273}.
\doiurl{10.1093/mnras/staa3664}.
\adsurl{https://ui.adsabs.harvard.edu/abs/2021MNRAS.501.1273H}.
\end{barticle}
\endbibitem

\bibitem[\protect\citeauthoryear{{Inglis} and {Gilbert}}{2013}]{Inglis2013}
\begin{barticle}
\bauthor{\bsnm{{Inglis}}, \binits{A.R.}},
\bauthor{\bsnm{{Gilbert}}, \binits{H.R.}}:
\byear{2013},
\batitle{{Hard X-Ray and Ultraviolet Emission during the 2011 June 7 Solar
  Flare}}.
\bjtitle{\apj}
\bvolume{777},
\bfpage{30}.
\doiurl{10.1088/0004-637X/777/1/30}.
\adsurl{2013ApJ...777...30I}.
\end{barticle}
\endbibitem

\bibitem[\protect\citeauthoryear{{Isobe}, {Takasaki}, and
  {Shibata}}{2005}]{Isobe2005}
\begin{barticle}
\bauthor{\bsnm{{Isobe}}, \binits{H.}},
\bauthor{\bsnm{{Takasaki}}, \binits{H.}},
\bauthor{\bsnm{{Shibata}}, \binits{K.}}:
\byear{2005},
\batitle{{Measurement of the Energy Release Rate and the Reconnection Rate in
  Solar Flares}}.
\bjtitle{\apj}
\bvolume{632},
\bfpage{1184}.
\doiurl{10.1086/444490}.
\adsurl{2005ApJ...632.1184I}.
\end{barticle}
\endbibitem

\bibitem[\protect\citeauthoryear{{Ji} \textit{et~al.}}{2006}]{Ji2006}
\begin{barticle}
\bauthor{\bsnm{{Ji}}, \binits{H.}},
\bauthor{\bsnm{{Huang}}, \binits{G.}},
\bauthor{\bsnm{{Wang}}, \binits{H.}},
\bauthor{\bsnm{{Zhou}}, \binits{T.}},
\bauthor{\bsnm{{Li}}, \binits{Y.}},
\bauthor{\bsnm{{Zhang}}, \binits{Y.}},
\bauthor{\bsnm{{Song}}, \binits{M.}}:
\byear{2006},
\batitle{{Converging Motion of H{\ensuremath{\alpha}} Conjugate Kernels: The
  Signature of Fast Relaxation of a Sheared Magnetic Field}}.
\bjtitle{\apjl}
\bvolume{636}(\bissue{2}),
\bfpage{L173}.
\doiurl{10.1086/500203}.
\adsurl{https://ui.adsabs.harvard.edu/abs/2006ApJ...636L.173J}.
\end{barticle}
\endbibitem

\bibitem[\protect\citeauthoryear{{Jiang} \textit{et~al.}}{2021a}]{Jiang2021a}
\begin{barticle}
\bauthor{\bsnm{{Jiang}}, \binits{C.}},
\bauthor{\bsnm{{Feng}}, \binits{X.}},
\bauthor{\bsnm{{Liu}}, \binits{R.}},
\bauthor{\bsnm{{Yan}}, \binits{X.}},
\bauthor{\bsnm{{Hu}}, \binits{Q.}},
\bauthor{\bsnm{{Moore}}, \binits{R.L.}},
\bauthor{\bsnm{{Duan}}, \binits{A.}},
\bauthor{\bsnm{{Cui}}, \binits{J.}},
\bauthor{\bsnm{{Zuo}}, \binits{P.}},
\bauthor{\bsnm{{Wang}}, \binits{Y.}},
\bauthor{\bsnm{{Wei}}, \binits{F.}}:
\byear{2021}a,
\batitle{{A fundamental mechanism of solar eruption initiation}}.
\bjtitle{Nature Astronomy}
\bvolume{5},
\bfpage{1126}.
\doiurl{10.1038/s41550-021-01414-z}.
\adsurl{https://ui.adsabs.harvard.edu/abs/2021NatAs...5.1126J}.
\end{barticle}
\endbibitem

\bibitem[\protect\citeauthoryear{{Jiang} \textit{et~al.}}{2021b}]{Jiang2021}
\begin{barticle}
\bauthor{\bsnm{{Jiang}}, \binits{C.}},
\bauthor{\bsnm{{Chen}}, \binits{J.}},
\bauthor{\bsnm{{Duan}}, \binits{A.}},
\bauthor{\bsnm{{Bian}}, \binits{X.}},
\bauthor{\bsnm{{Wang}}, \binits{X.}},
\bauthor{\bsnm{{Li}}, \binits{J.}},
\bauthor{\bsnm{{Zou}}, \binits{P.}},
\bauthor{\bsnm{{Feng}}, \binits{X.}}:
\byear{2021}b,
\batitle{{Formation of Magnetic Flux Rope during Solar Eruption. I. Evolution
  of Toroidal Flux and Reconnection Flux}}.
\bjtitle{Frontiers in Physics}
\bvolume{9},
\bfpage{575}.
\doiurl{10.3389/fphy.2021.746576}.
\adsurl{https://ui.adsabs.harvard.edu/abs/2021FrP.....9..575J}.
\end{barticle}
\endbibitem

\bibitem[\protect\citeauthoryear{{Jing}, {Chae}, and {Wang}}{2008}]{Jing2008}
\begin{barticle}
\bauthor{\bsnm{{Jing}}, \binits{J.}},
\bauthor{\bsnm{{Chae}}, \binits{J.}},
\bauthor{\bsnm{{Wang}}, \binits{H.}}:
\byear{2008},
\batitle{{Spatial Distribution of Magnetic Reconnection in the 2006 December 13
  Solar Flare as Observed by Hinode}}.
\bjtitle{\apjl}
\bvolume{672}(\bissue{1}),
\bfpage{L73}.
\doiurl{10.1086/526339}.
\adsurl{https://ui.adsabs.harvard.edu/abs/2008ApJ...672L..73J}.
\end{barticle}
\endbibitem

\bibitem[\protect\citeauthoryear{{Jing} \textit{et~al.}}{2005}]{Jing2005}
\begin{barticle}
\bauthor{\bsnm{{Jing}}, \binits{J.}},
\bauthor{\bsnm{{Qiu}}, \binits{J.}},
\bauthor{\bsnm{{Lin}}, \binits{J.}},
\bauthor{\bsnm{{Qu}}, \binits{M.}},
\bauthor{\bsnm{{Xu}}, \binits{Y.}},
\bauthor{\bsnm{{Wang}}, \binits{H.}}:
\byear{2005},
\batitle{{Magnetic Reconnection Rate and Flux-Rope Acceleration of Two-Ribbon
  Flares}}.
\bjtitle{\apj}
\bvolume{620}(\bissue{2}),
\bfpage{1085}.
\doiurl{10.1086/427165}.
\adsurl{https://ui.adsabs.harvard.edu/abs/2005ApJ...620.1085J}.
\end{barticle}
\endbibitem

\bibitem[\protect\citeauthoryear{{Joshi} \textit{et~al.}}{2017}]{Joshi2017}
\begin{barticle}
\bauthor{\bsnm{{Joshi}}, \binits{N.C.}},
\bauthor{\bsnm{{Sterling}}, \binits{A.C.}},
\bauthor{\bsnm{{Moore}}, \binits{R.L.}},
\bauthor{\bsnm{{Magara}}, \binits{T.}},
\bauthor{\bsnm{{Moon}}, \binits{Y.-J.}}:
\byear{2017},
\batitle{{Onset of a Large Ejective Solar Eruption from a Typical
  Coronal-jet-base Field Configuration}}.
\bjtitle{\apj}
\bvolume{845}(\bissue{1}),
\bfpage{26}.
\doiurl{10.3847/1538-4357/aa7c1b}.
\adsurl{https://ui.adsabs.harvard.edu/abs/2017ApJ...845...26J}.
\end{barticle}
\endbibitem

\bibitem[\protect\citeauthoryear{{Kazachenko}
  \textit{et~al.}}{2017}]{Kazachenko2017}
\begin{barticle}
\bauthor{\bsnm{{Kazachenko}}, \binits{M.D.}},
\bauthor{\bsnm{{Lynch}}, \binits{B.J.}},
\bauthor{\bsnm{{Welsch}}, \binits{B.T.}},
\bauthor{\bsnm{{Sun}}, \binits{X.}}:
\byear{2017},
\batitle{{A Database of Flare Ribbon Properties from the Solar Dynamics
  Observatory. I. Reconnection Flux}}.
\bjtitle{\apj}
\bvolume{845}(\bissue{1}),
\bfpage{49}.
\doiurl{10.3847/1538-4357/aa7ed6}.
\adsurl{https://ui.adsabs.harvard.edu/abs/2017ApJ...845...49K}.
\end{barticle}
\endbibitem

\bibitem[\protect\citeauthoryear{{Kosugi} \textit{et~al.}}{1991}]{Kosugi1991}
\begin{barticle}
\bauthor{\bsnm{{Kosugi}}, \binits{T.}},
\bauthor{\bsnm{{Makishima}}, \binits{K.}},
\bauthor{\bsnm{{Murakami}}, \binits{T.}},
\bauthor{\bsnm{{Sakao}}, \binits{T.}},
\bauthor{\bsnm{{Dotani}}, \binits{T.}},
\bauthor{\bsnm{{Inda}}, \binits{M.}},
\bauthor{\bsnm{{Kai}}, \binits{K.}},
\bauthor{\bsnm{{Masuda}}, \binits{S.}},
\bauthor{\bsnm{{Nakajima}}, \binits{H.}},
\bauthor{\bsnm{{Ogawara}}, \binits{Y.}},
\bauthor{\bsnm{{Sawa}}, \binits{M.}},
\bauthor{\bsnm{{Shibasaki}}, \binits{K.}}:
\byear{1991},
\batitle{{The Hard X-ray Telescope (HXT) for the SOLAR-A mission}}.
\bjtitle{\solphys}
\bvolume{136}(\bissue{1}),
\bfpage{17}.
\doiurl{10.1007/BF00151693}.
\adsurl{https://ui.adsabs.harvard.edu/abs/1991SoPh..136...17K}.
\end{barticle}
\endbibitem

\bibitem[\protect\citeauthoryear{{Krucker}, {Fivian}, and
  {Lin}}{2005}]{Krucker2005}
\begin{barticle}
\bauthor{\bsnm{{Krucker}}, \binits{S.}},
\bauthor{\bsnm{{Fivian}}, \binits{M.D.}},
\bauthor{\bsnm{{Lin}}, \binits{R.P.}}:
\byear{2005},
\batitle{{Hard X-ray footpoint motions in solar flares: Comparing magnetic
  reconnection models with observations}}.
\bjtitle{Advances in Space Research}
\bvolume{35}(\bissue{10}),
\bfpage{1707}.
\doiurl{10.1016/j.asr.2005.05.054}.
\adsurl{https://ui.adsabs.harvard.edu/abs/2005AdSpR..35.1707K}.
\end{barticle}
\endbibitem

\bibitem[\protect\citeauthoryear{{Krucker}, {Hurford}, and
  {Lin}}{2003}]{Krucker2003}
\begin{barticle}
\bauthor{\bsnm{{Krucker}}, \binits{S.}},
\bauthor{\bsnm{{Hurford}}, \binits{G.J.}},
\bauthor{\bsnm{{Lin}}, \binits{R.P.}}:
\byear{2003},
\batitle{{Hard X-Ray Source Motions in the 2002 July 23 Gamma-Ray Flare}}.
\bjtitle{\apjl}
\bvolume{595},
\bfpage{L103}.
\doiurl{10.1086/378840}.
\adsurl{2003ApJ...595L.103K}.
\end{barticle}
\endbibitem

\bibitem[\protect\citeauthoryear{{Krucker} \textit{et~al.}}{2011}]{Krucker2011}
\begin{barticle}
\bauthor{\bsnm{{Krucker}}, \binits{S.}},
\bauthor{\bsnm{{Hudson}}, \binits{H.S.}},
\bauthor{\bsnm{{Jeffrey}}, \binits{N.L.S.}},
\bauthor{\bsnm{{Battaglia}}, \binits{M.}},
\bauthor{\bsnm{{Kontar}}, \binits{E.P.}},
\bauthor{\bsnm{{Benz}}, \binits{A.O.}},
\bauthor{\bsnm{{Csillaghy}}, \binits{A.}},
\bauthor{\bsnm{{Lin}}, \binits{R.P.}}:
\byear{2011},
\batitle{{High-resolution Imaging of Solar Flare Ribbons and Its Implication on
  the Thick-target Beam Model}}.
\bjtitle{\apj}
\bvolume{739}(\bissue{2}),
\bfpage{96}.
\doiurl{10.1088/0004-637X/739/2/96}.
\adsurl{https://ui.adsabs.harvard.edu/abs/2011ApJ...739...96K}.
\end{barticle}
\endbibitem

\bibitem[\protect\citeauthoryear{{Lee} and {Gary}}{2008}]{Lee2008}
\begin{barticle}
\bauthor{\bsnm{{Lee}}, \binits{J.}},
\bauthor{\bsnm{{Gary}}, \binits{D.E.}}:
\byear{2008},
\batitle{{Parallel Motions of Coronal Hard X-Ray Source and H{$\alpha$}
  Ribbons}}.
\bjtitle{\apjl}
\bvolume{685},
\bfpage{L87}.
\doiurl{10.1086/592292}.
\adsurl{2008ApJ...685L..87L}.
\end{barticle}
\endbibitem

\bibitem[\protect\citeauthoryear{{Lee}, {Gary}, and {Choe}}{2006}]{Lee2006}
\begin{barticle}
\bauthor{\bsnm{{Lee}}, \binits{J.}},
\bauthor{\bsnm{{Gary}}, \binits{D.E.}},
\bauthor{\bsnm{{Choe}}, \binits{G.S.}}:
\byear{2006},
\batitle{{Magnetic Energy Release during the 2002 September 9 Solar Flare}}.
\bjtitle{\apj}
\bvolume{647}(\bissue{1}),
\bfpage{638}.
\doiurl{10.1086/505416}.
\adsurl{https://ui.adsabs.harvard.edu/abs/2006ApJ...647..638L}.
\end{barticle}
\endbibitem

\bibitem[\protect\citeauthoryear{{Lemen} \textit{et~al.}}{2012}]{Lemen2012}
\begin{barticle}
\bauthor{\bsnm{{Lemen}}, \binits{J.R.}},
\bauthor{\bsnm{{Title}}, \binits{A.M.}},
\bauthor{\bsnm{{Akin}}, \binits{D.J.}},
\bauthor{\bsnm{{Boerner}}, \binits{P.F.}},
\bauthor{\bsnm{{Chou}}, \binits{C.}},
\bauthor{\bsnm{{Drake}}, \binits{J.F.}},
\bauthor{\bsnm{{Duncan}}, \binits{D.W.}},
\bauthor{\bsnm{{Edwards}}, \binits{C.G.}},
\bauthor{\bsnm{{Friedlaender}}, \binits{F.M.}},
\bauthor{\bsnm{{Heyman}}, \binits{G.F.}},
\bauthor{\bsnm{{Hurlburt}}, \binits{N.E.}},
\bauthor{\bsnm{{Katz}}, \binits{N.L.}},
\bauthor{\bsnm{{Kushner}}, \binits{G.D.}},
\bauthor{\bsnm{{Levay}}, \binits{M.}},
\bauthor{\bsnm{{Lindgren}}, \binits{R.W.}},
\bauthor{\bsnm{{Mathur}}, \binits{D.P.}},
\bauthor{\bsnm{{McFeaters}}, \binits{E.L.}},
\bauthor{\bsnm{{Mitchell}}, \binits{S.}},
\bauthor{\bsnm{{Rehse}}, \binits{R.A.}},
\bauthor{\bsnm{{Schrijver}}, \binits{C.J.}},
\bauthor{\bsnm{{Springer}}, \binits{L.A.}},
\bauthor{\bsnm{{Stern}}, \binits{R.A.}},
\bauthor{\bsnm{{Tarbell}}, \binits{T.D.}},
\bauthor{\bsnm{{Wuelser}}, \binits{J.-P.}},
\bauthor{\bsnm{{Wolfson}}, \binits{C.J.}},
\bauthor{\bsnm{{Yanari}}, \binits{C.}},
\bauthor{\bsnm{{Bookbinder}}, \binits{J.A.}},
\bauthor{\bsnm{{Cheimets}}, \binits{P.N.}},
\bauthor{\bsnm{{Caldwell}}, \binits{D.}},
\bauthor{\bsnm{{Deluca}}, \binits{E.E.}},
\bauthor{\bsnm{{Gates}}, \binits{R.}},
\bauthor{\bsnm{{Golub}}, \binits{L.}},
\bauthor{\bsnm{{Park}}, \binits{S.}},
\bauthor{\bsnm{{Podgorski}}, \binits{W.A.}},
\bauthor{\bsnm{{Bush}}, \binits{R.I.}},
\bauthor{\bsnm{{Scherrer}}, \binits{P.H.}},
\bauthor{\bsnm{{Gummin}}, \binits{M.A.}},
\bauthor{\bsnm{{Smith}}, \binits{P.}},
\bauthor{\bsnm{{Auker}}, \binits{G.}},
\bauthor{\bsnm{{Jerram}}, \binits{P.}},
\bauthor{\bsnm{{Pool}}, \binits{P.}},
\bauthor{\bsnm{{Soufli}}, \binits{R.}},
\bauthor{\bsnm{{Windt}}, \binits{D.L.}},
\bauthor{\bsnm{{Beardsley}}, \binits{S.}},
\bauthor{\bsnm{{Clapp}}, \binits{M.}},
\bauthor{\bsnm{{Lang}}, \binits{J.}},
\bauthor{\bsnm{{Waltham}}, \binits{N.}}:
\byear{2012},
\batitle{{The Atmospheric Imaging Assembly (AIA) on the Solar Dynamics
  Observatory (SDO)}}.
\bjtitle{\solphys}
\bvolume{275},
\bfpage{17}.
\doiurl{10.1007/s11207-011-9776-8}.
\adsurl{2012SoPh..275...17L}.
\end{barticle}
\endbibitem

\bibitem[\protect\citeauthoryear{{Lin} \textit{et~al.}}{2002}]{Lin2002}
\begin{barticle}
\bauthor{\bsnm{{Lin}}, \binits{R.P.}},
\bauthor{\bsnm{{Dennis}}, \binits{B.R.}},
\bauthor{\bsnm{{Hurford}}, \binits{G.J.}},
\bauthor{\bsnm{{Smith}}, \binits{D.M.}},
\bauthor{\bsnm{{Zehnder}}, \binits{A.}},
\bauthor{\bsnm{{Harvey}}, \binits{P.R.}},
\bauthor{\bsnm{{Curtis}}, \binits{D.W.}},
\bauthor{\bsnm{{Pankow}}, \binits{D.}},
\bauthor{\bsnm{{Turin}}, \binits{P.}},
\bauthor{\bsnm{{Bester}}, \binits{M.}},
\bauthor{\bsnm{{Csillaghy}}, \binits{A.}},
\bauthor{\bsnm{{Lewis}}, \binits{M.}},
\bauthor{\bsnm{{Madden}}, \binits{N.}},
\bauthor{\bsnm{{van Beek}}, \binits{H.F.}},
\bauthor{\bsnm{{Appleby}}, \binits{M.}},
\bauthor{\bsnm{{Raudorf}}, \binits{T.}},
\bauthor{\bsnm{{McTiernan}}, \binits{J.}},
\bauthor{\bsnm{{Ramaty}}, \binits{R.}},
\bauthor{\bsnm{{Schmahl}}, \binits{E.}},
\bauthor{\bsnm{{Schwartz}}, \binits{R.}},
\bauthor{\bsnm{{Krucker}}, \binits{S.}},
\bauthor{\bsnm{{Abiad}}, \binits{R.}},
\bauthor{\bsnm{{Quinn}}, \binits{T.}},
\bauthor{\bsnm{{Berg}}, \binits{P.}},
\bauthor{\bsnm{{Hashii}}, \binits{M.}},
\bauthor{\bsnm{{Sterling}}, \binits{R.}},
\bauthor{\bsnm{{Jackson}}, \binits{R.}},
\bauthor{\bsnm{{Pratt}}, \binits{R.}},
\bauthor{\bsnm{{Campbell}}, \binits{R.D.}},
\bauthor{\bsnm{{Malone}}, \binits{D.}},
\bauthor{\bsnm{{Landis}}, \binits{D.}},
\bauthor{\bsnm{{Barrington-Leigh}}, \binits{C.P.}},
\bauthor{\bsnm{{Slassi-Sennou}}, \binits{S.}},
\bauthor{\bsnm{{Cork}}, \binits{C.}},
\bauthor{\bsnm{{Clark}}, \binits{D.}},
\bauthor{\bsnm{{Amato}}, \binits{D.}},
\bauthor{\bsnm{{Orwig}}, \binits{L.}},
\bauthor{\bsnm{{Boyle}}, \binits{R.}},
\bauthor{\bsnm{{Banks}}, \binits{I.S.}},
\bauthor{\bsnm{{Shirey}}, \binits{K.}},
\bauthor{\bsnm{{Tolbert}}, \binits{A.K.}},
\bauthor{\bsnm{{Zarro}}, \binits{D.}},
\bauthor{\bsnm{{Snow}}, \binits{F.}},
\bauthor{\bsnm{{Thomsen}}, \binits{K.}},
\bauthor{\bsnm{{Henneck}}, \binits{R.}},
\bauthor{\bsnm{{McHedlishvili}}, \binits{A.}},
\bauthor{\bsnm{{Ming}}, \binits{P.}},
\bauthor{\bsnm{{Fivian}}, \binits{M.}},
\bauthor{\bsnm{{Jordan}}, \binits{J.}},
\bauthor{\bsnm{{Wanner}}, \binits{R.}},
\bauthor{\bsnm{{Crubb}}, \binits{J.}},
\bauthor{\bsnm{{Preble}}, \binits{J.}},
\bauthor{\bsnm{{Matranga}}, \binits{M.}},
\bauthor{\bsnm{{Benz}}, \binits{A.}},
\bauthor{\bsnm{{Hudson}}, \binits{H.}},
\bauthor{\bsnm{{Canfield}}, \binits{R.C.}},
\bauthor{\bsnm{{Holman}}, \binits{G.D.}},
\bauthor{\bsnm{{Crannell}}, \binits{C.}},
\bauthor{\bsnm{{Kosugi}}, \binits{T.}},
\bauthor{\bsnm{{Emslie}}, \binits{A.G.}},
\bauthor{\bsnm{{Vilmer}}, \binits{N.}},
\bauthor{\bsnm{{Brown}}, \binits{J.C.}},
\bauthor{\bsnm{{Johns-Krull}}, \binits{C.}},
\bauthor{\bsnm{{Aschwanden}}, \binits{M.}},
\bauthor{\bsnm{{Metcalf}}, \binits{T.}},
\bauthor{\bsnm{{Conway}}, \binits{A.}}:
\byear{2002},
\batitle{{The Reuven Ramaty High-Energy Solar Spectroscopic Imager (RHESSI)}}.
\bjtitle{\solphys}
\bvolume{210},
\bfpage{3}.
\doiurl{10.1023/A:1022428818870}.
\adsurl{2002SoPh..210....3L}.
\end{barticle}
\endbibitem

\bibitem[\protect\citeauthoryear{{Litvinenko}}{1996}]{Litvinenko1996}
\begin{barticle}
\bauthor{\bsnm{{Litvinenko}}, \binits{Y.E.}}:
\byear{1996},
\batitle{{Particle Acceleration in Reconnecting Current Sheets with a Nonzero
  Magnetic Field}}.
\bjtitle{\apj}
\bvolume{462},
\bfpage{997}.
\doiurl{10.1086/177213}.
\adsurl{https://ui.adsabs.harvard.edu/abs/1996ApJ...462..997L}.
\end{barticle}
\endbibitem

\bibitem[\protect\citeauthoryear{{Liu} and {Wang}}{2009}]{Liu2009a}
\begin{barticle}
\bauthor{\bsnm{{Liu}}, \binits{C.}},
\bauthor{\bsnm{{Wang}}, \binits{H.}}:
\byear{2009},
\batitle{{Reconnection Electric Field and Hardness of X-Ray Emission of Solar
  Flares}}.
\bjtitle{\apjl}
\bvolume{696}(\bissue{1}),
\bfpage{L27}.
\doiurl{10.1088/0004-637X/696/1/L27}.
\adsurl{https://ui.adsabs.harvard.edu/abs/2009ApJ...696L..27L}.
\end{barticle}
\endbibitem

\bibitem[\protect\citeauthoryear{{Liu} \textit{et~al.}}{2016}]{Liu2016}
\begin{barticle}
\bauthor{\bsnm{{Liu}}, \binits{R.}},
\bauthor{\bsnm{{Kliem}}, \binits{B.}},
\bauthor{\bsnm{{Titov}}, \binits{V.S.}},
\bauthor{\bsnm{{Chen}}, \binits{J.}},
\bauthor{\bsnm{{Wang}}, \binits{Y.}},
\bauthor{\bsnm{{Wang}}, \binits{H.}},
\bauthor{\bsnm{{Liu}}, \binits{C.}},
\bauthor{\bsnm{{Xu}}, \binits{Y.}},
\bauthor{\bsnm{{Wiegelmann}}, \binits{T.}}:
\byear{2016},
\batitle{{Structure, Stability, and Evolution of Magnetic Flux Ropes from the
  Perspective of Magnetic Twist}}.
\bjtitle{\apj}
\bvolume{818}(\bissue{2}),
\bfpage{148}.
\doiurl{10.3847/0004-637X/818/2/148}.
\adsurl{https://ui.adsabs.harvard.edu/abs/2016ApJ...818..148L}.
\end{barticle}
\endbibitem

\bibitem[\protect\citeauthoryear{{Liu} \textit{et~al.}}{2013}]{Liu2013}
\begin{barticle}
\bauthor{\bsnm{{Liu}}, \binits{W.-J.}},
\bauthor{\bsnm{{Qiu}}, \binits{J.}},
\bauthor{\bsnm{{Longcope}}, \binits{D.W.}},
\bauthor{\bsnm{{Caspi}}, \binits{A.}}:
\byear{2013},
\batitle{{Determining Heating Rates in Reconnection Formed Flare Loops of the
  M8.0 Flare on 2005 May 13}}.
\bjtitle{\apj}
\bvolume{770},
\bfpage{111}.
\doiurl{10.1088/0004-637X/770/2/111}.
\adsurl{2013ApJ...770..111L}.
\end{barticle}
\endbibitem

\bibitem[\protect\citeauthoryear{{Liu} \textit{et~al.}}{2009}]{Liu2009b}
\begin{barticle}
\bauthor{\bsnm{{Liu}}, \binits{W.}},
\bauthor{\bsnm{{Petrosian}}, \binits{V.}},
\bauthor{\bsnm{{Dennis}}, \binits{B.R.}},
\bauthor{\bsnm{{Holman}}, \binits{G.D.}}:
\byear{2009},
\batitle{{Conjugate Hard X-Ray Footpoints in the 2003 October 29 X10 Flare:
  Unshearing Motions, Correlations, and Asymmetries}}.
\bjtitle{\apj}
\bvolume{693}(\bissue{1}),
\bfpage{847}.
\doiurl{10.1088/0004-637X/693/1/847}.
\adsurl{https://ui.adsabs.harvard.edu/abs/2009ApJ...693..847L}.
\end{barticle}
\endbibitem

\bibitem[\protect\citeauthoryear{{Longcope}, {Qiu}, and
  {Brewer}}{2016}]{Longcope2016}
\begin{barticle}
\bauthor{\bsnm{{Longcope}}, \binits{D.}},
\bauthor{\bsnm{{Qiu}}, \binits{J.}},
\bauthor{\bsnm{{Brewer}}, \binits{J.}}:
\byear{2016},
\batitle{{A Reconnection-driven Model of the Hard X-Ray Loop-top Source from
  Flare 2004-Feb-26}}.
\bjtitle{\apj}
\bvolume{833}(\bissue{2}),
\bfpage{211}.
\doiurl{10.3847/1538-4357/833/2/211}.
\adsurl{https://ui.adsabs.harvard.edu/abs/2016ApJ...833..211L}.
\end{barticle}
\endbibitem

\bibitem[\protect\citeauthoryear{{Masson} \textit{et~al.}}{2009}]{Masson2009}
\begin{barticle}
\bauthor{\bsnm{{Masson}}, \binits{S.}},
\bauthor{\bsnm{{Pariat}}, \binits{E.}},
\bauthor{\bsnm{{Aulanier}}, \binits{G.}},
\bauthor{\bsnm{{Schrijver}}, \binits{C.J.}}:
\byear{2009},
\batitle{{The Nature of Flare Ribbons in Coronal Null-Point Topology}}.
\bjtitle{\apj}
\bvolume{700},
\bfpage{559}.
\doiurl{10.1088/0004-637X/700/1/559}.
\adsurl{2009ApJ...700..559M}.
\end{barticle}
\endbibitem

\bibitem[\protect\citeauthoryear{{Miklenic}
  \textit{et~al.}}{2007}]{Miklenic2007}
\begin{barticle}
\bauthor{\bsnm{{Miklenic}}, \binits{C.H.}},
\bauthor{\bsnm{{Veronig}}, \binits{A.M.}},
\bauthor{\bsnm{{Vr{\v{s}}nak}}, \binits{B.}},
\bauthor{\bsnm{{Hanslmeier}}, \binits{A.}}:
\byear{2007},
\batitle{{Reconnection and energy release rates in a two-ribbon flare}}.
\bjtitle{\aap}
\bvolume{461}(\bissue{2}),
\bfpage{697}.
\doiurl{10.1051/0004-6361:20065751}.
\adsurl{https://ui.adsabs.harvard.edu/abs/2007A&A...461..697M}.
\end{barticle}
\endbibitem

\bibitem[\protect\citeauthoryear{{Moore} \textit{et~al.}}{2001}]{Moore2001}
\begin{barticle}
\bauthor{\bsnm{{Moore}}, \binits{R.L.}},
\bauthor{\bsnm{{Sterling}}, \binits{A.C.}},
\bauthor{\bsnm{{Hudson}}, \binits{H.S.}},
\bauthor{\bsnm{{Lemen}}, \binits{J.R.}}:
\byear{2001},
\batitle{{Onset of the Magnetic Explosion in Solar Flares and Coronal Mass
  Ejections}}.
\bjtitle{\apj}
\bvolume{552}(\bissue{2}),
\bfpage{833}.
\doiurl{10.1086/320559}.
\adsurl{https://ui.adsabs.harvard.edu/abs/2001ApJ...552..833M}.
\end{barticle}
\endbibitem

\bibitem[\protect\citeauthoryear{{Naus} \textit{et~al.}}{2021}]{Naus2021}
\begin{botherref}
\oauthor{\bsnm{{Naus}}, \binits{S.J.}},
\oauthor{\bsnm{{Qiu}}, \binits{J.}},
\oauthor{\bsnm{{DeVore}}, \binits{C.R.}},
\oauthor{\bsnm{{Antiochos}}, \binits{S.K.}},
\oauthor{\bsnm{{Dahlin}}, \binits{J.T.}},
\oauthor{\bsnm{{Drake}}, \binits{J.F.}},
\oauthor{\bsnm{{Swisdak}}, \binits{M.}}:
2021,
{Correlated Spatiotemporal Evolution of Extreme-Ultraviolet Ribbons and Hard
  X-rays in a Solar Flare}.
\textit{arXiv e-prints},
arXiv:2109.15314.
\adsurl{https://ui.adsabs.harvard.edu/abs/2021arXiv210915314N}.
\end{botherref}
\endbibitem

\bibitem[\protect\citeauthoryear{{Pesnell}, {Thompson}, and
  {Chamberlin}}{2012}]{Pesnell2012}
\begin{barticle}
\bauthor{\bsnm{{Pesnell}}, \binits{W.D.}},
\bauthor{\bsnm{{Thompson}}, \binits{B.J.}},
\bauthor{\bsnm{{Chamberlin}}, \binits{P.C.}}:
\byear{2012},
\batitle{{The Solar Dynamics Observatory (SDO)}}.
\bjtitle{\solphys}
\bvolume{275}(\bissue{1-2}),
\bfpage{3}.
\doiurl{10.1007/s11207-011-9841-3}.
\adsurl{https://ui.adsabs.harvard.edu/abs/2012SoPh..275....3P}.
\end{barticle}
\endbibitem

\bibitem[\protect\citeauthoryear{{Poletto} and {Kopp}}{1986}]{Poletto1986}
\begin{bchapter}
\bauthor{\bsnm{{Poletto}}, \binits{G.}},
\bauthor{\bsnm{{Kopp}}, \binits{R.A.}}:
\byear{1986},
\bctitle{{Macroscopic electric fields during two-ribbon flares}}.
In: \beditor{\bsnm{{Neidig}}, \binits{D.F.}} (ed.)
\bbtitle{The lower atmosphere of solar flares; Proceedings of the Solar Maximum
  Mission Symposium, Sunspot, NM, Aug. 20-24, 1985 (A87-26201 10-92). Sunspot,
  NM, National Solar Observatory, 1986, p. 453-465. DOE-sponsored research.},
\bfpage{453}.
\adsurl{1986lasf.conf..453P}.
\end{bchapter}
\endbibitem

\bibitem[\protect\citeauthoryear{{Qiu}}{2009}]{Qiu2009}
\begin{barticle}
\bauthor{\bsnm{{Qiu}}, \binits{J.}}:
\byear{2009},
\batitle{{Observational Analysis of Magnetic Reconnection Sequence}}.
\bjtitle{\apj}
\bvolume{692},
\bfpage{1110}.
\doiurl{10.1088/0004-637X/692/2/1110}.
\adsurl{2009ApJ...692.1110Q}.
\end{barticle}
\endbibitem

\bibitem[\protect\citeauthoryear{{Qiu}}{2021}]{Qiu2021}
\begin{barticle}
\bauthor{\bsnm{{Qiu}}, \binits{J.}}:
\byear{2021},
\batitle{{The Neupert Effect of Flare Ultraviolet and Soft X-Ray Emissions}}.
\bjtitle{\apj}
\bvolume{909}(\bissue{2}),
\bfpage{99}.
\doiurl{10.3847/1538-4357/abe0b3}.
\adsurl{https://ui.adsabs.harvard.edu/abs/2021ApJ...909...99Q}.
\end{barticle}
\endbibitem

\bibitem[\protect\citeauthoryear{{Qiu}, {Liu}, and {Longcope}}{2012}]{Qiu2012}
\begin{barticle}
\bauthor{\bsnm{{Qiu}}, \binits{J.}},
\bauthor{\bsnm{{Liu}}, \binits{W.-J.}},
\bauthor{\bsnm{{Longcope}}, \binits{D.W.}}:
\byear{2012},
\batitle{{Heating of Flare Loops with Observationally Constrained Heating
  Functions}}.
\bjtitle{\apj}
\bvolume{752}(\bissue{2}),
\bfpage{124}.
\doiurl{10.1088/0004-637X/752/2/124}.
\adsurl{https://ui.adsabs.harvard.edu/abs/2012ApJ...752..124Q}.
\end{barticle}
\endbibitem

\bibitem[\protect\citeauthoryear{{Qiu} \textit{et~al.}}{2002}]{Qiu2002}
\begin{barticle}
\bauthor{\bsnm{{Qiu}}, \binits{J.}},
\bauthor{\bsnm{{Lee}}, \binits{J.}},
\bauthor{\bsnm{{Gary}}, \binits{D.E.}},
\bauthor{\bsnm{{Wang}}, \binits{H.}}:
\byear{2002},
\batitle{{Motion of Flare Footpoint Emission and Inferred Electric Field in
  Reconnecting Current Sheets}}.
\bjtitle{\apj}
\bvolume{565},
\bfpage{1335}.
\doiurl{10.1086/324706}.
\adsurl{2002ApJ...565.1335Q}.
\end{barticle}
\endbibitem

\bibitem[\protect\citeauthoryear{{Qiu} \textit{et~al.}}{2004}]{Qiu2004}
\begin{barticle}
\bauthor{\bsnm{{Qiu}}, \binits{J.}},
\bauthor{\bsnm{{Wang}}, \binits{H.}},
\bauthor{\bsnm{{Cheng}}, \binits{C.Z.}},
\bauthor{\bsnm{{Gary}}, \binits{D.E.}}:
\byear{2004},
\batitle{{Magnetic Reconnection and Mass Acceleration in Flare-Coronal Mass
  Ejection Events}}.
\bjtitle{\apj}
\bvolume{604}(\bissue{2}),
\bfpage{900}.
\doiurl{10.1086/382122}.
\adsurl{https://ui.adsabs.harvard.edu/abs/2004ApJ...604..900Q}.
\end{barticle}
\endbibitem

\bibitem[\protect\citeauthoryear{{Qiu} \textit{et~al.}}{2007}]{Qiu2007}
\begin{barticle}
\bauthor{\bsnm{{Qiu}}, \binits{J.}},
\bauthor{\bsnm{{Hu}}, \binits{Q.}},
\bauthor{\bsnm{{Howard}}, \binits{T.A.}},
\bauthor{\bsnm{{Yurchyshyn}}, \binits{V.B.}}:
\byear{2007},
\batitle{{On the Magnetic Flux Budget in Low-Corona Magnetic Reconnection and
  Interplanetary Coronal Mass Ejections}}.
\bjtitle{\apj}
\bvolume{659}(\bissue{1}),
\bfpage{758}.
\doiurl{10.1086/512060}.
\adsurl{https://ui.adsabs.harvard.edu/abs/2007ApJ...659..758Q}.
\end{barticle}
\endbibitem

\bibitem[\protect\citeauthoryear{{Qiu} \textit{et~al.}}{2010}]{Qiu2010}
\begin{barticle}
\bauthor{\bsnm{{Qiu}}, \binits{J.}},
\bauthor{\bsnm{{Liu}}, \binits{W.}},
\bauthor{\bsnm{{Hill}}, \binits{N.}},
\bauthor{\bsnm{{Kazachenko}}, \binits{M.}}:
\byear{2010},
\batitle{{Reconnection and Energetics in Two-ribbon Flares: A Revisit of the
  Bastille-day Flare}}.
\bjtitle{\apj}
\bvolume{725},
\bfpage{319}.
\doiurl{10.1088/0004-637X/725/1/319}.
\adsurl{2010ApJ...725..319Q}.
\end{barticle}
\endbibitem

\bibitem[\protect\citeauthoryear{{Qiu} \textit{et~al.}}{2017}]{Qiu2017}
\begin{barticle}
\bauthor{\bsnm{{Qiu}}, \binits{J.}},
\bauthor{\bsnm{{Longcope}}, \binits{D.W.}},
\bauthor{\bsnm{{Cassak}}, \binits{P.A.}},
\bauthor{\bsnm{{Priest}}, \binits{E.R.}}:
\byear{2017},
\batitle{{Elongation of Flare Ribbons}}.
\bjtitle{\apj}
\bvolume{838}(\bissue{1}),
\bfpage{17}.
\doiurl{10.3847/1538-4357/aa6341}.
\adsurl{https://ui.adsabs.harvard.edu/abs/2017ApJ...838...17Q}.
\end{barticle}
\endbibitem

\bibitem[\protect\citeauthoryear{{Saba}, {Gaeng}, and
  {Tarbell}}{2006}]{Saba2006}
\begin{barticle}
\bauthor{\bsnm{{Saba}}, \binits{J.L.R.}},
\bauthor{\bsnm{{Gaeng}}, \binits{T.}},
\bauthor{\bsnm{{Tarbell}}, \binits{T.D.}}:
\byear{2006},
\batitle{{Analysis of Solar Flare Ribbon Evolution: A Semiautomated Approach}}.
\bjtitle{\apj}
\bvolume{641}(\bissue{2}),
\bfpage{1197}.
\doiurl{10.1086/500631}.
\adsurl{https://ui.adsabs.harvard.edu/abs/2006ApJ...641.1197S}.
\end{barticle}
\endbibitem

\bibitem[\protect\citeauthoryear{{Schou} \textit{et~al.}}{2012}]{Schou2012}
\begin{barticle}
\bauthor{\bsnm{{Schou}}, \binits{J.}},
\bauthor{\bsnm{{Scherrer}}, \binits{P.H.}},
\bauthor{\bsnm{{Bush}}, \binits{R.I.}},
\bauthor{\bsnm{{Wachter}}, \binits{R.}},
\bauthor{\bsnm{{Couvidat}}, \binits{S.}},
\bauthor{\bsnm{{Rabello-Soares}}, \binits{M.C.}},
\bauthor{\bsnm{{Bogart}}, \binits{R.S.}},
\bauthor{\bsnm{{Hoeksema}}, \binits{J.T.}},
\bauthor{\bsnm{{Liu}}, \binits{Y.}},
\bauthor{\bsnm{{Duvall}}, \binits{T.L.}},
\bauthor{\bsnm{{Akin}}, \binits{D.J.}},
\bauthor{\bsnm{{Allard}}, \binits{B.A.}},
\bauthor{\bsnm{{Miles}}, \binits{J.W.}},
\bauthor{\bsnm{{Rairden}}, \binits{R.}},
\bauthor{\bsnm{{Shine}}, \binits{R.A.}},
\bauthor{\bsnm{{Tarbell}}, \binits{T.D.}},
\bauthor{\bsnm{{Title}}, \binits{A.M.}},
\bauthor{\bsnm{{Wolfson}}, \binits{C.J.}},
\bauthor{\bsnm{{Elmore}}, \binits{D.F.}},
\bauthor{\bsnm{{Norton}}, \binits{A.A.}},
\bauthor{\bsnm{{Tomczyk}}, \binits{S.}}:
\byear{2012},
\batitle{{Design and Ground Calibration of the Helioseismic and Magnetic Imager
  (HMI) Instrument on the Solar Dynamics Observatory (SDO)}}.
\bjtitle{\solphys}
\bvolume{275}(\bissue{1-2}),
\bfpage{229}.
\doiurl{10.1007/s11207-011-9842-2}.
\adsurl{https://ui.adsabs.harvard.edu/abs/2012SoPh..275..229S}.
\end{barticle}
\endbibitem

\bibitem[\protect\citeauthoryear{{Schwartz}
  \textit{et~al.}}{1992}]{Schwartz1992}
\begin{bchapter}
\bauthor{\bsnm{{Schwartz}}, \binits{R.A.}},
\bauthor{\bsnm{{Dennis}}, \binits{B.R.}},
\bauthor{\bsnm{{Fishman}}, \binits{G.J.}},
\bauthor{\bsnm{{Meegan}}, \binits{C.A.}},
\bauthor{\bsnm{{Wilson}}, \binits{R.B.}},
\bauthor{\bsnm{{Paciesas}}, \binits{W.S.}}:
\byear{1992},
\bctitle{{BATSE flare observations in solar cycle 22.}}
In: \bbtitle{NASA Conference Publication},
\bsertitle{NASA Conference Publication}
\bseriesno{3137},
\bfpage{457}.
\adsurl{https://ui.adsabs.harvard.edu/abs/1992NASCP3137..457S}.
\end{bchapter}
\endbibitem

\bibitem[\protect\citeauthoryear{{Sharykin}
  \textit{et~al.}}{2018}]{Sharykin2018}
\begin{barticle}
\bauthor{\bsnm{{Sharykin}}, \binits{I.N.}},
\bauthor{\bsnm{{Zimovets}}, \binits{I.V.}},
\bauthor{\bsnm{{Myshyakov}}, \binits{I.I.}},
\bauthor{\bsnm{{Meshalkina}}, \binits{N.S.}}:
\byear{2018},
\batitle{{Flare Energy Release at the Magnetic Field Polarity Inversion Line
  during the M1.2 Solar Flare of 2015 March 15. I. Onset of Plasma Heating and
  Electron Acceleration}}.
\bjtitle{\apj}
\bvolume{864}(\bissue{2}),
\bfpage{156}.
\doiurl{10.3847/1538-4357/aada15}.
\adsurl{https://ui.adsabs.harvard.edu/abs/2018ApJ...864..156S}.
\end{barticle}
\endbibitem

\bibitem[\protect\citeauthoryear{{Su}, {Golub}, and {Van
  Ballegooijen}}{2007}]{Su2007}
\begin{barticle}
\bauthor{\bsnm{{Su}}, \binits{Y.}},
\bauthor{\bsnm{{Golub}}, \binits{L.}},
\bauthor{\bsnm{{Van Ballegooijen}}, \binits{A.A.}}:
\byear{2007},
\batitle{{A Statistical Study of Shear Motion of the Footpoints in Two-Ribbon
  Flares}}.
\bjtitle{\apj}
\bvolume{655},
\bfpage{606}.
\doiurl{10.1086/510065}.
\adsurl{2007ApJ...655..606S}.
\end{barticle}
\endbibitem

\bibitem[\protect\citeauthoryear{{Su} \textit{et~al.}}{2006}]{Su2006}
\begin{barticle}
\bauthor{\bsnm{{Su}}, \binits{Y.N.}},
\bauthor{\bsnm{{Golub}}, \binits{L.}},
\bauthor{\bsnm{{van Ballegooijen}}, \binits{A.A.}},
\bauthor{\bsnm{{Gros}}, \binits{M.}}:
\byear{2006},
\batitle{{Analysis of Magnetic Shear in An X17 Solar Flare on October 28,
  2003}}.
\bjtitle{\solphys}
\bvolume{236}(\bissue{2}),
\bfpage{325}.
\doiurl{10.1007/s11207-006-0039-z}.
\adsurl{https://ui.adsabs.harvard.edu/abs/2006SoPh..236..325S}.
\end{barticle}
\endbibitem

\bibitem[\protect\citeauthoryear{{Temmer} \textit{et~al.}}{2007}]{Temmer2007}
\begin{barticle}
\bauthor{\bsnm{{Temmer}}, \binits{M.}},
\bauthor{\bsnm{{Veronig}}, \binits{A.M.}},
\bauthor{\bsnm{{Vr{\v s}nak}}, \binits{B.}},
\bauthor{\bsnm{{Miklenic}}, \binits{C.}}:
\byear{2007},
\batitle{{Energy Release Rates along H{$\alpha$} Flare Ribbons and the Location
  of Hard X-Ray Sources}}.
\bjtitle{\apj}
\bvolume{654},
\bfpage{665}.
\doiurl{10.1086/509634}.
\adsurl{2007ApJ...654..665T}.
\end{barticle}
\endbibitem

\bibitem[\protect\citeauthoryear{{Wang} \textit{et~al.}}{2017a}]{Wang2017b}
\begin{barticle}
\bauthor{\bsnm{{Wang}}, \binits{J.}},
\bauthor{\bsnm{{Yan}}, \binits{X.}},
\bauthor{\bsnm{{Qu}}, \binits{Z.}},
\bauthor{\bsnm{{Xue}}, \binits{Z.}},
\bauthor{\bsnm{{Yang}}, \binits{L.}}:
\byear{2017}a,
\batitle{{Formation and Eruption Process of a Filament in Active Region NOAA
  12241}}.
\bjtitle{\apj}
\bvolume{839}(\bissue{2}),
\bfpage{128}.
\doiurl{10.3847/1538-4357/aa6bf3}.
\adsurl{https://ui.adsabs.harvard.edu/abs/2017ApJ...839..128W}.
\end{barticle}
\endbibitem

\bibitem[\protect\citeauthoryear{{Wang} \textit{et~al.}}{2017b}]{Wang2017a}
\begin{barticle}
\bauthor{\bsnm{{Wang}}, \binits{J.}},
\bauthor{\bsnm{{Sim{\~o}es}}, \binits{P.J.A.}},
\bauthor{\bsnm{{Jeffrey}}, \binits{N.L.S.}},
\bauthor{\bsnm{{Fletcher}}, \binits{L.}},
\bauthor{\bsnm{{Wright}}, \binits{P.J.}},
\bauthor{\bsnm{{Hannah}}, \binits{I.G.}}:
\byear{2017}b,
\batitle{{Observations of Reconnection Flows in a Flare on the Solar Disk}}.
\bjtitle{\apjl}
\bvolume{847}(\bissue{1}),
\bfpage{L1}.
\doiurl{10.3847/2041-8213/aa8904}.
\adsurl{https://ui.adsabs.harvard.edu/abs/2017ApJ...847L...1W}.
\end{barticle}
\endbibitem

\bibitem[\protect\citeauthoryear{{Warmuth} and {Mann}}{2020}]{Warmuth2020}
\begin{barticle}
\bauthor{\bsnm{{Warmuth}}, \binits{A.}},
\bauthor{\bsnm{{Mann}}, \binits{G.}}:
\byear{2020},
\batitle{{Thermal-nonthermal energy partition in solar flares derived from
  X-ray, EUV, and bolometric observations. Discussion of recent studies}}.
\bjtitle{\aap}
\bvolume{644},
\bfpage{A172}.
\doiurl{10.1051/0004-6361/202039529}.
\adsurl{https://ui.adsabs.harvard.edu/abs/2020A&A...644A.172W}.
\end{barticle}
\endbibitem

\bibitem[\protect\citeauthoryear{{Warren} and {Warshall}}{2001}]{Warren2001}
\begin{barticle}
\bauthor{\bsnm{{Warren}}, \binits{H.P.}},
\bauthor{\bsnm{{Warshall}}, \binits{A.D.}}:
\byear{2001},
\batitle{{Ultraviolet Flare Ribbon Brightenings and the Onset of Hard X-Ray
  Emission}}.
\bjtitle{\apjl}
\bvolume{560}(\bissue{1}),
\bfpage{L87}.
\doiurl{10.1086/324060}.
\adsurl{https://ui.adsabs.harvard.edu/abs/2001ApJ...560L..87W}.
\end{barticle}
\endbibitem

\bibitem[\protect\citeauthoryear{{Xu} \textit{et~al.}}{2004}]{Xu2004}
\begin{barticle}
\bauthor{\bsnm{{Xu}}, \binits{Y.}},
\bauthor{\bsnm{{Cao}}, \binits{W.}},
\bauthor{\bsnm{{Liu}}, \binits{C.}},
\bauthor{\bsnm{{Yang}}, \binits{G.}},
\bauthor{\bsnm{{Qiu}}, \binits{J.}},
\bauthor{\bsnm{{Jing}}, \binits{J.}},
\bauthor{\bsnm{{Denker}}, \binits{C.}},
\bauthor{\bsnm{{Wang}}, \binits{H.}}:
\byear{2004},
\batitle{{Near-Infrared Observations at 1.56 Microns of the 2003 October 29 X10
  White-Light Flare}}.
\bjtitle{\apjl}
\bvolume{607}(\bissue{2}),
\bfpage{L131}.
\doiurl{10.1086/422099}.
\adsurl{https://ui.adsabs.harvard.edu/abs/2004ApJ...607L.131X}.
\end{barticle}
\endbibitem

\bibitem[\protect\citeauthoryear{{Yang} \textit{et~al.}}{2009}]{Yang2009}
\begin{barticle}
\bauthor{\bsnm{{Yang}}, \binits{Y.-H.}},
\bauthor{\bsnm{{Cheng}}, \binits{C.Z.}},
\bauthor{\bsnm{{Krucker}}, \binits{S.}},
\bauthor{\bsnm{{Lin}}, \binits{R.P.}},
\bauthor{\bsnm{{Ip}}, \binits{W.H.}}:
\byear{2009},
\batitle{{A Statistical Study of Hard X-Ray Footpoint Motions in Large Solar
  Flares}}.
\bjtitle{\apj}
\bvolume{693},
\bfpage{132}.
\doiurl{10.1088/0004-637X/693/1/132}.
\adsurl{2009ApJ...693..132Y}.
\end{barticle}
\endbibitem

\bibitem[\protect\citeauthoryear{{Yang} \textit{et~al.}}{2011}]{Yang2011}
\begin{barticle}
\bauthor{\bsnm{{Yang}}, \binits{Y.-H.}},
\bauthor{\bsnm{{Cheng}}, \binits{C.Z.}},
\bauthor{\bsnm{{Krucker}}, \binits{S.}},
\bauthor{\bsnm{{Hsieh}}, \binits{M.-S.}}:
\byear{2011},
\batitle{{Estimation of the Reconnection Electric Field in the 2003 October 29
  X10 Flare}}.
\bjtitle{\apj}
\bvolume{732}(\bissue{1}),
\bfpage{15}.
\doiurl{10.1088/0004-637X/732/1/15}.
\adsurl{https://ui.adsabs.harvard.edu/abs/2011ApJ...732...15Y}.
\end{barticle}
\endbibitem

\bibitem[\protect\citeauthoryear{{Zhong} \textit{et~al.}}{2019}]{Zhong2019}
\begin{barticle}
\bauthor{\bsnm{{Zhong}}, \binits{Z.}},
\bauthor{\bsnm{{Guo}}, \binits{Y.}},
\bauthor{\bsnm{{Ding}}, \binits{M.D.}},
\bauthor{\bsnm{{Fang}}, \binits{C.}},
\bauthor{\bsnm{{Hao}}, \binits{Q.}}:
\byear{2019},
\batitle{{Transition from Circular-ribbon to Parallel-ribbon Flares Associated
  with a Bifurcated Magnetic Flux Rope}}.
\bjtitle{\apj}
\bvolume{871}(\bissue{1}),
\bfpage{105}.
\doiurl{10.3847/1538-4357/aaf863}.
\adsurl{https://ui.adsabs.harvard.edu/abs/2019ApJ...871..105Z}.
\end{barticle}
\endbibitem

\bibitem[\protect\citeauthoryear{{Zhu}, {Qiu}, and {Longcope}}{2018}]{Zhu2018}
\begin{barticle}
\bauthor{\bsnm{{Zhu}}, \binits{C.}},
\bauthor{\bsnm{{Qiu}}, \binits{J.}},
\bauthor{\bsnm{{Longcope}}, \binits{D.W.}}:
\byear{2018},
\batitle{{Two-phase Heating in Flaring Loops}}.
\bjtitle{\apj}
\bvolume{856}(\bissue{1}),
\bfpage{27}.
\doiurl{10.3847/1538-4357/aaad10}.
\adsurl{https://ui.adsabs.harvard.edu/abs/2018ApJ...856...27Z}.
\end{barticle}
\endbibitem

\bibitem[\protect\citeauthoryear{{Zimovets}, {Sharykin}, and
  {Gan}}{2020}]{Zimovets2020}
\begin{barticle}
\bauthor{\bsnm{{Zimovets}}, \binits{I.V.}},
\bauthor{\bsnm{{Sharykin}}, \binits{I.N.}},
\bauthor{\bsnm{{Gan}}, \binits{W.Q.}}:
\byear{2020},
\batitle{{Relationships between Photospheric Vertical Electric Currents and
  Hard X-Ray Sources in Solar Flares: Statistical Study}}.
\bjtitle{\apj}
\bvolume{891}(\bissue{2}),
\bfpage{138}.
\doiurl{10.3847/1538-4357/ab75be}.
\adsurl{https://ui.adsabs.harvard.edu/abs/2020ApJ...891..138Z}.
\end{barticle}
\endbibitem

\end{thebibliography}
\end{article}
\end{document}